\documentclass[twocolumn,usenatbib]{mn2e}

\usepackage{amssymb}
\usepackage{subfigure}
\usepackage{graphicx}
\usepackage{graphicx,natbib}
\usepackage{multirow}
\usepackage{natbib}
\usepackage{amssymb,amsmath}

\usepackage{tablefootnote}

\usepackage{lineno}
\usepackage{epstopdf}
\usepackage{appendix}
\usepackage{color}

\usepackage{multicol}        
\usepackage{bm}		
\usepackage{pdflscape}	
\usepackage{supertabular}	

\title[Variability Monitoring of Blazars]{Multicolour Optical Variability Monitoring of Blazars with High Time Resolution}
\author[X. Chang et al.]{\parbox{\linewidth}{
X. Chang,$^{1}$ T. F. Yi,$^{2, 3}$\thanks{E-mail: yitingfeng98@163.com} D. R. Xiong,$^{4}$\thanks{E-mail: xiongdingrong@ynao.ac.cn} C. X. Liu,$^{1}$
X. Yang,$^{3, 5}$  H. Z. Li,$^{6}$  Y. L. Gong,$^{1}$ W. W. Na,$^{6}$  Y. Li,$^{2}$ Z. H. Chen,$^{2}$ J. P. Chen,$^{2}$  L. S. Mao$^{2}$}\\\\
$^{1}$ South-Western Institute for Astronomy Research, Yunnan University, Kunming 650500, People's Republic of China\\
$^{2}$ Key Laboratory of Colleges and Universities in Yunnan Province for High-energy Astrophysics, Department of Physics, Yunnan Normal University, \\Kunming 650500, People's Republic of China\\
$^{3}$ Guangxi Key Laboratory for the Relativistic Astrophysics, Nanning 530004, People's Republic of China\\
$^{4}$ Yunnan Observatories, Chinese Academy of Sciences, 396 Yangfangwang, Guandu District, Kunming, 650216, People's Republic of China\\
$^{5}$ GXU-NAOC Center for Astrophysics and Space Sciences, Department of Physics, Guangxi University, Nanning 530004, People's Republic of China\\
$^{6}$ Physics Department, Yuxi Normal University, Yuxi 653100, People's Republic of China\\
}
\begin{document}
\maketitle

\begin{abstract}
We carried out a high time-resolution, multicolour optical observing campaign for eight $\gamma$-ray detected blazars during 2010-2020. We analyze flux variations, correlations between magnitudes and colours on different timescales. Intraday variability (IDV) is detected in all eight sources of our sample. A bluer-when-brighter (BWB) chromatic trend is dominant on intraday timescales. On the short timescales, the BWB trend only shows up in ON 231, 3C 279, BL Lacertae and 1E 1458.8+2249. There is a BWB trend in 3C 279 on the long timescale. We estimate the upper limits of black hole mass for three blazars (i.e. ON 321, 1ES 1426+42.8, PKS 1510-089) using variability timescales. On April 13, 2010 a potential quasi-periodic oscillation (QPO) with the period of $P=48.67\pm13.90$ minutes is found in 1ES 1426+42.8. The light curve on March 16, 2021 further shows the existence of the QPO phenomenon. The QPO in this target deserves further observation and confirmation.

\end{abstract}


\section{Introduction}
 Active galactic nuclei (AGNs) are the brightest long-live objects in the Universe, which are powered by the accretion process of Super-Massive Black Holes (SMBHs) in the center of galaxies \citep{Espo15,Xion17}. Blazars are the most extreme subclass of AGNs, with relativistic jets pointing in the direction of the observer. They are characterized by rapid optical variation, high luminosity, high and variable polarization, apparent superluminal motion, non-thermal continuous radiation, and high-energy gamma-ray radiation \citep{Urry95}. The two sub-classes of blazars are BL Lacertae objects (BL Lacs) and flat-spectrum radio quasars (FSRQs). BL Lacs have featureless optical spectra, while FSRQs have broad emission lines in their optical spectra \citep{Stoc91,Marc96}. The topical spectral energy distributions (SEDs) of blazars have two peaks \citep{Foss98}. The low-frequency peak ranges from radio to UV or X-ray, and the high-frequency peak extends from X-ray to $\gamma$-ray \citep{Abdo10a}. The former is dominated by the synchrotron emission of relativistic electrons, while the latter can be explained by the inverse Compton scattering (SSC or EC), synchrotron radiation of protons, lepto-hadronic (P$\gamma$) models, or hadronic (PP) models \citep[e.g., ][]{Ghis10,Derm12,Bott13}. According to different positions of synchrotron peak frequencies, blazars can be further divided into high-synchrotron-peaked blazars (HSP), intermediate-synchrotron-peaked blazars (ISP), and low-synchrotron-peaked blazars (LSP) \citep{Abdo10a}.

 The variability is a useful tool to explore the nature of blazars. The variability of blazars can be broadly divided into intraday variability (IDV), short-term variability (STV), and long-term variability (LTV). Variations in the flux of a few tenths or hundredths of magnitudes over a timescale of tens of minutes to a few hours are often called as IDV \citep{Wagn95}. The timescale of STV ranges from days to months, and LTV ranges from months to years \citep{Gupt08a,Dai15}. Some blazars show high amplitudes of variability on timescales as short as several minutes in different wavebands \citep{Dai01,Xie02,Saga04,Fan09}. The minimum variability timescale is related to the mass of the central black hole. Therefore, the mass of black holes can be constrained by the observed minimum timescales of rapid variations in the optical regimes \citep{Liu15}. Observations of the IDV can be used to derive the central sizes and the black hole masses in blazars \citep{Gupt09}. Models related to jets and accretion disks have been proposed to explain the variability behaviors under different timescales, but many details of the models are still under discussion \citep{Bhat21}.

The optical brightness/flux variations are associated with the colour/spectral variations for blazars \citep{Gu11,Xion16}. The correlations between magnitudes and colours are often used to constrain the origin of flux variations \citep{Agar16}. The bluer-when-brighter (BWB) trend was found in most BL Lacs, while FSRQs show the redder-when-brighter (RWB) trend \citep{Gu06}. There are even more complicated correlations than the BWB trend and RWB trend between magnitudes and colours \citep[e.g., ][]{Bonn12}. However, a consolidated framework of correlations between magnitudes and colours in different blazars has not yet been established. More multi-colour observations are still needed to study the correlations between magnitudes and colours.

The periodic or quasi-periodic oscillations (QPOs) of flux variations were reported in some blazars \citep{Sill88}. The QPOs of IDV were only detected in a few blazars \citep{Gupt09,Hong18}. \citet{Urry93} found that PKS 2155-304 showed a potential QPO in 0.7 day in UV and optical bands. \citet{Lach09} reported a QPO of $\sim$4.6 hours in the XMM-Newton X-ray light curve of PKS 2155-304. \citet{Espa08} reported a QPO of $\sim$55 minutes in the XMM-Newton light curve of the quasar 3C 273. \citet{Gupt09} detected high probabilities QPOs of IDV with timescales between $\sim$25 and $\sim$73 minutes for S5 0716+714, which was the first good evidence for quasi-periodic components in the optical IDV of blazars. \citet{Hong18} also found that S5 0716+714 had a possible QPO of about 50 minutes at 99\% confidence level. If the observed timescale of periodic variability indicates an innermost stable orbital period from the accretion disk, then the QPOs of IDV can be used to estimate/limit the black hole mass \citep[e.g., ][]{Gupt09,Dai15}. In order to find more QPOs of IDV and further understand them, new optical intraday observations need to be carried out.

In this work, we present multi-colour photometric data for a sample of eight sources from 2010 to 2020 with high time-resolution. We analyze flux variations, correlations between magnitudes and colours on different timescales. QPOs of IDV are detected in one of the eight blazars. Considering the available telescope time and the observation limitations of telescope, we chose the eight sources as our observed sample. All of the eight sources were detected with gamma-ray radiation \citep{Ajel20}. The sample includes all subclasses of blazars. Seven sources have been detected with TeV radiation (see Table 1) and the remaining one target is included in the third Fermi-LAT catalog of high-energy sources (3FHL, $>$ 100 GeV Fermi-LAT event)\citep{Ajel17}. Our sample in $R$ band ranges from 12 to 17.

 This paper is organized as follows. Section 2 describes the observations and data reduction. Section 3 presents the results. Discussion and conclusions are reported in Sections 4 and 5.

\section{Observations and Data Reduction}
\label{sect:Observations and Data Reduction}
We observed the eight blazars with the 1.02 m optical telescope administered by Yunnan Astronomical Observatories (YNAO) of China. During our observation periods, the 1.02 m telescope was equipped with an Andor DW436 CCD (2048 $\times$ 2048 pixels) camera at the Cassegrain focus ($f$ = 13.3 m). The field of view (FOV) of the CCD image is 7.3 $\times$ 7.3 arcmin$^2$. The pixel scale is 0.21 arcsec/pixel \citep{Dai15,Xion16}. The readout noise and gain are 6.33 electrons/pixel and 2.0 electrons/ADU, respectively \citep{Dai15,Liao14}. The standard Johnson broadband filters are used for all frames.

Our multi-band photometry observations are performed through a cyclic mode in the $V$, $R$ and $I$ bands. The exposure time ranges from 1 to 6 minutes. Therefore, the data can be considered as quasi-simultaneous measurements ($<$10 minutes). Different exposure time is set depending on seeing, weather conditions, and the brightness of sources. The twilight sky flat-field images are taken in good weather conditions. Several bias frames are taken at the beginning of the night's observation. The data processing is carried out by using standard Image Reduction and Analysis Facility (IRAF\footnote{IRAF is distributed by the National Optical Astronomy Observatories, which are operated by the Association of Universities for Research in Astronomy, Inc., under cooperative agreement with the National Science Foundation.}) software. Aperture photometry is carried out by using APPHOT task of the IRAF after the flat-field and bias are corrected. We tried different photometry apertures every night and then selected the best aperture radius from 0.5-2.0 Full Width at Half Maximum (FWHM) to obtain the best signal-to-noise ratio. For all of the eight blazars, at least three local comparison stars are required in the same frame. The magnitudes and finding charts of comparison stars are obtained from the Web page (Finding Charts for AGN\footnote{https://www.lsw.uni-heidelberg.de/projects/extragalactic/charts/}). A comparison star with colours similar to the blazar is chosen for flux calibration. We transformed the instrumental magnitude of the blazar to the apparent magnitude using the differential photometry \citep{Bai98,Zhan04,Fan14}. The star with the smallest variation in differential magnitudes compared to the comparison star is chosen as the check star. The rms errors of the photometry of a specific night are derived based on the comparison star and the check star as follows:
\begin{equation}
\sigma=\sqrt\frac{\sum^N_{i=1}(m_i-\overline{m})^2}{N-1},
\label{eq:LebsequeIp1}
\end{equation}
where ${m_i}$ is the differential magnitude of the comparison star and check star, $\overline {m}$ is the average differential magnitude for one night, and $N$ is the number of observations in a given night. Intraday variability amplitude (Amp) is introduced by \citet{Heid96}, and defined as:
\begin{equation}
 Amp=100\times\sqrt{(A_{max}-A_{min})^2-2\sigma^2} \; percent,
\label{eq:Lebseque2}
\end{equation}
where $A_{max}$ and $A_{min}$ are the maximum and minimum magnitudes of the light curve for the night being considered, respectively, and $\sigma$ is the rms error.

We obtained the optical multi-band observation data of eight targets in a total of 36 nights from 2010 to 2020. Table 1 lists the names, redshifts, types and TeV radiations. The light curves of all eight sources in IDV, STV, and LTV are given in Figures 1-3, respectively. Our observational data is listed in Table 2.

\begin{table*}
   \centering
   \caption{List of the sources.}
   \label{tab:tab1}
   \renewcommand\arraystretch{1.2}
  \setlength{\tabcolsep}{8.2mm}
   \begin{tabular}{cccccc} 
      \hline
Object & Other name & z & SED type & Optical type & TeV\\\hline
0851+203 &	OJ 287 & 0.306 & LSP& BLLac & Yes\\
1219+285 &	ON 231 & 0.102 & ISP & BLLac & Yes\\
1253-055 &	3C 279 & 0.538 & LSP & FSRQ & Yes\\
1426+428 &	1ES 1426+42.8 & 0.129 & HSP& BLLac & Yes\\
1458+228 &	1E 1458.8+2249 & 0.235 & ISP & BLLac & No\\
1510-089 &	PKS 1510-089 & 0.361 & LSP & FSRQ & Yes\\
1727+502 &	OT 546 & 0.055 & HSP& BLLac & Yes\\
2200+420 &	BL Lacertae & 0.069 & LSP& BLLac & Yes\\\hline
\multicolumn{4}{l}{}
   \end{tabular}
\end{table*}

\begin{table*}
   \centering
   \caption{The log of observations.}
   \label{tab:tab2}
   \renewcommand\arraystretch{1.2}
  \setlength{\tabcolsep}{6.2mm}
   \begin{tabular}{cccccc} 
      \hline
Object & Date (UT) & JD & Mag & $\sigma$ & Band\\\hline
OJ 287 & 2013 Apr 03	&	2456386.01409	&	14.141	&	0.020	&	I	\\
& 2013 Apr 03	&	2456386.02665	&	14.181	&	0.020&	I	\\
... &  ...	&	...	&	...	&	...	&	...	\\
ON 231 & 2014 May 08	&	2456786.04872	&	14.924	&	0.016	&	I	\\
& 2014 May 08	&	2456786.05225	&	14.906	&	0.016	&	I	\\
... &  ...	&	...	&	...	&	...	&	...	\\
3C 279	&	2016 May 07	&	2457516.04973	&	15.101	&	0.016	&	I	\\
	&	2016 May 07	&	2457516.05574	&	15.058	&	0.016	&	I	\\
... &  ...	&	...	&	...	&	...	&	...	\\
1ES 1426+42.8 &	2010 Apr 13	&	2455300.16457	&	15.609	&	0.004	&	I	\\
& 2010 Apr 13	&	2455300.16811	&	15.645	&	0.004	&	I	\\
... &  ...	&	...	&	...	&	...	&	...	\\
1E 1458.8+2249 & 2011 May 26	&	2455708.03137	&	16.947	&	0.026	&	I	\\
& 2011 May 26	&	2455708.03755	&	16.744	&	0.026&	I	\\
... &  ...	&	...	&	...	&	...	&	...	\\
PKS 1510-089	&	2010 Apr 13	&	2455300.30559	&	16.405	&	0.010	&	R	\\
	&	2010 Apr 13	&	2455300.31144	&	16.620	&	0.010	&	R	\\
... &  ...	&	...	&	...	&	...	&	...	\\
OT 546 & 2010 May 06	&	2455323.35508	&	14.991	&	0.013	&	I	\\
& 2010 May 06	&	2455323.35772	&	14.944	&	0.013	&	I	\\
... &  ...	&	...	&	...	&	...	&	...	\\
BL Lacertae & 2019 Sept 16	&	2458742.99949	&	15.075	&	0.015	&	B	\\
& 2019 Sept 16	&	2458743.00429	&	15.097	&	0.015	&	B	\\
... &  ...	&	...	&	...	&	...	&	...	\\\hline
\multicolumn{6}{l}{Notes:}\\
\multicolumn{6}{l}{Column 1 is the name of the object;}\\
\multicolumn{6}{l}{Column 2 is the universal time (UT) of the observation;}\\
\multicolumn{6}{l}{Column 3 is the corresponding Julian day (JD);}\\
\multicolumn{6}{l}{Column 4 is the magnitude;}\\
\multicolumn{6}{l}{Column 5 is the rms error;}\\
\multicolumn{6}{l}{Column 6 is the observed band.}\\
\multicolumn{6}{l}{(The full Table 2 can be accessed electronically in machine readable format.)}\\
   \end{tabular}
\end{table*}

\begin{table*}
   \centering
   \caption{Results of the IDV analysis.}
   \label{tab:tab1}
   \renewcommand\arraystretch{1.2}
  \setlength{\tabcolsep}{2.2mm}
   \begin{tabular}{ccccccccccccc} 
      \hline
Object & Date	&	Band	&	N	&	Normal	&	F	&	Fc(99)	&	Fa	&	Fa(99)	&	V/N	&	A\%	&	Ave(mag)	\\\hline
OJ 287	&	2013 Apr 03	&	I	&	11	&	Y	&	0.71	&	4.85	&	0.34	&	8.65	&	N	&	2.83	&	14.17		\\
	&	2013 Apr 03	&	R	&	11	&	N	&	1.34	&	4.85	&	4.08	&	8.65	&	N	&	3.33	&	14.76		\\
	&	...	&	...	&	...	&	...	&	...	&	...	&	...	&	...	&	...	&	...	&		...		\\
ON 231	&	2014 May 08	&	I	&	46	&	Y	&	20.01	&	2.01	&	9.55	&	3.04	&	V	&	30.31	&	14.89		\\
	&	2020 Apr 14	&	I	&	57	&	Y	&	27.86	&	1.87	&	11.19	&	2.73	&	V	&	9.98	&	14.68		\\
	&	...	&	...	&	...	&	...	&	...	&	...	&	...	&	...	&	...	&	...	&		...		\\
3C 279	&	2016 May 07	&	I	&	12	&	Y	&	1.03	&	4.46	&	2.02	&	7.56	&	N	&	5.57	&	15.08		\\
	&	2016 May 07	&	R	&	13	&	Y	&	2.17	&	4.16	&	0.30	&	7.21	&	N	&	11.60	&	15.76		\\
	&	...	&	...	&	...	&	...	&	...	&	...	&	...	&	...	&	...	&	...	&		...		\\
1ES 1426+42.8	&	2010 Apr 13	&	I	&	31	&	Y	&	130.61	&	2.39	&	5.33	&	3.86	&	V	&	14.89	&	15.60		\\
	&	...	&	...	&	...	&	...	&	...	&	...	&	...	&	...	&	...	&	...	&		...		\\
1E 1458.8+2249	&	2011 May 26	&	I	&	11	&	Y	&	141.88	&	4.85	&	29.82	&	8.65	&	V	&	96.53	&	16.43		\\
	&	2020 Apr 14	&	I	&	51	&	Y	&	2.07	&	1.95	&	5.48	&	2.88	&	V	&	8.70	&	14.93		\\
	&	...	&	...	&	...	&	...	&	...	&	...	&	...	&	...	&	...	&	...	&		...		\\
PKS 1510-089	&	2010 Apr 13	&	R	&	23	&	Y	&	116.36	&	2.72	&	2.47	&	4.58	&	PV	&	49.28	&	16.55		\\
	&	2010 May 06	&	R	&	45	&	Y	&	257.05	&	2.04	&	8.21	&	3.05	&	V	&	42.39	&	16.60		\\
	&	...	&	...	&	...	&	...	&	...	&	...	&	...	&	...	&	...	&	...	&		...		\\
OT 546	&	2010 May 06	&	I	&	17	&	N	&	8.73	&	3.41	&	0.34	&	5.41	&	N	&	12.06	&	14.97		\\
	&	2010 May 06	&	R	&	16	&	Y	&	3.82	&	3.52	&	1.27	&	5.67	&	PV	&	26.39	&	15.47		\\
	&	...	&	...	&	...	&	...	&	...	&	...	&	...	&	...	&	...	&	...	&		...		\\
BL Lacertae	&	2019 Sep 16	&	B	&	33	&	Y	&	2.60	&	2.32	&	4.68	&	3.79	&	V	&	9.68	&	15.04		\\
	&	2019 Sep 16	&	I	&	30	&	Y	&	0.53	&	2.42	&	3.39	&	3.90	&	N	&	4.43	&	12.65		\\
	&	...	&	...	&	...	&	...	&	...	&	...	&	...	&	...	&	...	&	...	&		...		\\\hline
\multicolumn{12}{l}{Notes:}\\
\multicolumn{12}{l}{Column 1 is the name of the object;}\\
\multicolumn{12}{l}{Column 2 is the date of the observation;}\\
\multicolumn{12}{l}{Column 3 is the observed band;}\\
\multicolumn{12}{l}{Column 4 is the number of data points;}\\
\multicolumn{12}{l}{Column 5 is the result of the normal distribution of the light curve;}\\
\multicolumn{12}{l}{Column 6 is the average $F$ value;}\\
\multicolumn{12}{l}{Column 7 is the critical $F$ value with 99\% confidence level;}\\
\multicolumn{12}{l}{Column 8 is the $F$ value of ANOVA;}\\
\multicolumn{12}{l}{Column 9 is the critical $F$ value of ANOVA with 99\% confidence level;}\\
\multicolumn{12}{l}{Column 10 is the variability status (V: variable, PV: probably variable, N: non-variable);}\\
\multicolumn{12}{l}{Column 11 is the variability amplitude;}\\
\multicolumn{12}{l}{Column 12 is daily average magnitudes.}\\
\multicolumn{12}{l}{(The full Table 3 can be accessed electronically in machine readable format.)}\\
   \end{tabular}
\end{table*}

\begin{table*}
   \centering
   \caption{Results of error-weighted linear regression analysis.}
   \label{tab:tab1}
   \renewcommand\arraystretch{1.2}
  \setlength{\tabcolsep}{6.2mm}
   \begin{tabular}{cccc} 
      \hline
      \hline
Object & Date & r & p \\\hline
OJ 287	&	2013 Apr 03	&	0.930	&	$<0.0001$	\\
OJ 287	&	2020 Mar 14	&	0.821	&	$<0.0001$	\\
OJ 287	&	2020 Mar 15	&	0.501	&	0.0001	\\
OJ 287	&	2020 Mar 16	&	0.786	&	0.0010	\\
ON 231	&	2020 Apr 14	&	0.614	&	$<0.0001$	\\
ON 231	&	2020 Apr 15	&	0.815	&	$<0.0001$	\\
3C 279	&	2016 May 07	&	0.948	&	$<0.0001$	\\
3C 279	&	2016 May 27	&	0.866	&	$<0.0001$	\\
3C 279	&	2016 May 28	&	0.954	&	$<0.0001$	\\
3C 279	&	2016 May 31	&	0.724	&	$<0.0001$	\\
3C 279	&	2018 May 07	&	0.639	&	$<0.0001$	\\
3C 279	&	2018 May 08	&	0.628	&	$<0.0001$	\\
3C 279	&	2018 May 09	&	0.290	&	0.1600	\\
3C 279	&	2018 May 10	&	0.390	&	0.0060	\\
3C 279	&	2018 Jun 06	&	0.839	&	0.0006	\\
3C 279	&	2018 Jun 07	&	0.532	&	0.0900	\\
1E 1458.8+2249	&	2020 Apr 14	&	0.842	&	$<0.0001$	\\
1E 1458.8+2249	&	2020 Apr 15	&	0.798	&	$<0.0001$	\\
PKS 1510-089	&	2013 Apr 02	&	0.816	&	0.0002	\\
PKS 1510-089	&	2016 May 07	&	0.792	&	0.0100	\\
PKS 1510-089	&	2016 May 08	&	0.767	&	0.0160	\\
PKS 1510-089	&	2016 May 09	&	0.796	&	0.0030	\\
OT 546	&	2010 May 06	&	0.740	&	0.0020	\\
OT 546	&	2016 May 07	&	0.510	&	0.0009	\\
OT 546	&	2016 May 08	&	0.677	&	0.0002	\\
OT 546	&	2016 May 09	&	0.562	&	0.0050	\\
OT 546	&	2016 May 12	&	0.418	&	0.1200	\\
BL Lacertae	&	2019 Sep 16	&	0.549	&	0.0030	\\
BL Lacertae	&	2019 Sep 17	&	0.577	&	0.0490	\\
BL Lacertae	&	2019 Sep 19	&	0.670	&	$<0.0001$	\\
OJ 287	&	2020 Mar 13 - 2020 Mar 16	&	0.002	&	0.9830	\\
ON 231	&	2020 Apr 14 - 2020 Apr 15	&	0.852	&	$<0.0001$	\\
3C 279	&	2016 May 07 - 2016 May 31	&	0.740	&	$<0.0001$	\\
3C 279	&	2018 May 07 - 2018 Jun 07	&	0.208	&	0.0040	\\
1E 1458.8+2249	&	2020 Apr 14 - 2020 Apr 15	&	0.628	&	$<0.0001$	\\
PKS 1510-089	&	2016 May 07 - 2016 May 09	&	-0.431	&	0.0170	\\
OT 546	&	2016 May 07 - 2016 May 12	&	0.029	&	0.7740	\\
BL Lacertae	&	2019 Sep 16 - 2019 Sep 19	&	0.897	&	$<0.0001$	\\
3C 279	&	2016 - 2018	&	0.835	&	$<0.0001$	\\\hline
\multicolumn{4}{l}{Notes:}\\
\multicolumn{4}{l}{Column 1 is the name of the object;}\\
\multicolumn{4}{l}{Column 2 is the date of the observation;}\\
\multicolumn{4}{l}{Column 3 is the coefficient of correlation;}\\
\multicolumn{4}{l}{Column 4 is the chance probability.}\\
   \end{tabular}
\end{table*}

\section{Results}
\subsection{Variability Analysis}

In order to quantify the IDV/microvariation, we employed two different statistical methods: the F-test and the ANOVA-test (one-way analysis of variance).

The F-test is considered as a proper statistic to quantify the optical variability \citep{de10,Josh11,Hu14,Agar15}. The $F$ value is calculated as:
\begin{equation}
F_1=\frac{Var(BL-StarA)}{Var(StarA-StarB)},
\label{eq:LebsequeIp3}
\end{equation}
\begin{equation}
F_2=\frac{Var(BL-StarB)}{Var(StarA-StarB)},
\label{eq:LebsequeIp4}
\end{equation}
where $StarA$ is the comparison star, $StarB$ is the check star, $BL$ is the blazar, and $Var()$ is the variances of the differential instrumental magnitudes. The $F$ value from the average of $F$$_1$ and $F$$_2$ is compared with the critical value $F$$^{\alpha}_{\nu_{bl},\nu_\ast}$, where $\nu_{bl}$ and $\nu_\ast$ are the number of degrees of freedom for the blazar and comparison star, respectively ($\nu$ = $N$ $-$ 1), and $\alpha$ is the significance level set as 0.01 (2.6$\sigma$) \citep{Xion16}. If the average $F$ value is larger than the critical value, then the blazar is variable at a 99\% confidence level.

ANOVA is a robust and powerful estimator for microvariations \citep{de10}. We use ANOVA in the analysis because it relies on the expected variance from the subsamples of the data rather than the error measurement. According to the exposure time, we bin the data in groups of three or five observations \citep{de10,Xion16}. This method is only applicable for light curves with more than nine observations in a given night. If the measurements in the last group are less than three or five, then they are combined with the previous group. The $F$$^{\alpha}_{\nu_1,\nu_2}$ is the critical value of ANOVA, where $\nu_1$ = $k - 1$ ($k$ is the number of groups), $\nu_2$ = $N - k$ ($N$ is the number of measurements), and $\alpha$ is the significance level \citep{Hu14}. The results of IDV are shown in Table 3.

The blazar is considered to be in a variable status if the light curves of a night satisfy the above two criteria and follow the normal distribution. Whether the light curves follow normal distribution is presented with N/Y (meaning no or yes) in column 5 of Table 3. The blazar is considered to be in a probably variable status if one of the above two criteria is satisfied and the light curves conform to the normal distribution. The blazar is considered to be in a non-variable status if none of the criteria are satisfied, or the light curves do not conform to the normal distribution. The results of IDVs are shown in Table 3. IDVs were found in 15 nights. The corresponding light curves are given in Figure 1.

\subsection{Intraday Variability, Short-Term Variability and Long-Term Variability}
In this subsection, basing on previous literature, we will report some new observation results for each of our sources.

1. OJ 287. The BL Lac object OJ 287 is one of the most widely observed extragalactic objects. Its redshift is $z = 0.306$ \citep{Gupt08b}. \citet{Sill88} discovered that OJ 287 has a $\sim$12 years periodicity and proposed a model of a binary black hole to explain this period. OJ 287 reached its brightest state with $V = 12$ mag in 1972 \citep{Qian03,Fan09}. In 2015, there was an outburst of 12.9 mag in the optical $R$ band \citep{Valt16}. \citet{Fan09} detected IDV of OJ 287. The timescale was from 10 minutes to 2 hours and the magnitude variation was from 0.11 mag to 0.75 mag. Furthermore, OJ 287 usually shows a BWB trend in colour behavior \citep{Taka89,Cari92,Vagn03,Wu06,Vill10}.

We observed OJ 287 in the $V$, $R$, and $I$ bands on April 3, 2013, and March 13 through 16, 2020. The IDV was detected in one day. On March 13, 2020, it was monotonically brightened by $\Delta$$I$ = 0.069 mag in 265.28 minutes from JD = 2458922.036 to JD = 2458922.220, $\Delta$$R$ = 0.052 mag in 218.17 minutes from JD = 2458922.039 to JD = 2458922.191, and $\Delta$$V$ = 0.053 mag in 185.67 minutes from JD = 2458922.042 to JD = 2458922.171. The variability amplitudes on March 13, 2020 were 9.47\%, 5.15\%, and 5.23\% in $I$, $R$, and $V$ bands, respectively. The light curve of the short-term timescale showed an obvious brightening that the source brightened by 0.237 mag, 0.191 mag, 0.207 mag from March 13 to March 16, 2020 in $V$, $R$ and $I$ bands, respectively.

2. ON 231. ON 231 (1219+285) was classified as a BL Lac object and its redshift is $z = 0.102$ \citep{Weis85}. ON 231 has a variation with a characteristic timescale of an order of years \citep{Poll79}. From April to May 1988, an abnormal outburst was observed by \citet{Mass99}, and the magnitude of $R$ band reached the maximum brightness of 12.2 mag. \citet{Tost98} carried out multi-band optical monitoring of the source for three years (1994-1997). The light curve of the source from 1994 to 1997 showed the brightest ($\sim$ 13.5 mag) and faintest ($\sim$ 15.0 mag) states in the $R$ band. \citet{Gupt08b} observed ON 231 on January 11, 2007 in the $R$ band. In their observations, the source did not show IDV in one night. They observed the source at $R$ $\sim$ 15.0 mag. They argued that the source was in the low state, which was comparable to the faintest state observed by \citet{Tost98}.

We observed ON 231 in the $I$ band on May 8, 2014, and $V$, $R$, $I$ bands on April 14 and April 15, 2020. The IDVs were detected in all three days. On May 8, 2014, we detected that the source brightened by 0.304 mag within 16 minutes and then darkened by 0.3 mag within 27 minutes (see Figure 1) in the I band. The variability amplitude was 30.31\% on May 8, 2014. On April 14, 2020, the magnitude variations were $\Delta$$V$ = 0.085 mag, $\Delta$$R$ = 0.059 mag and $\Delta$$I$ = 0.123 mag, respectively. On April 15, 2020, the source brightened by 0.031 mag in the I band. From April 14 to April 15, 2020, the largest magnitude variations in the $V$, $R$, and $I$ bands were $\Delta$$V$ = 0.199 mag, $\Delta$$R$ = 0.152 mag, and $\Delta$$I$ = 0.145 mag, respectively. As can be seen from Figure 2, the light curve showed an obvious brightening.

3. 3C 279. The quasar 3C 279 ($z = 0.538$) is one of the most prominent blazars and extremely variable at all wavelengths \citep{Kata00}. In the optical band, it showed fast and significant outbursts on a single night \citep[e.g., ][]{Mill96}. Rapid variations have also been observed in $IR$, $UV$, $X$-ray, and $\gamma$-ray \citep[e.g., ][]{Fan99}. \citet{Xie99} reported the most rapid optical variation $\Delta$$V$ = 1.17 mag in 40 minutes on May 22, 1996. \citet{Webb90} reported that the source brightened by $\sim$2.0 mag within 24 hours. \citet{Gupt08b} reported a large variation $\Delta$$R$ = 1.5 mag of the source in the time span of 42 days (January-February 2007) which was a much larger variation than the previous $\Delta$$R$ = 0.91 mag in 49 days (April-June 2001) reported by \citet{Xie02}.

We observed 3C 279 for 11 days from 2016 to 2018. The IDVs were detected on two days. On May 31, 2016, the brightness variation of the source was $\Delta$$V$ = 0.195 mag (Amp=19.28\%). On May 08, 2018, the source brightened by $\Delta$$V$ = 0.118 mag, $\Delta$$R$ = 0.099 mag within 74.9 minutes and darkened by $\Delta$$V$ = 0.19 mag, $\Delta$$R$ = 0.126 mag within 117.75 and 123.42 minutes. The variability amplitudes of the V and R bands were 12.15\% and 9.59\%, respectively. The light curves of the long-term timescale are shown in Figure 3.

4. 1ES 1426+42.8. \citet{Abdo10b} classified 1ES 1426+42.8 ($z = 0.129$) as BL Lac object based on its spectral energy distributions. It is a TeV source. Its spectrum has been measured by the High Energy Gamma-Ray Astronomy (HEGRA) up to 10 TeV \citep{Cost03}. In addition, it is a low power, high synchrotron peak ($\nu_{peak} > 100keV$) BL Lac object \citep{Cost03}. Most of the researches of 1ES 1426+42.8 focus on the high-energy band, while the optical bands remain less explored. In the blazar monitoring program of \citet{Kurt09} from May 1997 to September 2006 (66 days), 1ES 1426+42.8 had a small variation of 0.10 $\pm$ 0.05 mag in the $R$ band. \citet{Leon09} observed 1ES 1426+42.8 in May-June 2008 in $R$ band. It is brightened by 20 \% during the observation.

The IDV is detected in our observations of 1ES 1426+42.8 in the $I$ band on April 13, 2010. 1ES 1426+42.8 had a rapid outburst of $\Delta$$R$ = 0.149 mag within 20.68 minutes from JD = 2455300.244 to JD = 2455300.258. The variability amplitude was 14.89\% on April 13, 2010. We also observed three consecutive rising and falling brightness variations of 1ES 1426+42.8 on April 13, 2010.

5. 1E 1458.8+2249. \citet{Heid98} classified 1E 1458.8 + 2249 as a BL Lac object. \citet{Fior98} reported that the magnitudes of this source are $V = 15.58 \pm 0.01$, $R = 15.06 \pm 0.01$, and $I = 14.60 \pm 0.01$, respectively. \citet{Mass03} carried out optical multi-band observations of 1E 1458.8+2249 from 1994 to 2001. The $R$ band magnitude varied between 15.5 mag and 16.5 mag from 1994 to 1998. They observed a flare with $R = 14.75$ mag in January 2000. On February 8, 2001, it reached 14.62 mag in the $R$ band. They observed the magnitudes on February 20, 2001 as follows: $V = 15.19 \pm 0.06$ mag, $B = 15.53 \pm 0.05$ mag, and $U = 14.99 \pm 0.05$ mag, and gave the magnitudes on February 21 as follows: $I = 14.29 \pm 0.03$ mag, $R = 14.79 \pm 0.03$ mag, and $V = 15.17 \pm 0.04$ mag, respectively.

We observed 1E 1458.8 + 2249 in the $I$ band on 2011 May 26, and in the $V$, $R$, and $I$ bands from April 14 to April 15, 2020. The IDVs were detected on May 26, 2011 and April 14, 2020 in the I band. On May 26, 2011, the I band magnitude ranges from 15.98 to 16.95 mag. Figure 1 shows that 1E 1458.8 + 2249 had a drastic outburst of 0.97 mag. On April 14, 2020, the brightness of the I band was between 14.99 mag and 15.09 mag, and the variability amplitude was 8.70\%.

6. PKS 1510-089. PKS 1510-089 ($z = 0.361$) is one of the most extreme AGNs, exhibiting strong and rapid variability in all wave bands \citep{Rani11}. It is a highly polarized quasar in the optical band \citep{Kata08}. It is observed with variations in the optical bands of 0.65 mag in 41 minutes on June 14, 1999 \citep{Dai01}. The optical variations were 2.00 mag in 41 minutes on May 29, 2000, and 0.85 mag in 44 minutes on April 16, 2001 \citep{Xie01,Xie02}. These variations had similar timescales ($\sim$ 42 min). \citet{Wu05} reported that on March 08, 2002, the source brightness in the $R$ band dropped from 16.62 $\pm$ 0.12 to 17.92 $\pm$ 0.15 mag within 19 minutes, and rose back to 16.57 $\pm$ 0.15 mag in 16 minutes. They found that the actual timescale of the minimum was 35 minutes.

We observed PKS 1510-089 for 11 days from 2010 to 2016. The IDVs were detected on five days. On May 6, 7, 8, 2010, May 6, 2011, and April 2, 2013, the variations of magnitude were $\Delta$$R$ = 0.424 mag, $\Delta$$R$ = 0.341 mag, $\Delta$$R$ = 0.187 mag, $\Delta$$R$ = 0.227 mag, and $\Delta$$I$ = 0.412 mag, respectively. The variability amplitudes were 42.39\%, 34.09\%, 18.69\%, 22.69\%, and 57.65\%, respectively. The light curves of the short-term timescale and the long-timescale are shown in Figure 2 and Figure 3.

7. OT 546. OT 546 (1727+502) was discovered by \citet{Zwic66} and classified as a BL Lac object by \citet{Ange80}. \citet{Oke78} measured the redshift of OT 546 as $z = 0.0554 \pm 0.0003$. \citet{Fan95} reported the mass of its SMBH as $M_{\bullet} = 10^{8.73} M_{\odot}$. From 1975 to 1987, \citet{Kinm76} reported that the average brightness in the $V$ band was around 16 mag. \citet{Pica88} reported an average magnitude of 16.70 mag in the $B$ band from 1975 to 1987. The monitoring of \citet{Kata00} displayed that the brightness varied between 15.76 mag and 16.12 mag in the $V$ band from 1996 to 1997. \citet{Guo14} detected that the $V$ band varied between 15.72 mag and 16.05 mag from 2009 February 16 to July 1.

We observed OT 546 in the $V$, $R$, $I$ bands on May 6, 2010, and from May 7 to May 12, 2016. The IDV was detected on May 07, 2016 in the I band with the magnitude variation of $\Delta$$I$ = 0.142 mag. The variability amplitude was 13.80\% on May 07, 2016. During the observation period between May 7 and May 12, 2016, the largest magnitude variations in the $V$, $R$ and $I$ bands were $\Delta$$V$ = 0.284 mag, $\Delta$$R$ = 0.355 mag, and $\Delta$$I$ = 0.423 mag, respectively.

8. BL Lacertae. BL Lacertae (2200+420) is the prototype of BL Lacs, and its redshift is $z$ = 0.069 \citep{Mill77}. \citet{Woo02} discovered that its black hole mass was $M_{\bullet} = 10^{8.23} M_{\odot}$. It is highly variable from radio to $\gamma$-ray. A rapid optical variation of 1.3 mag in 20 hours was detected in the $V$ band by \citet{Weis73}. On August 25, 1996, it faded $\Delta$ $B = 0.73$ mag within 53 minutes, and on October 19, 1996, it faded $\Delta$ $I = 0.55$ mag within 70 minutes \citep{Xie99}. \citet{Li21a} reported a violent variation on September 15, 2017 in the $B$ band with an variability amplitude of 16.5\% (0.17 mag). The brightest magnitude of $R$ band is 12.95 mag on September 17, 2017, while the faintest magnitude of $R$ band is 13.55 mag on October 28, 2018.

We observed BL Lacertae in the $B$, $V$, $R$, and $I$ bands from September 16 to September 19, 2019. The IDV was detected on September 16, 2019 in the B band with the magnitude variation of $\Delta$$B$ = 0.099 mag. The light curve of short-term timescale showed an obvious brightening from September 16 to September 19, 2019. The brightness variations of the source were $\Delta$$B$ = 0.549 mag, $\Delta$$V$ = 0.481 mag, $\Delta$$R$ = 0.426 mag and $\Delta$$I$ = 0.433 mag within 4 days, respectively.

\begin{figure*}
\centering
\includegraphics[scale=.16]{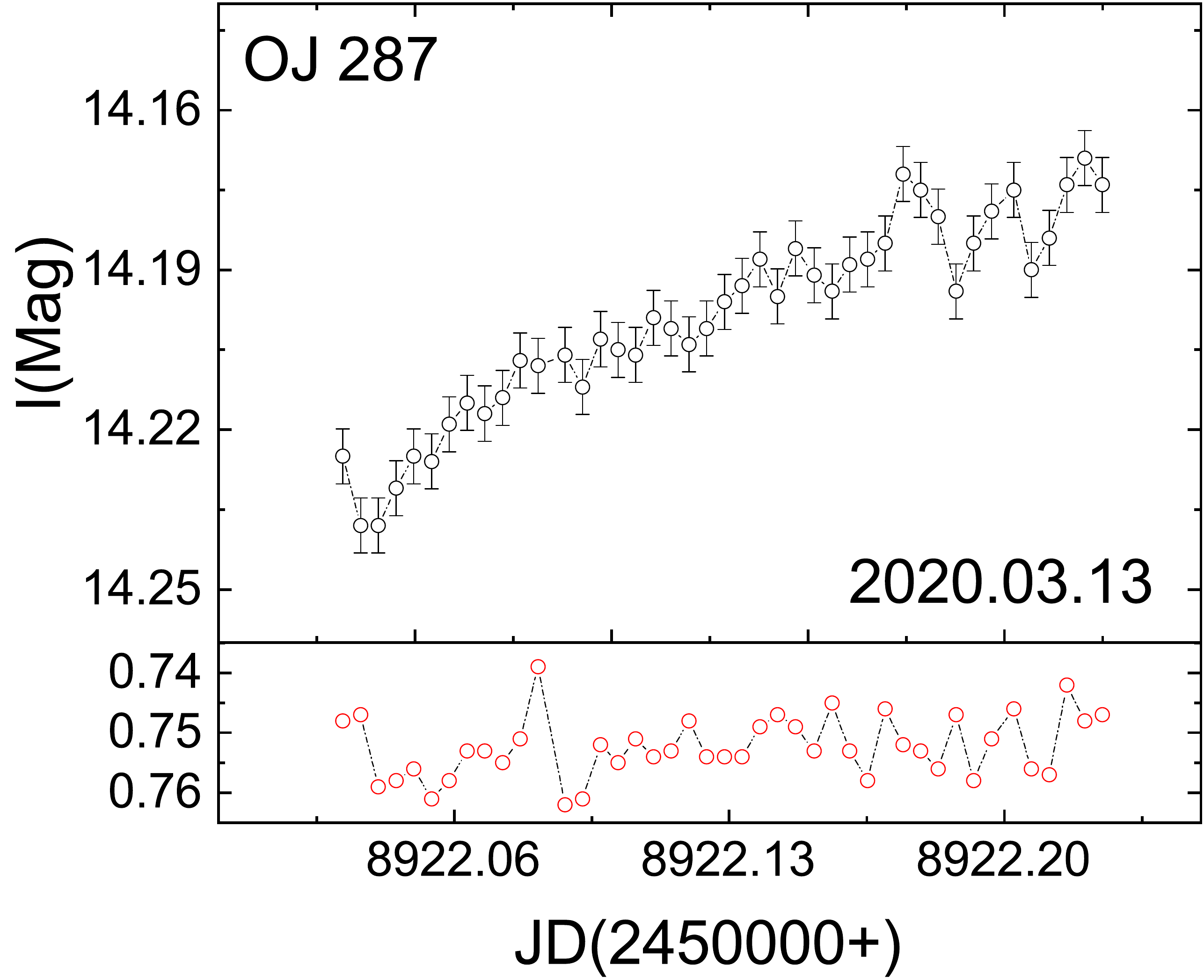}
\includegraphics[scale=.16]{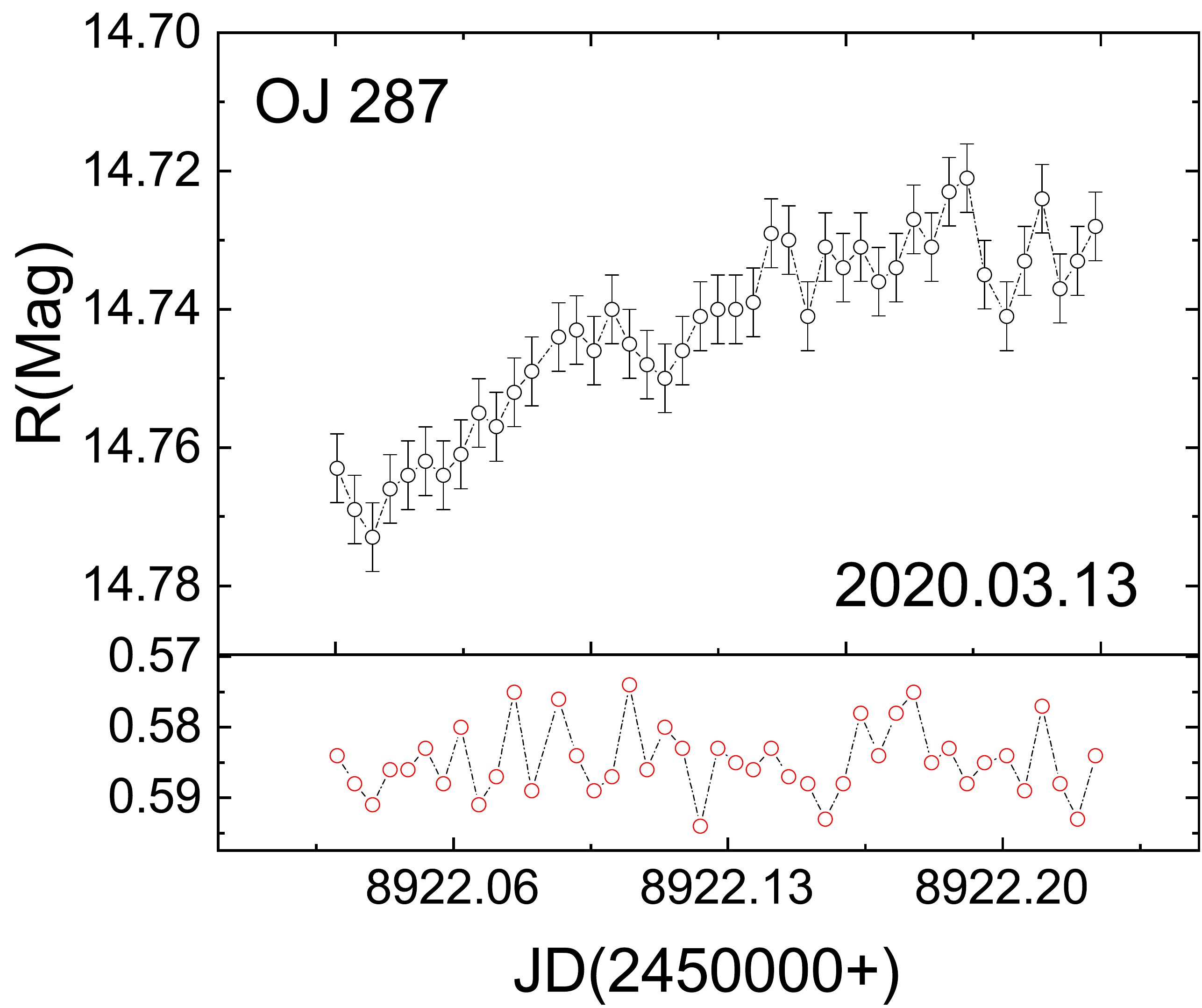}
\includegraphics[scale=.16]{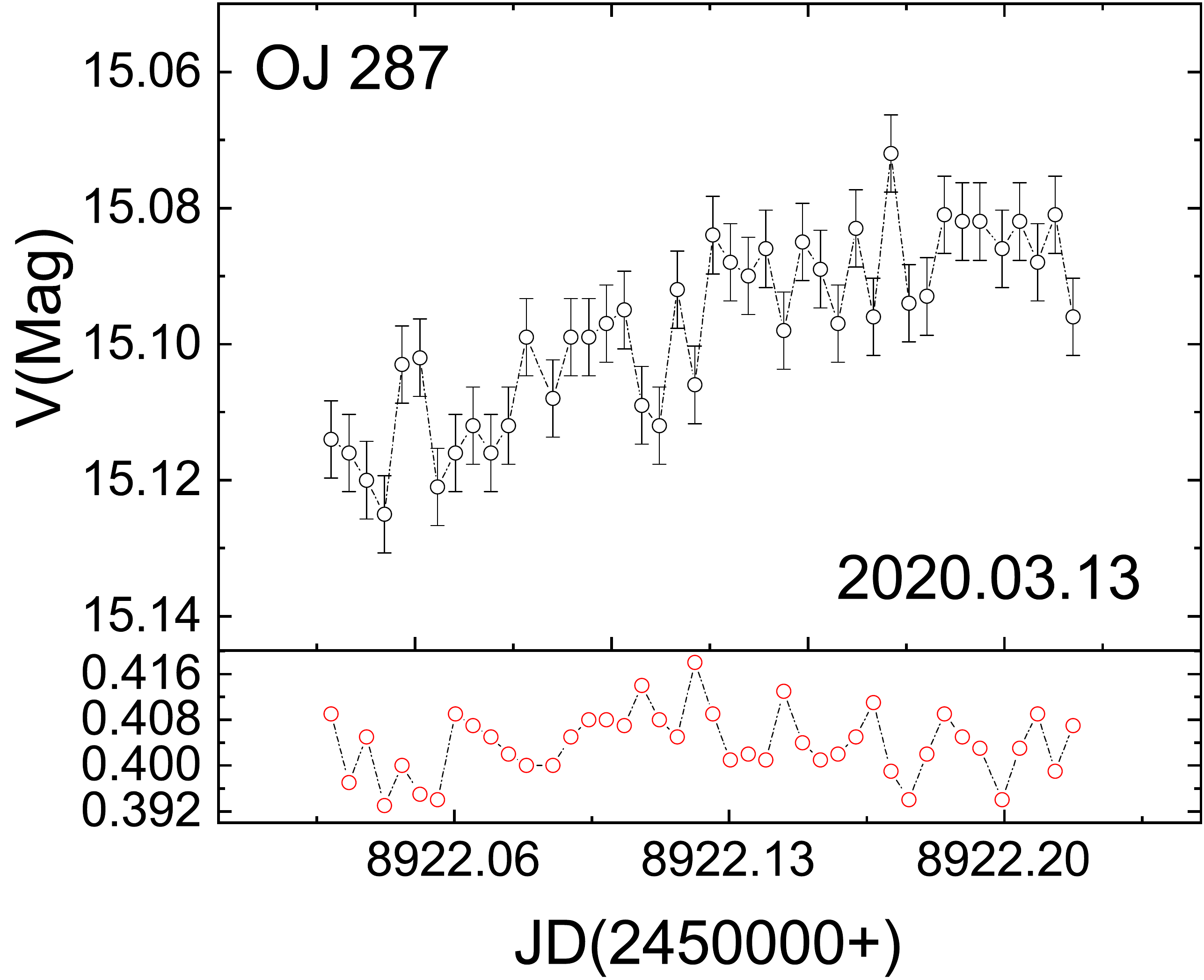}
\includegraphics[scale=.16]{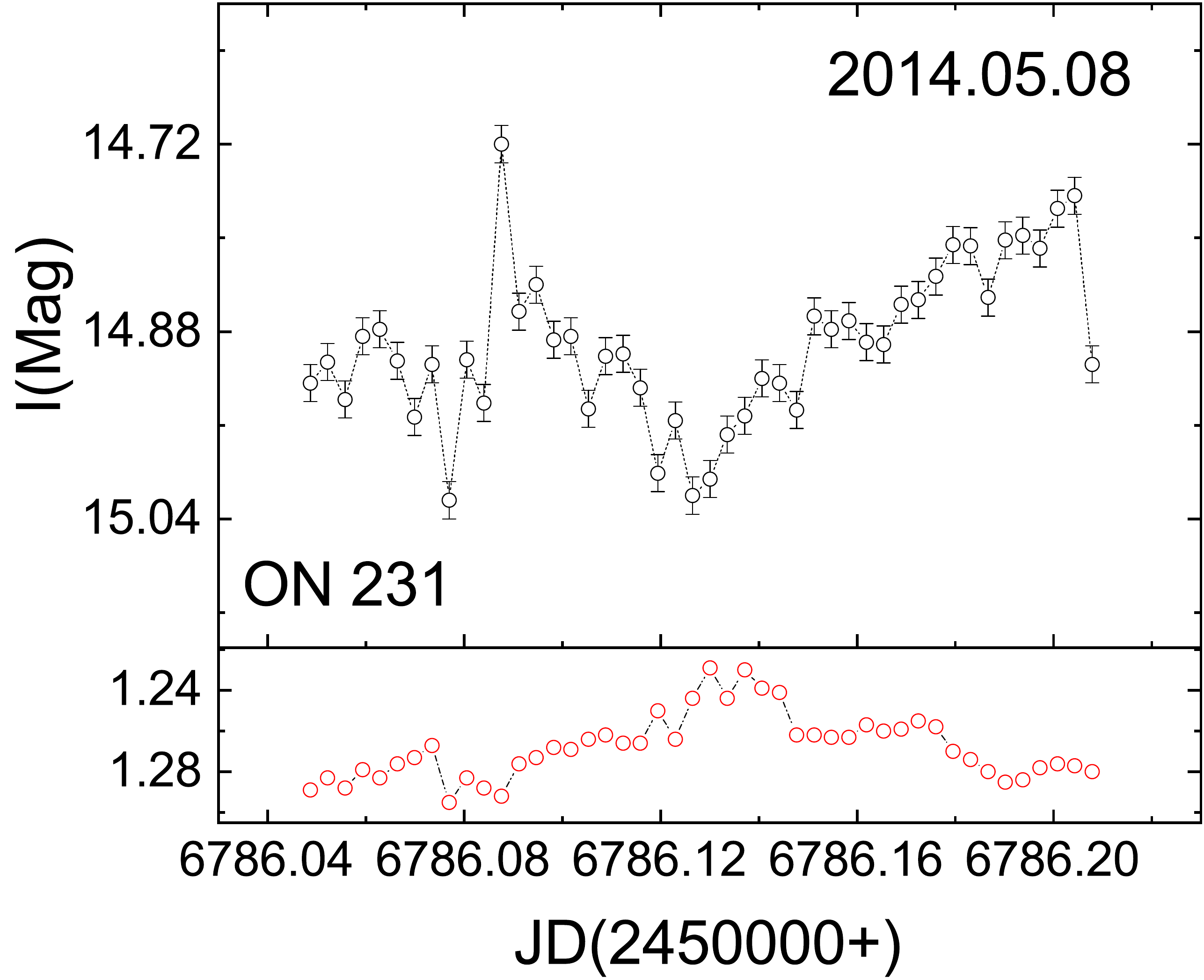}
\includegraphics[scale=.16]{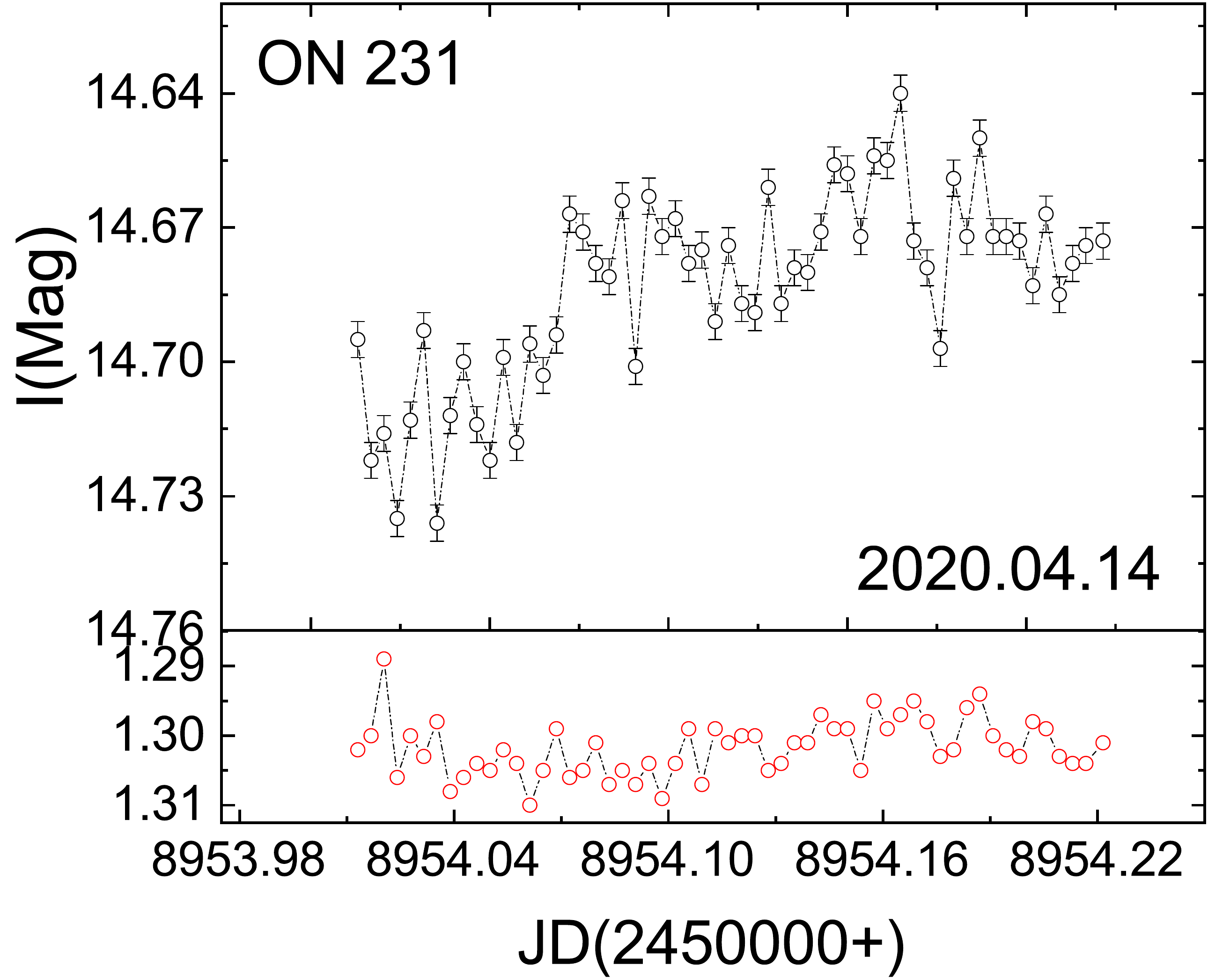}
\includegraphics[scale=.16]{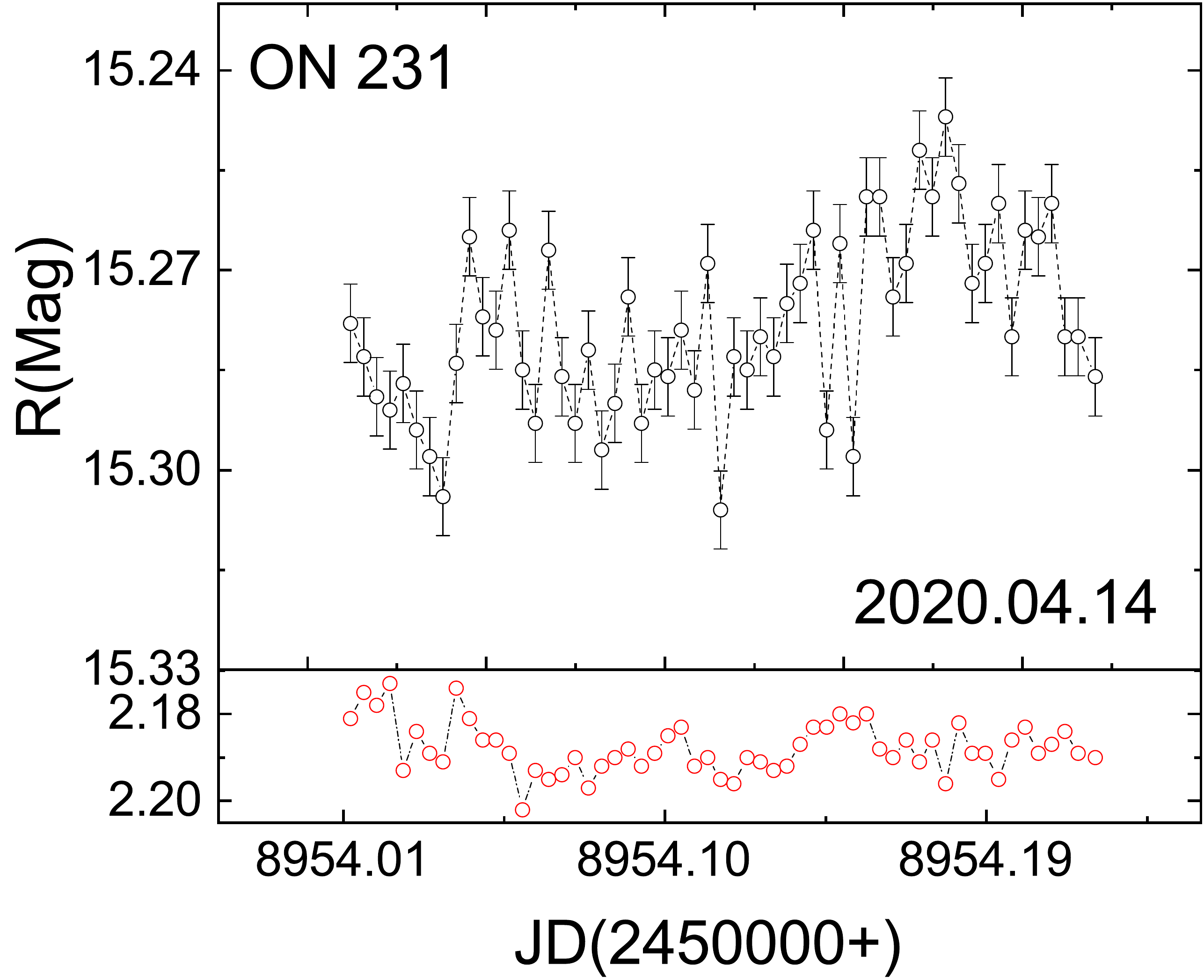}
\includegraphics[scale=.16]{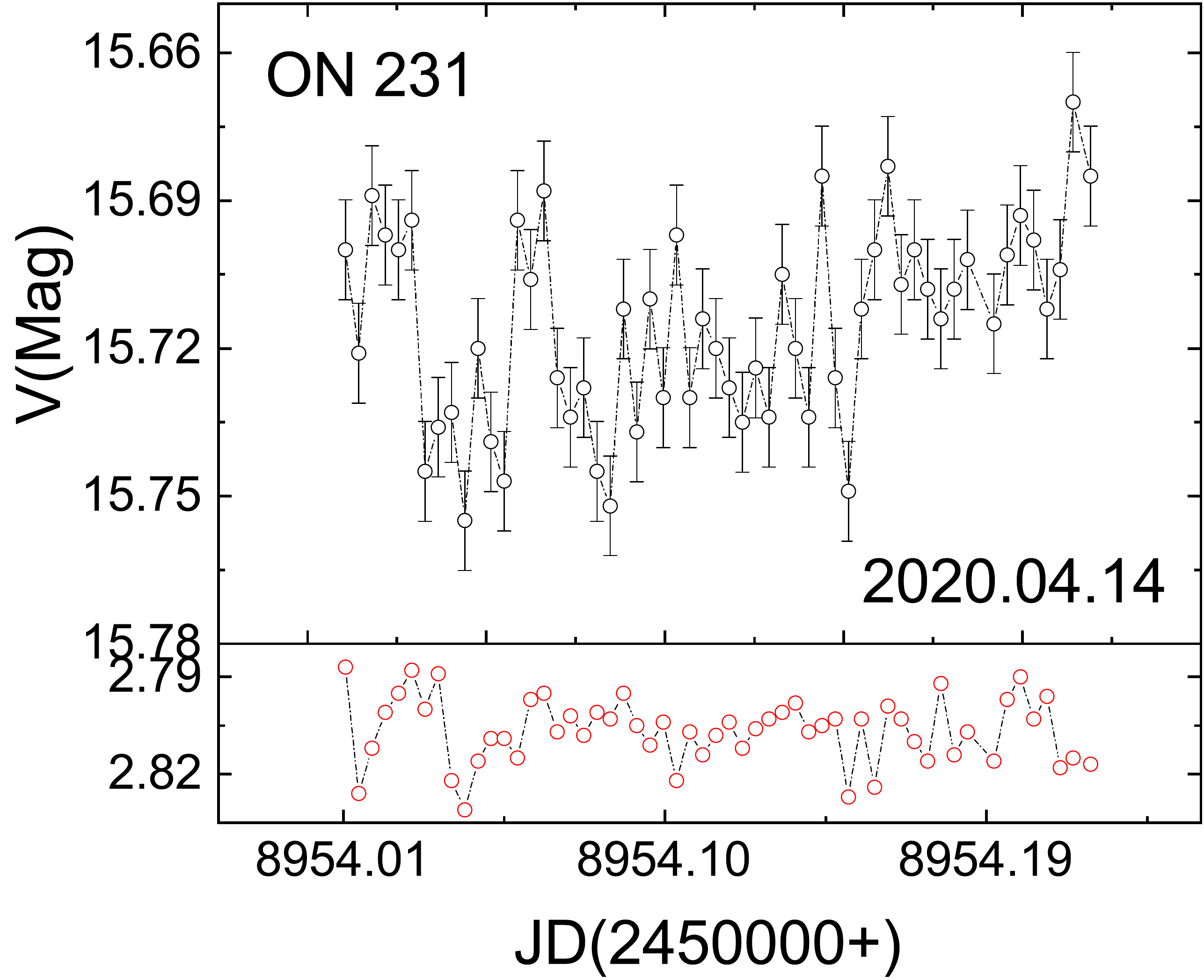}
\includegraphics[scale=.16]{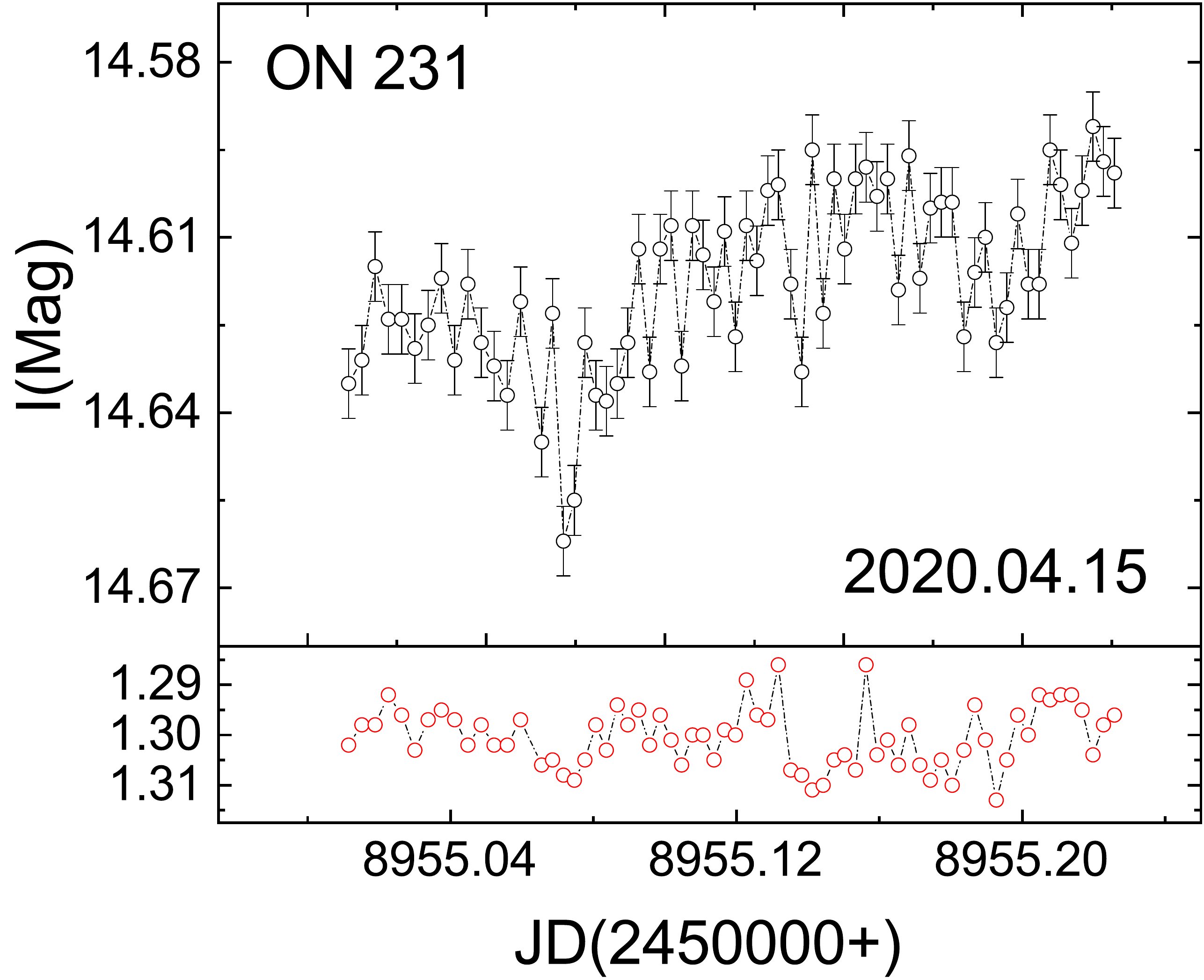}
\includegraphics[scale=.16]{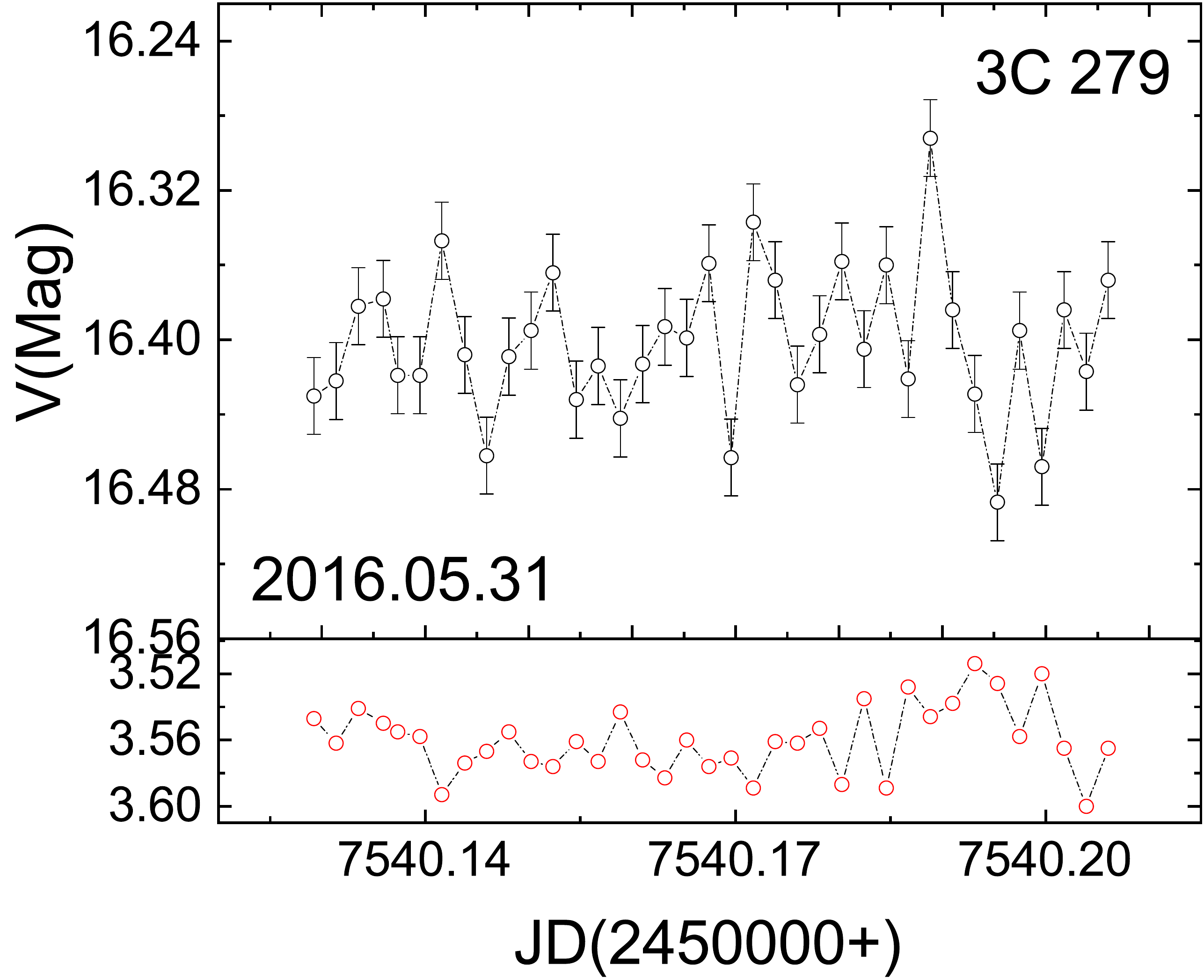}
\includegraphics[scale=.16]{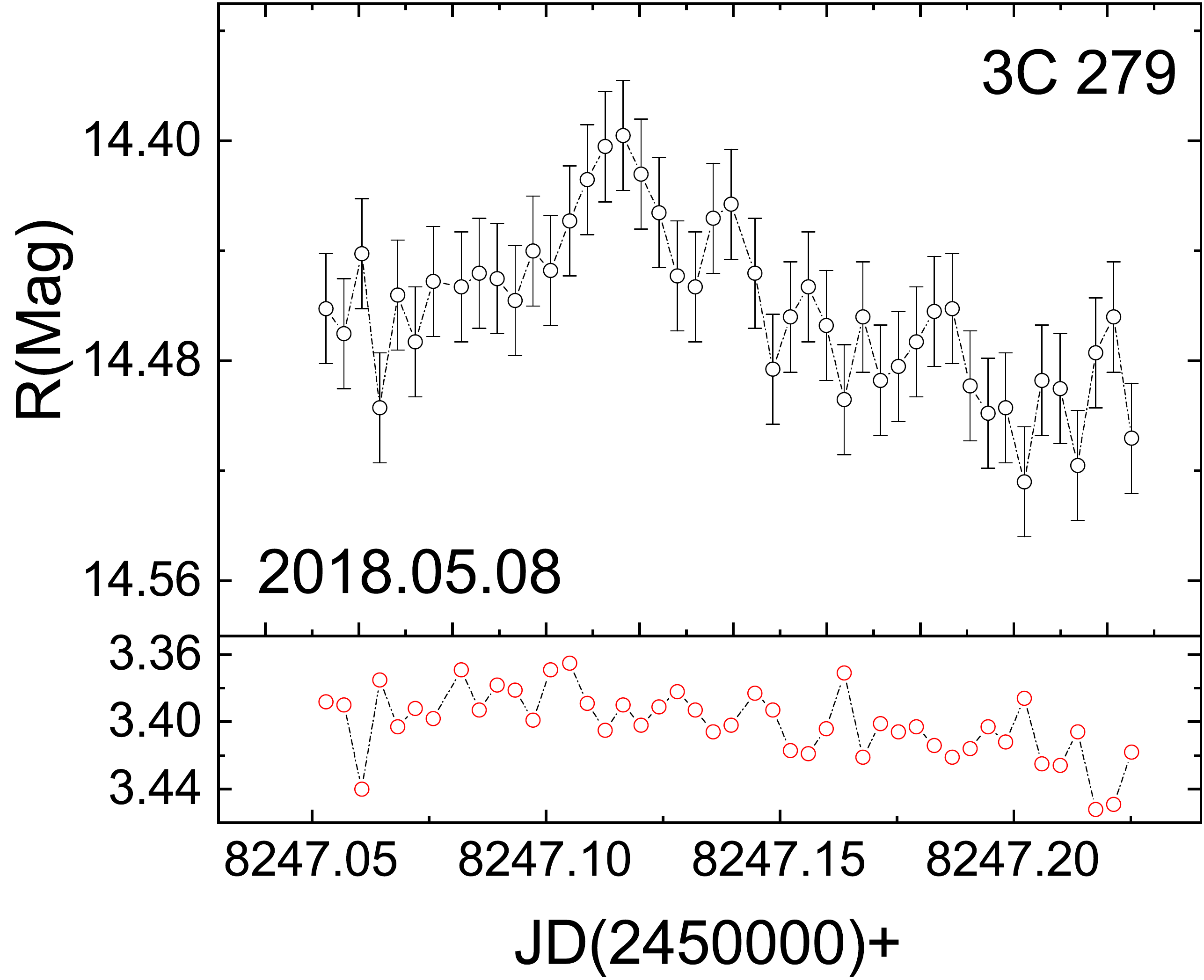}
\includegraphics[scale=.16]{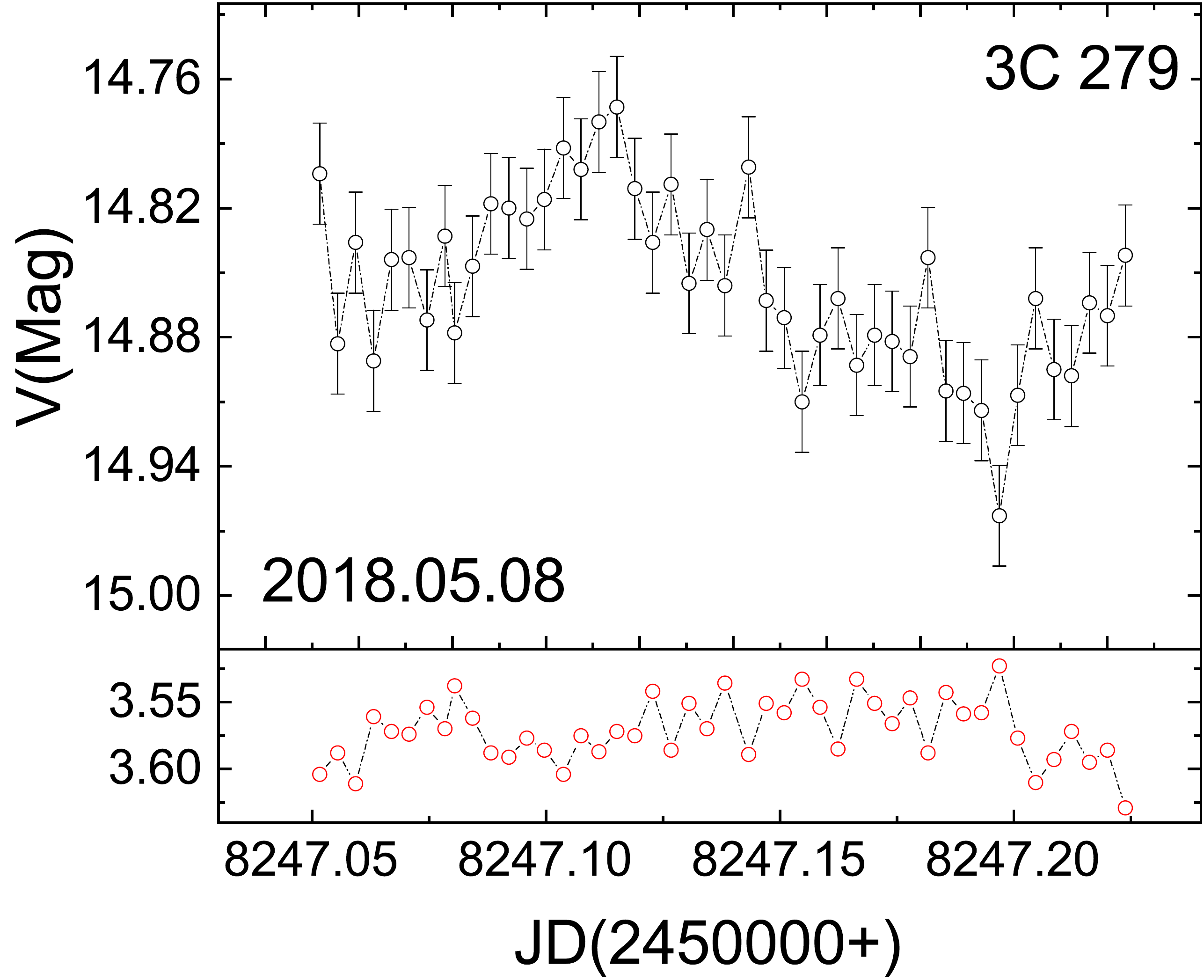}
\includegraphics[scale=.16]{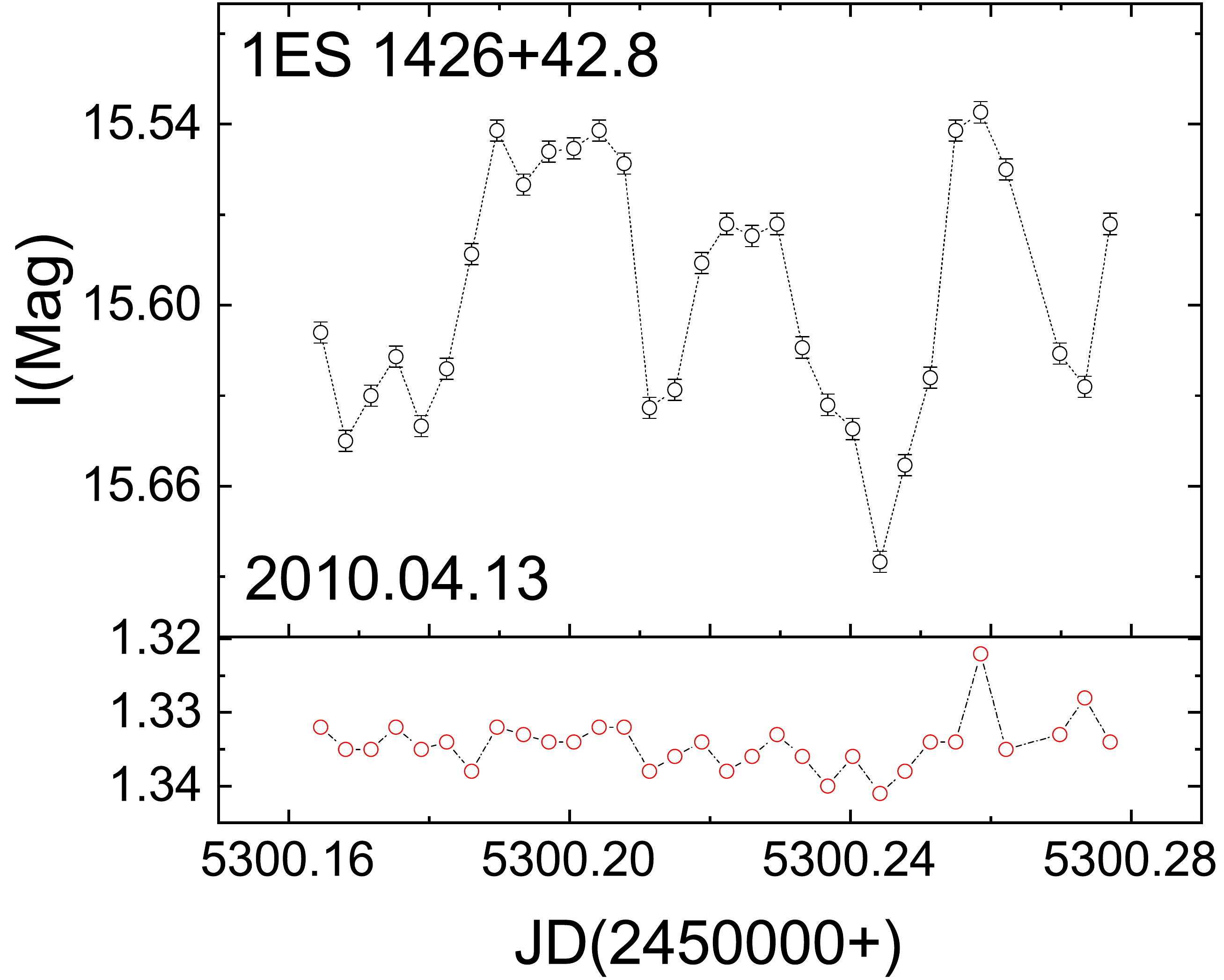}
\includegraphics[scale=.16]{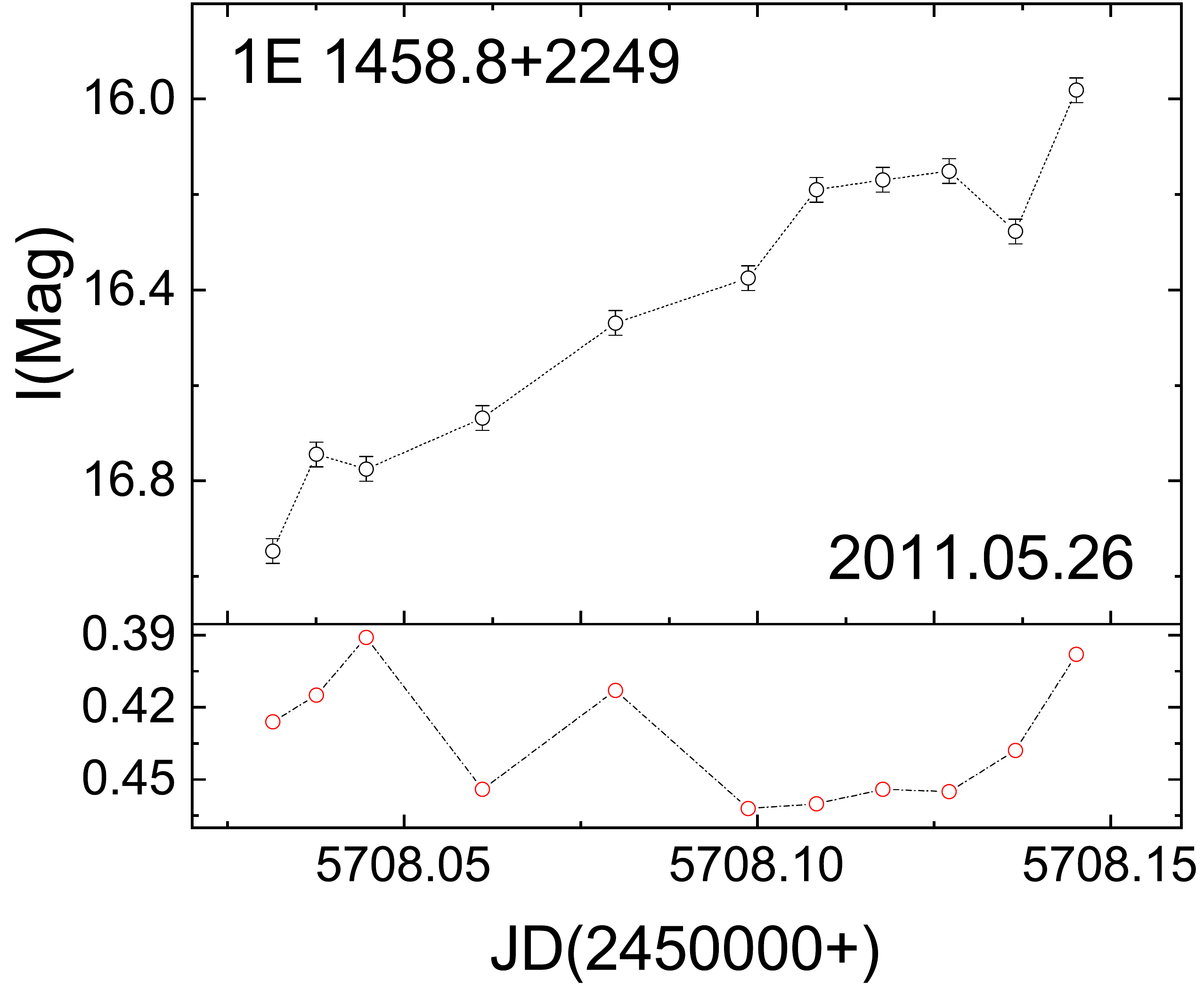}
\includegraphics[scale=.16]{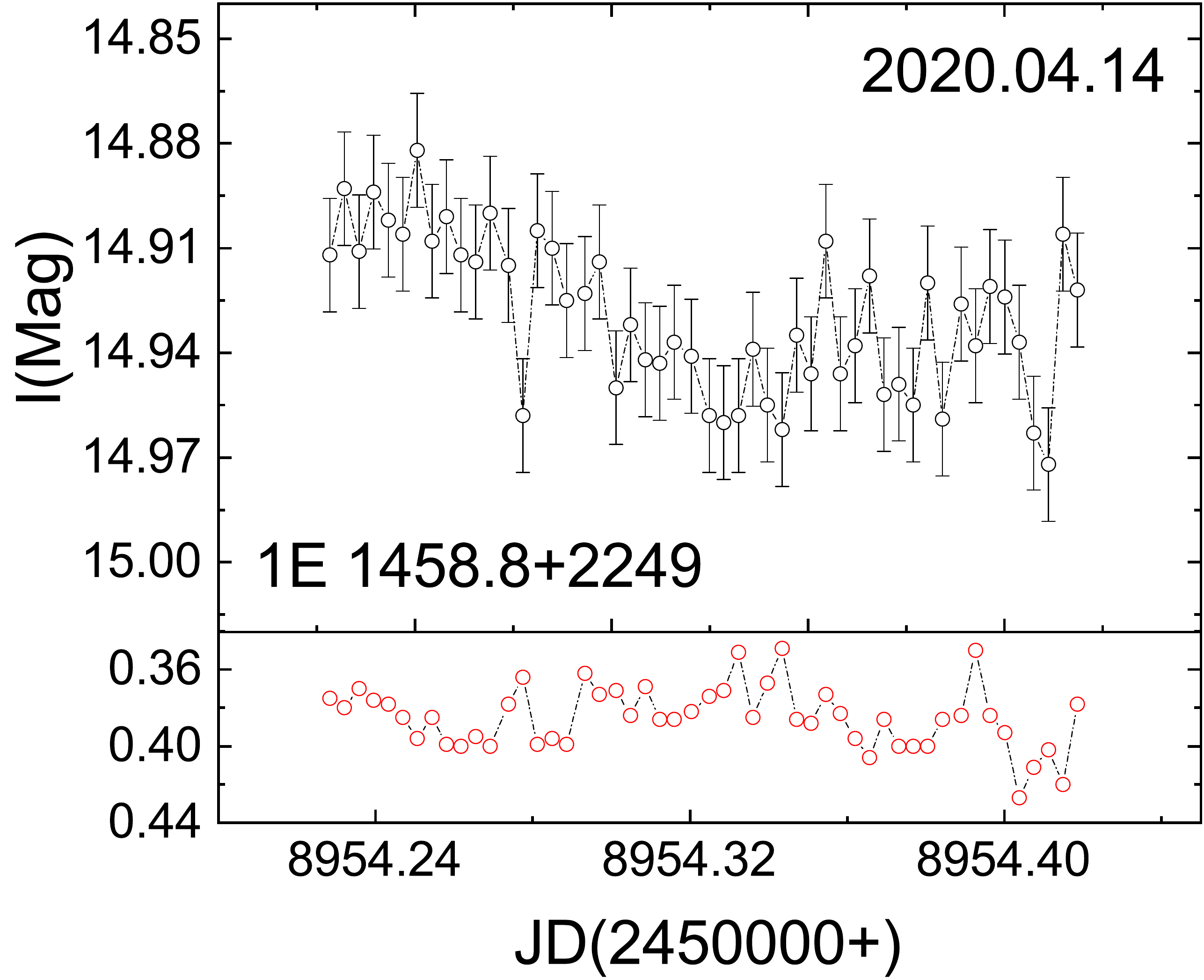}
\includegraphics[scale=.16]{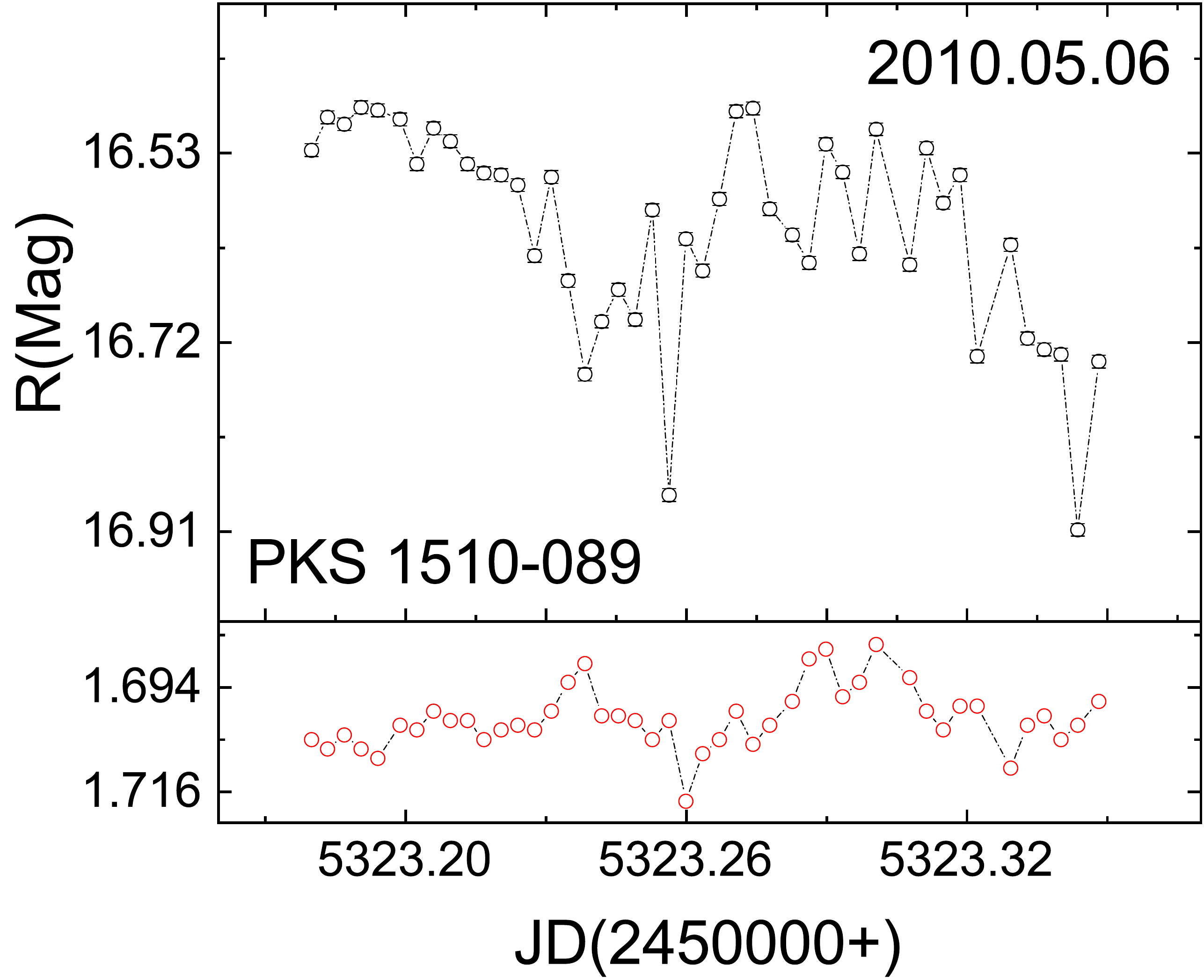}
\includegraphics[scale=.16]{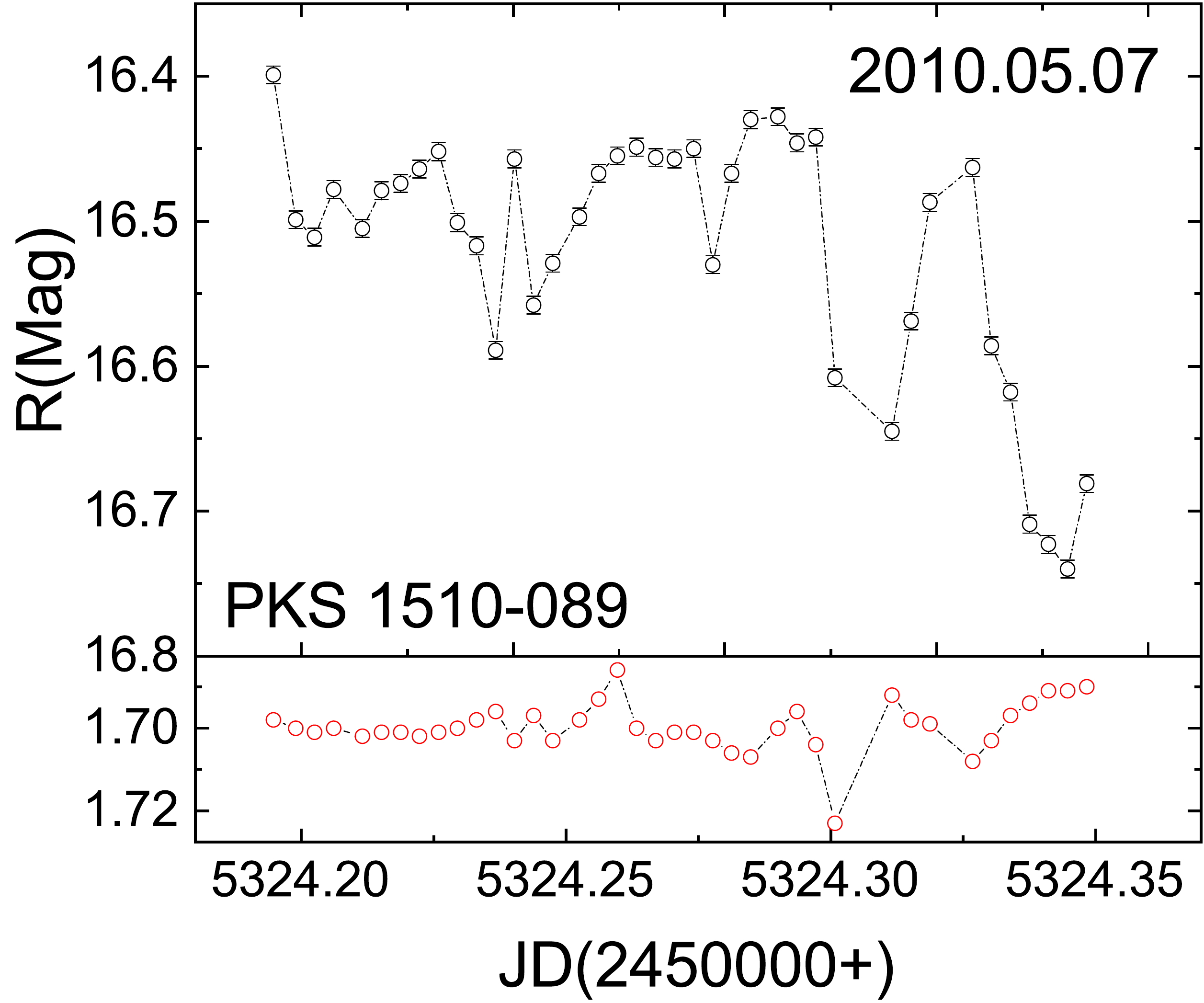}
\includegraphics[scale=.16]{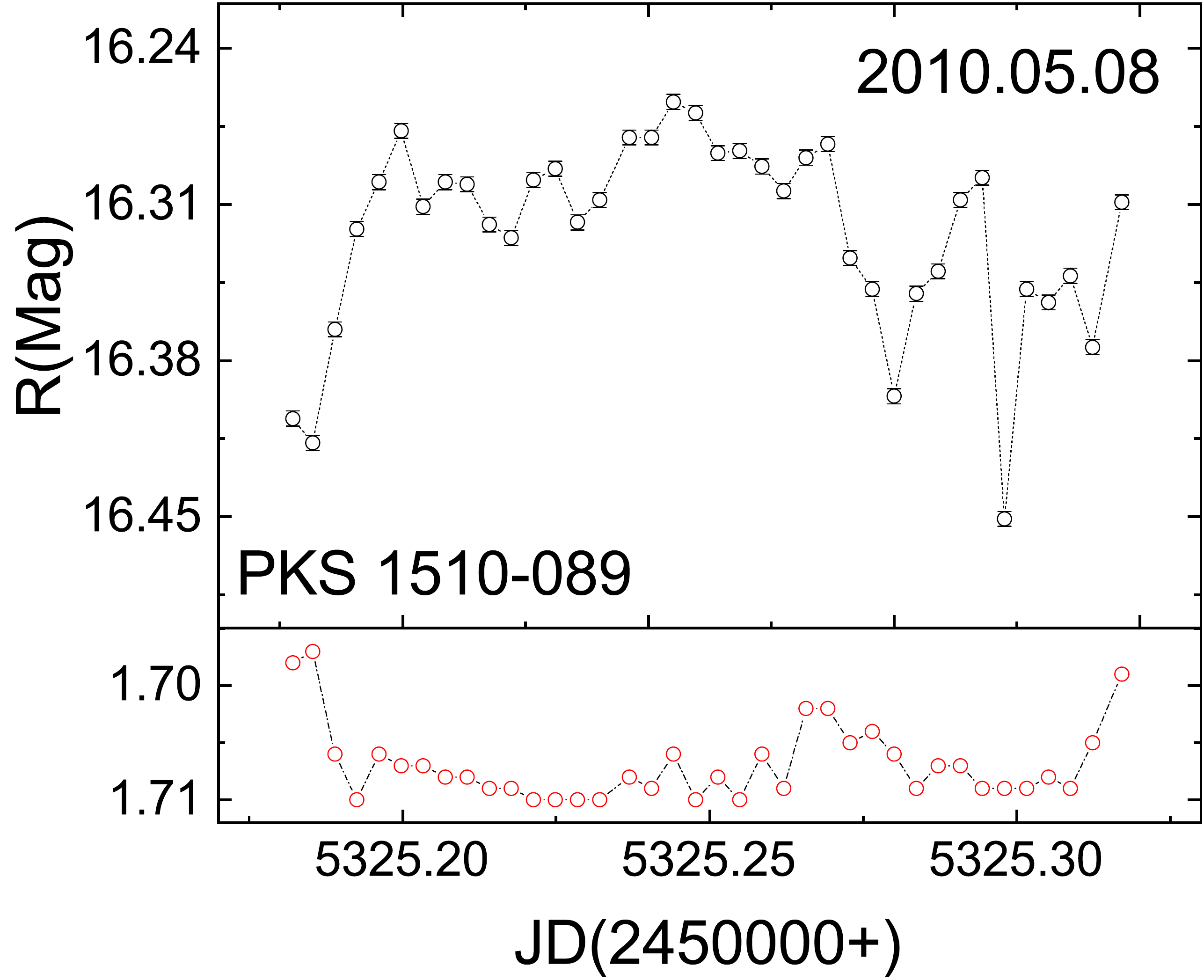}
\includegraphics[scale=.16]{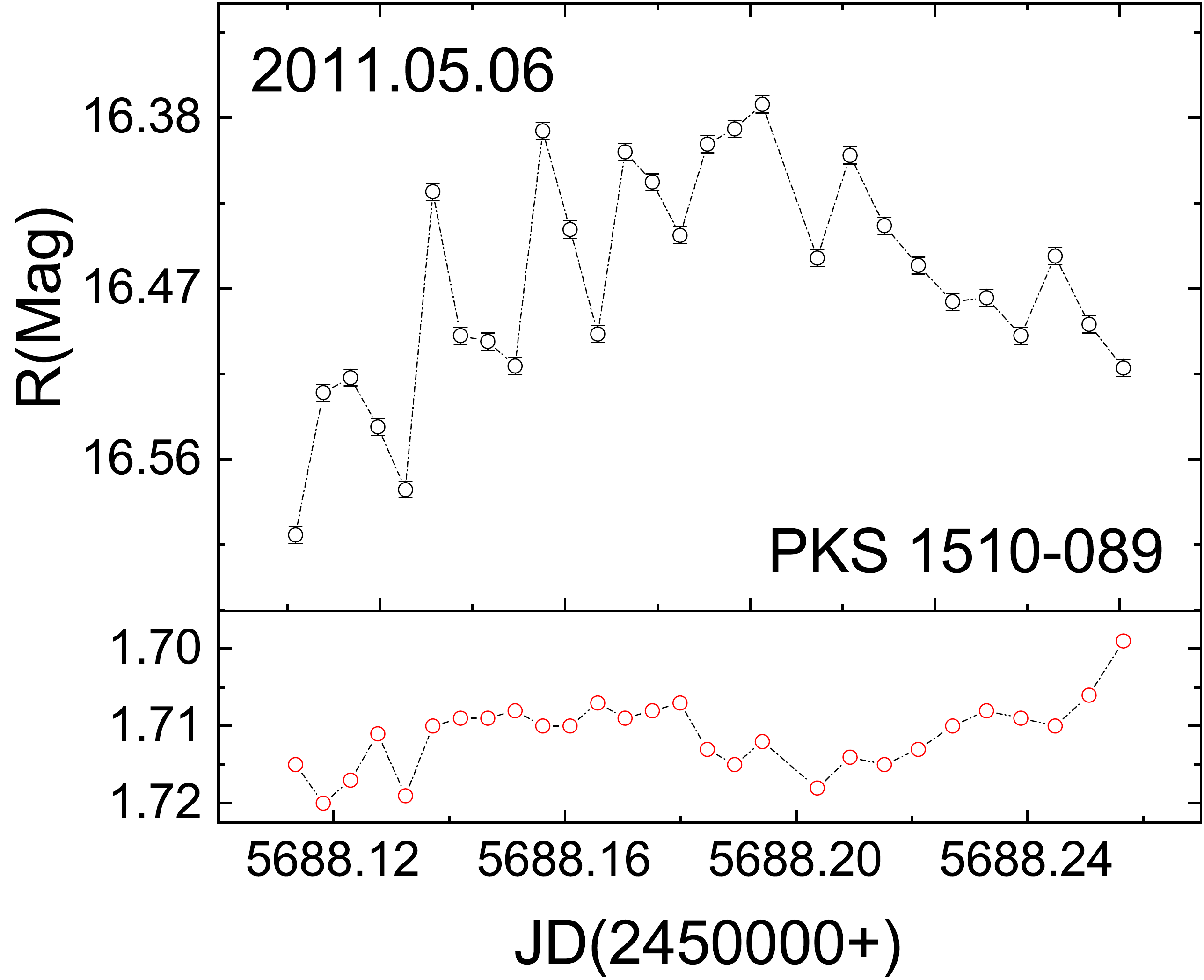}
\includegraphics[scale=.16]{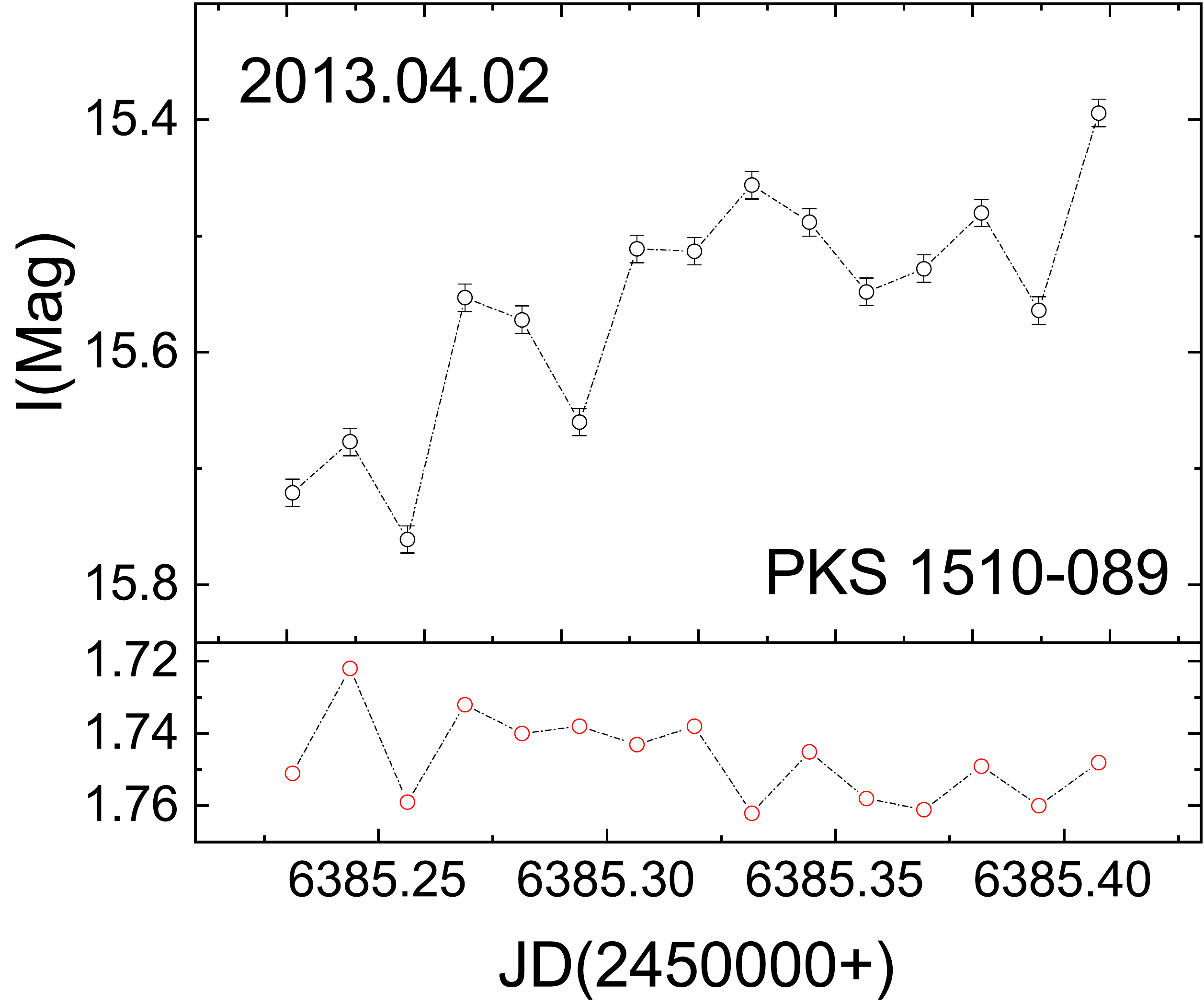}
\includegraphics[scale=.16]{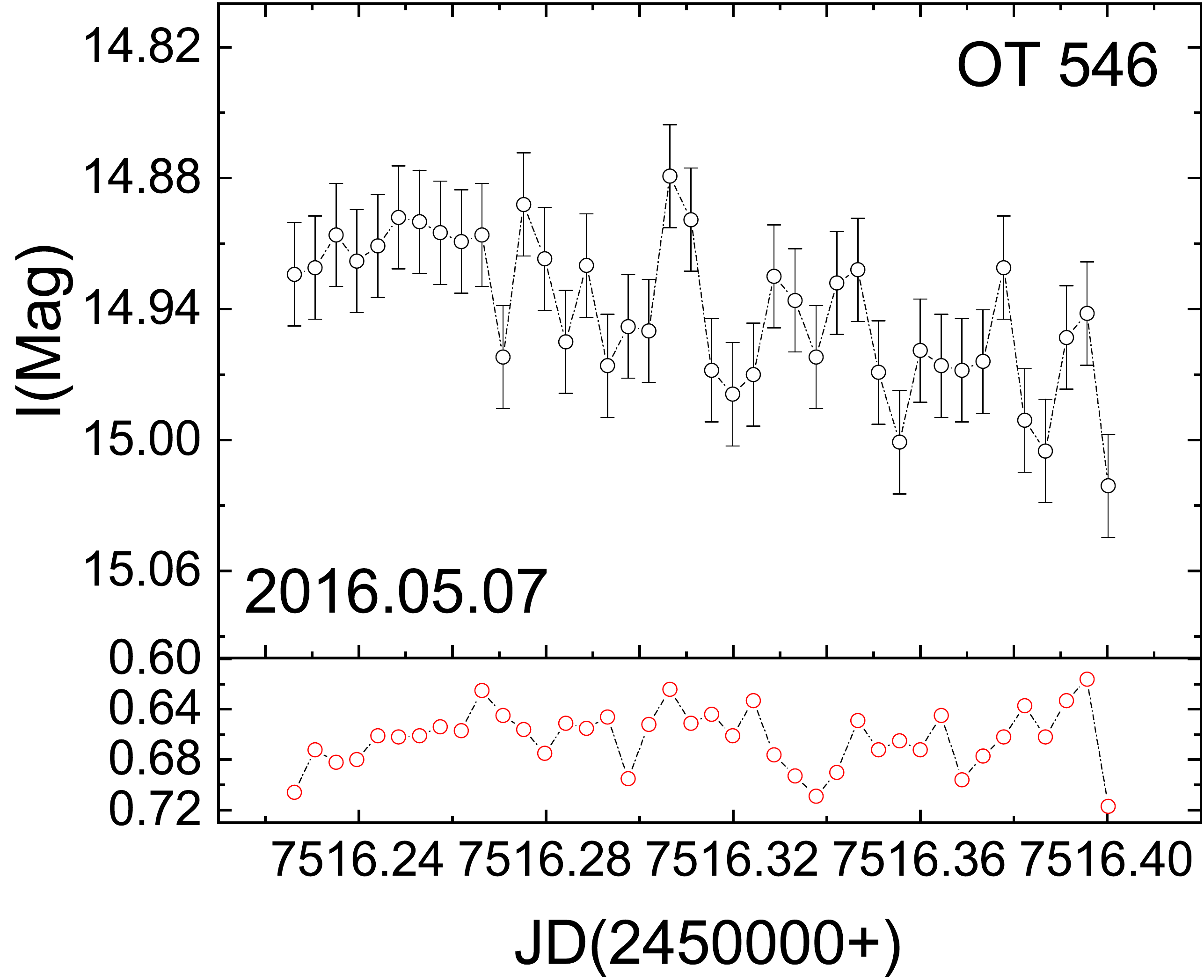}
\includegraphics[scale=.16]{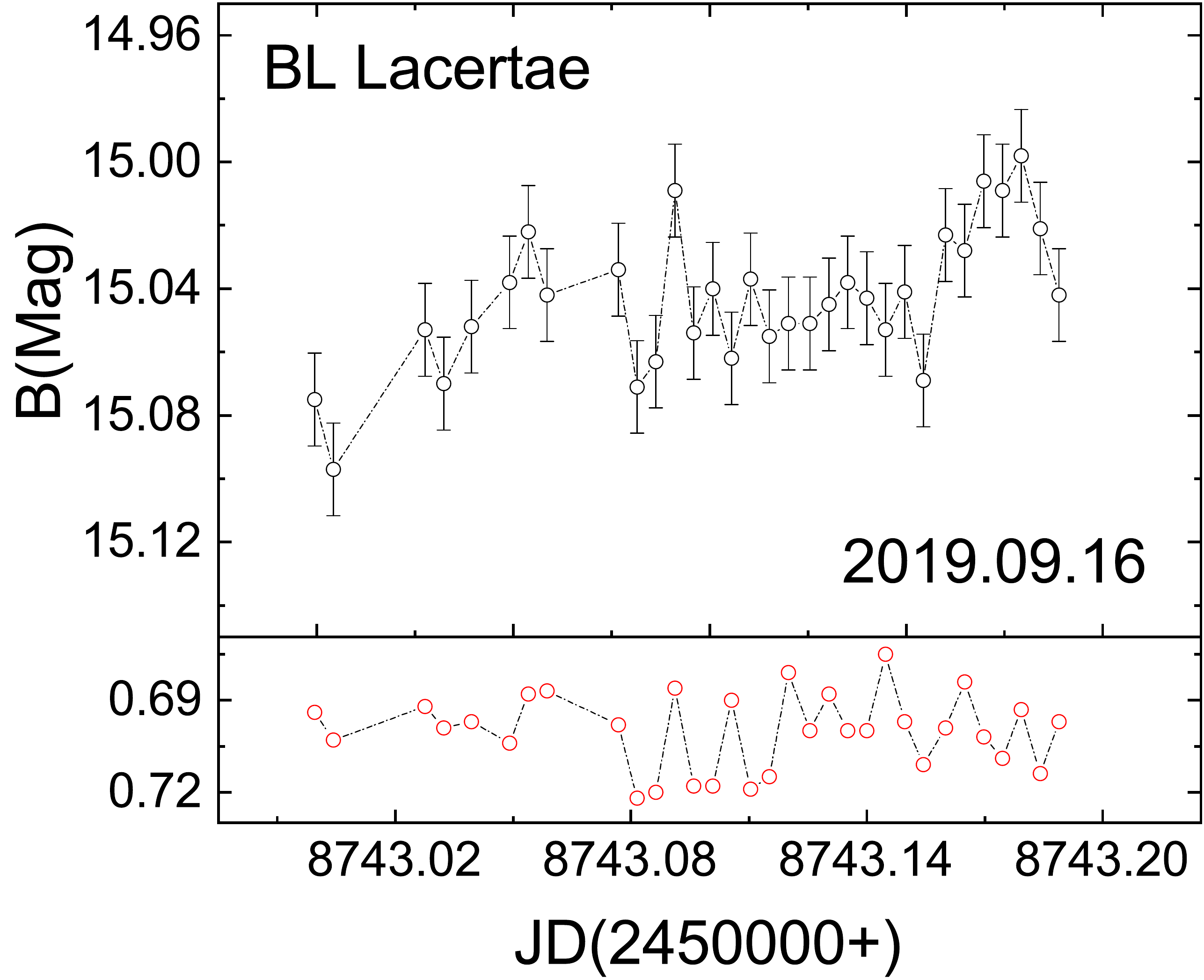}
\caption{Light curves of the IDV for the eight sources. The black open circles are the light curves for the sources. The red open circles are the magnitude difference between comparison stars and check stars in the same period. \label{}}
\end{figure*}

\begin{figure*}
\centering
\includegraphics[scale=.18]{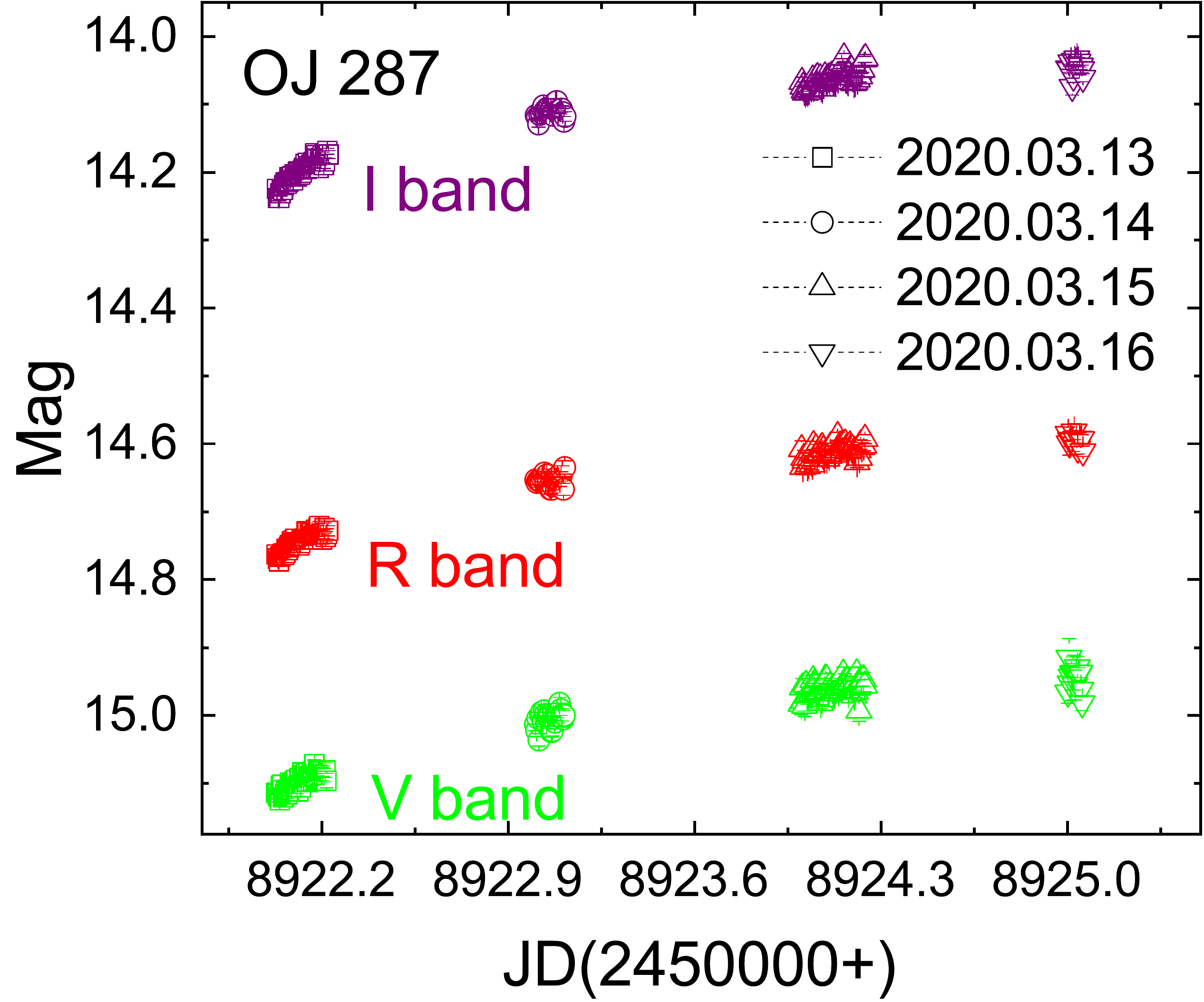}
\includegraphics[scale=.18]{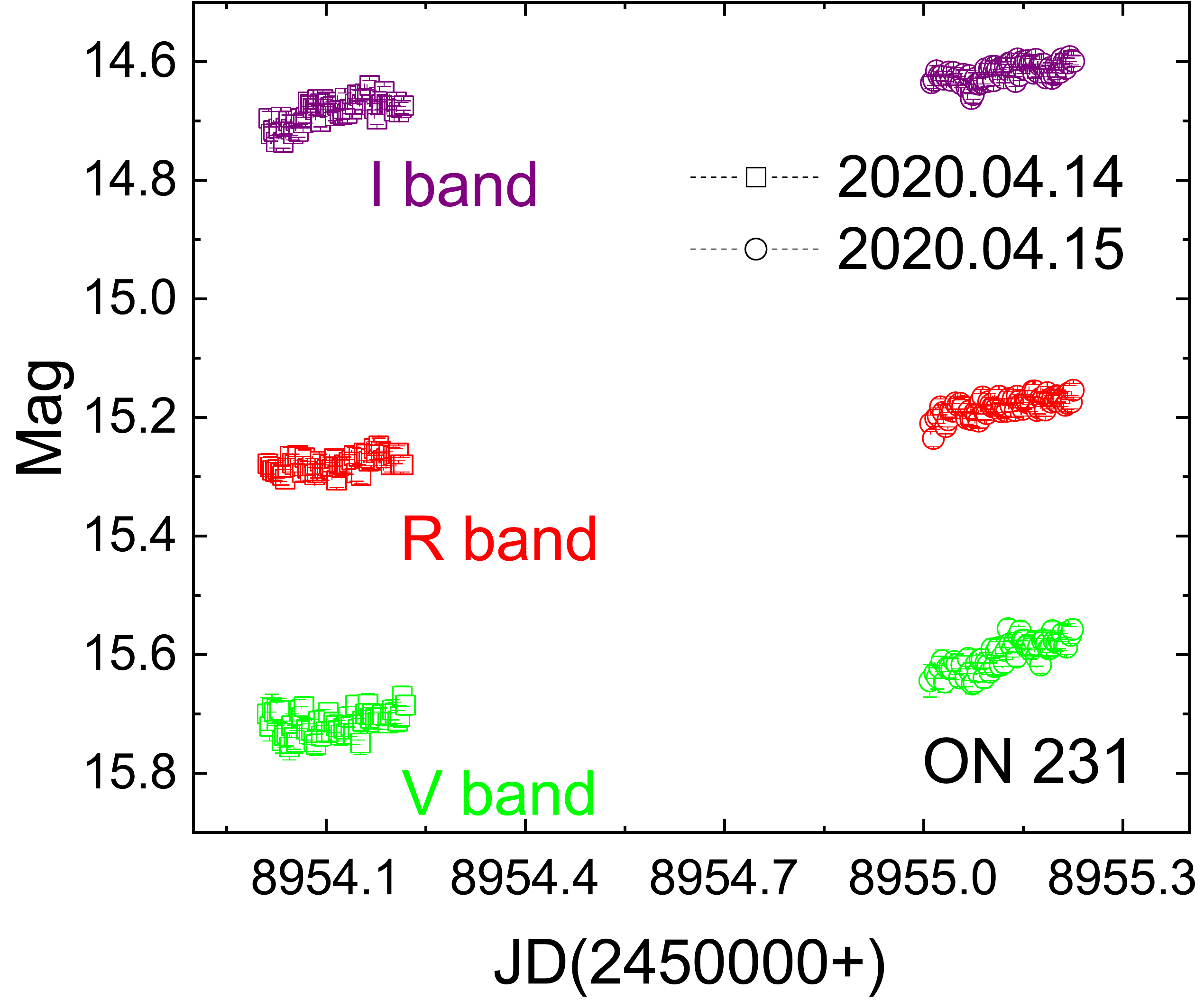}
\includegraphics[scale=.18]{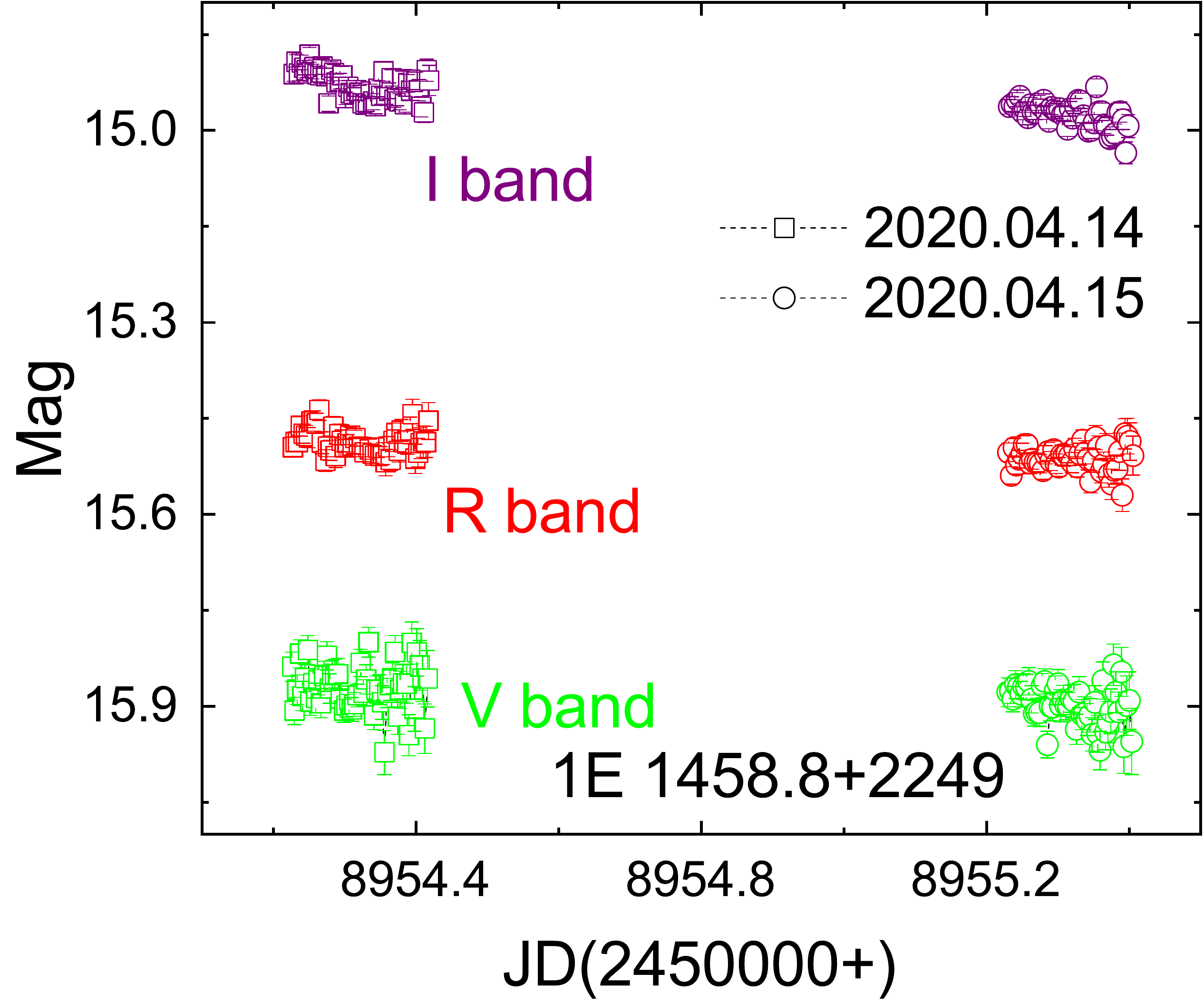}
\includegraphics[scale=.18]{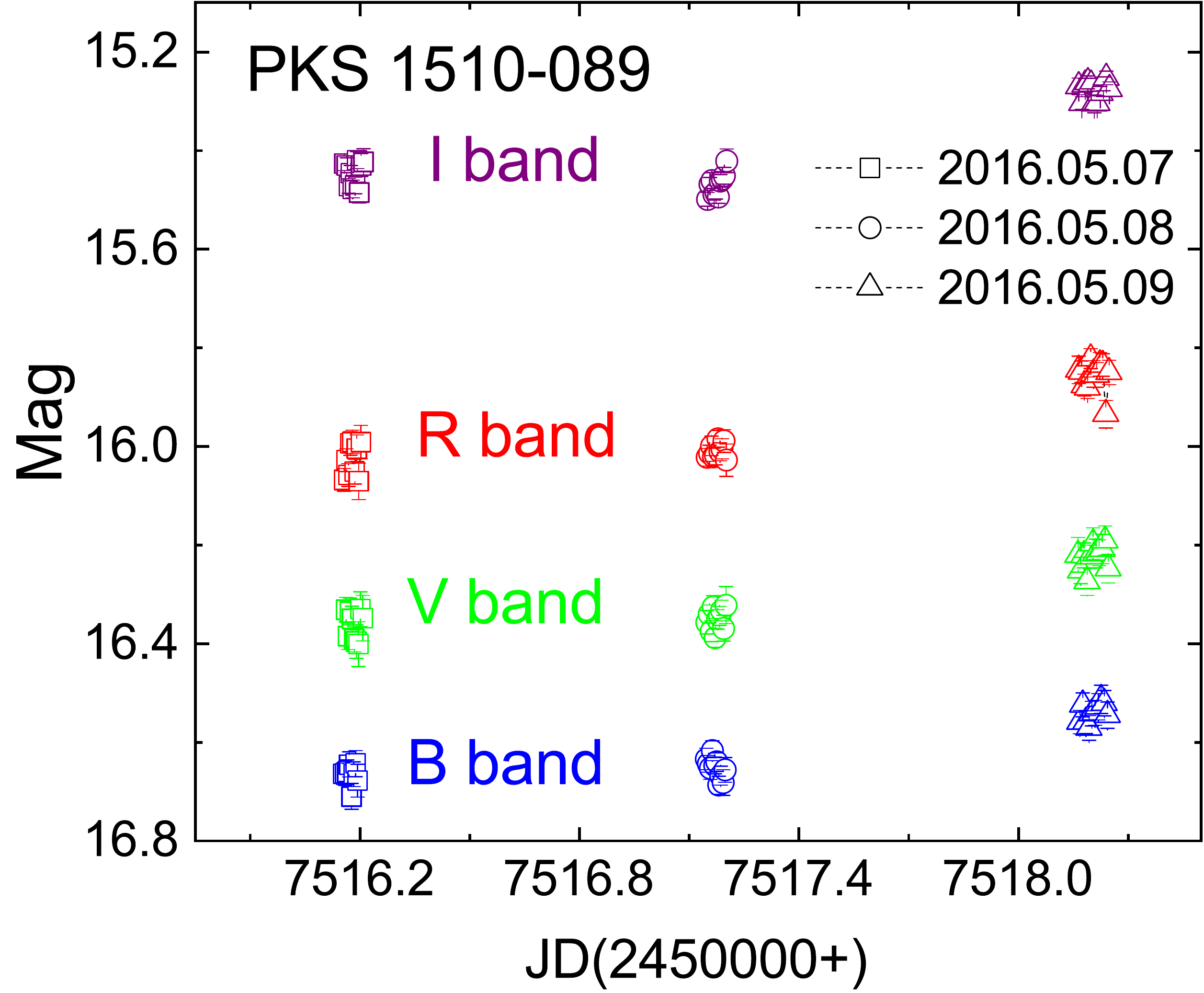}
\includegraphics[scale=.18]{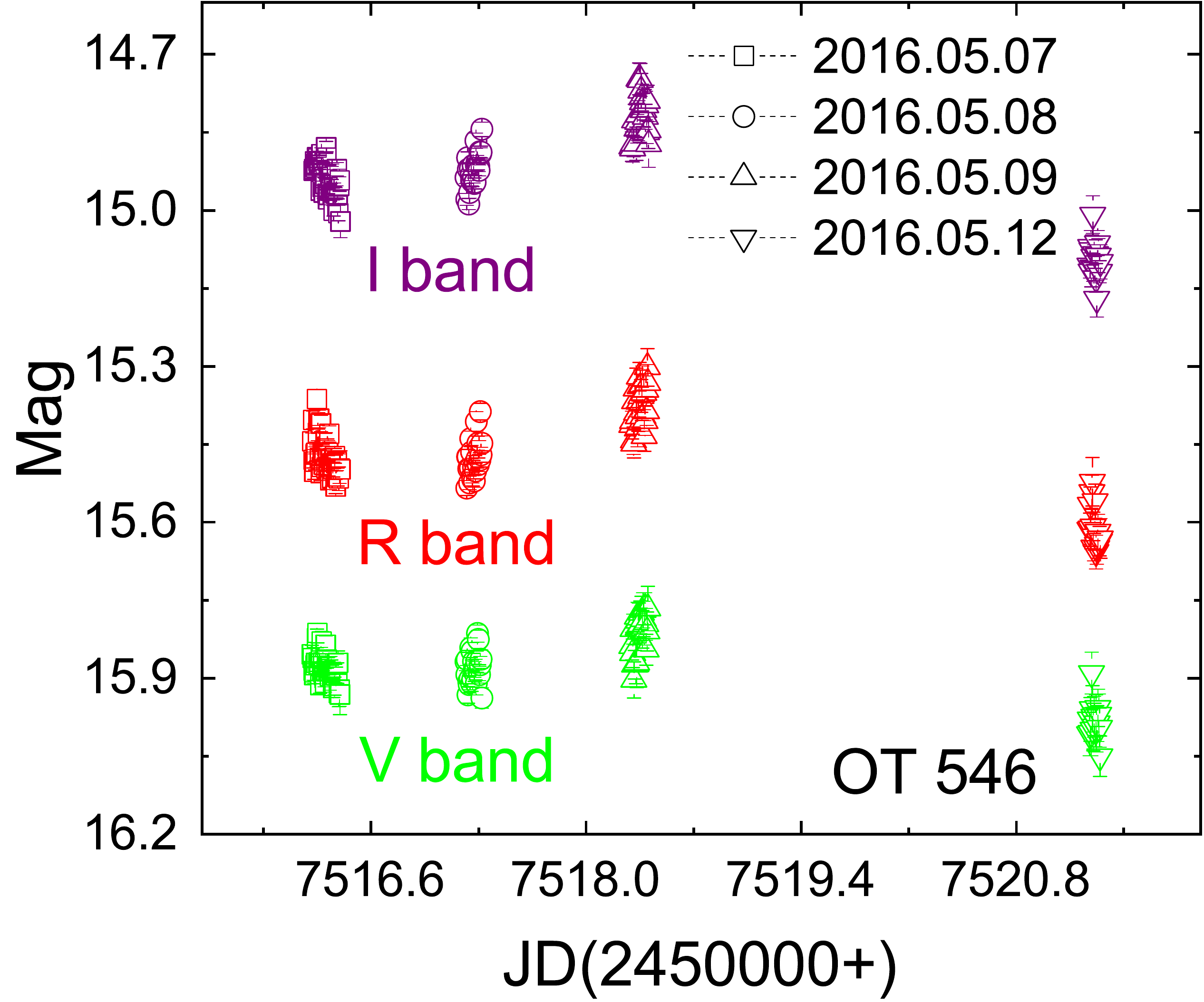}
\includegraphics[scale=.18]{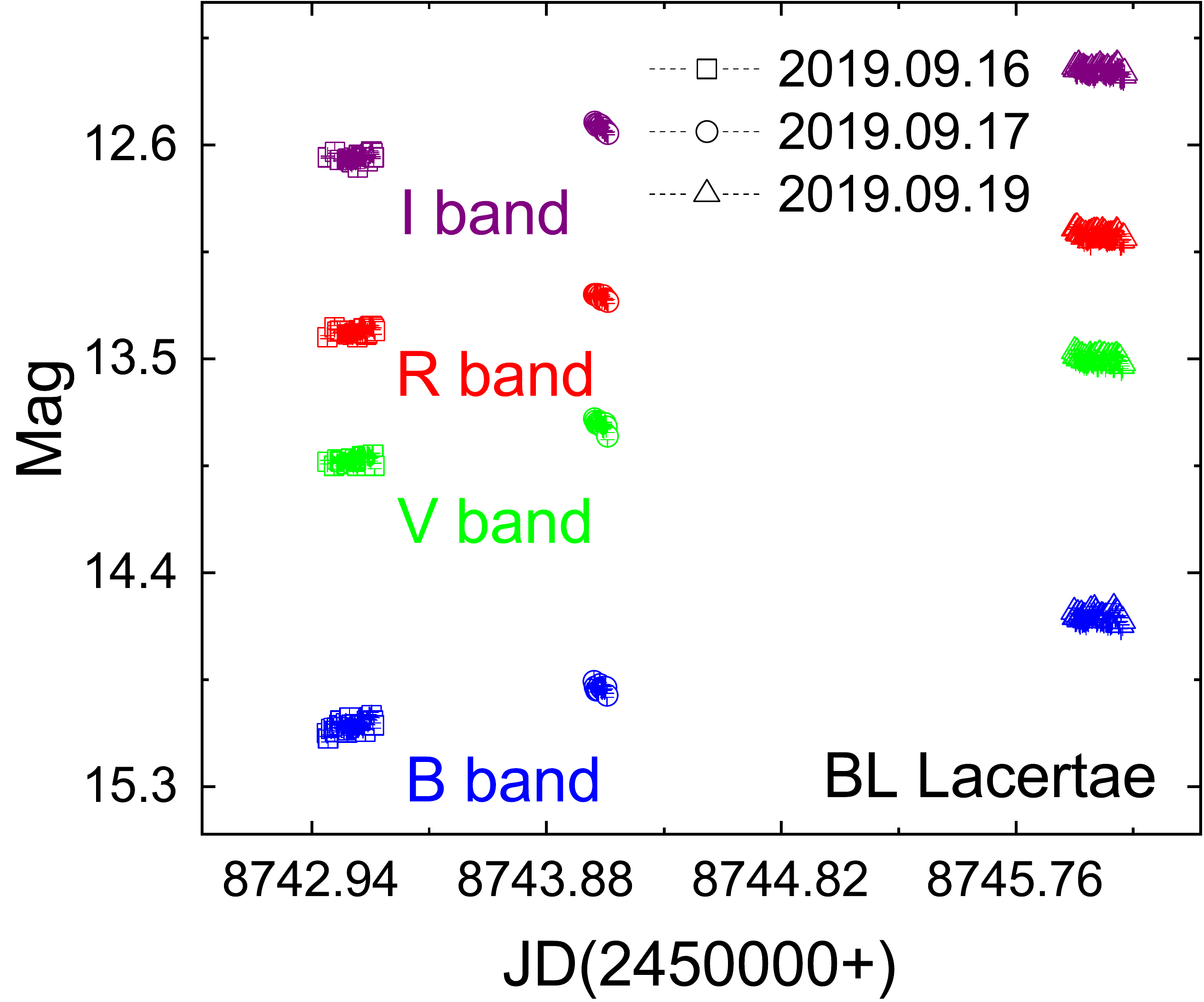}
\caption{Short-term light curves of the six sources. \label{}}
\end{figure*}

\begin{figure*}
\centering
\includegraphics[scale=.20]{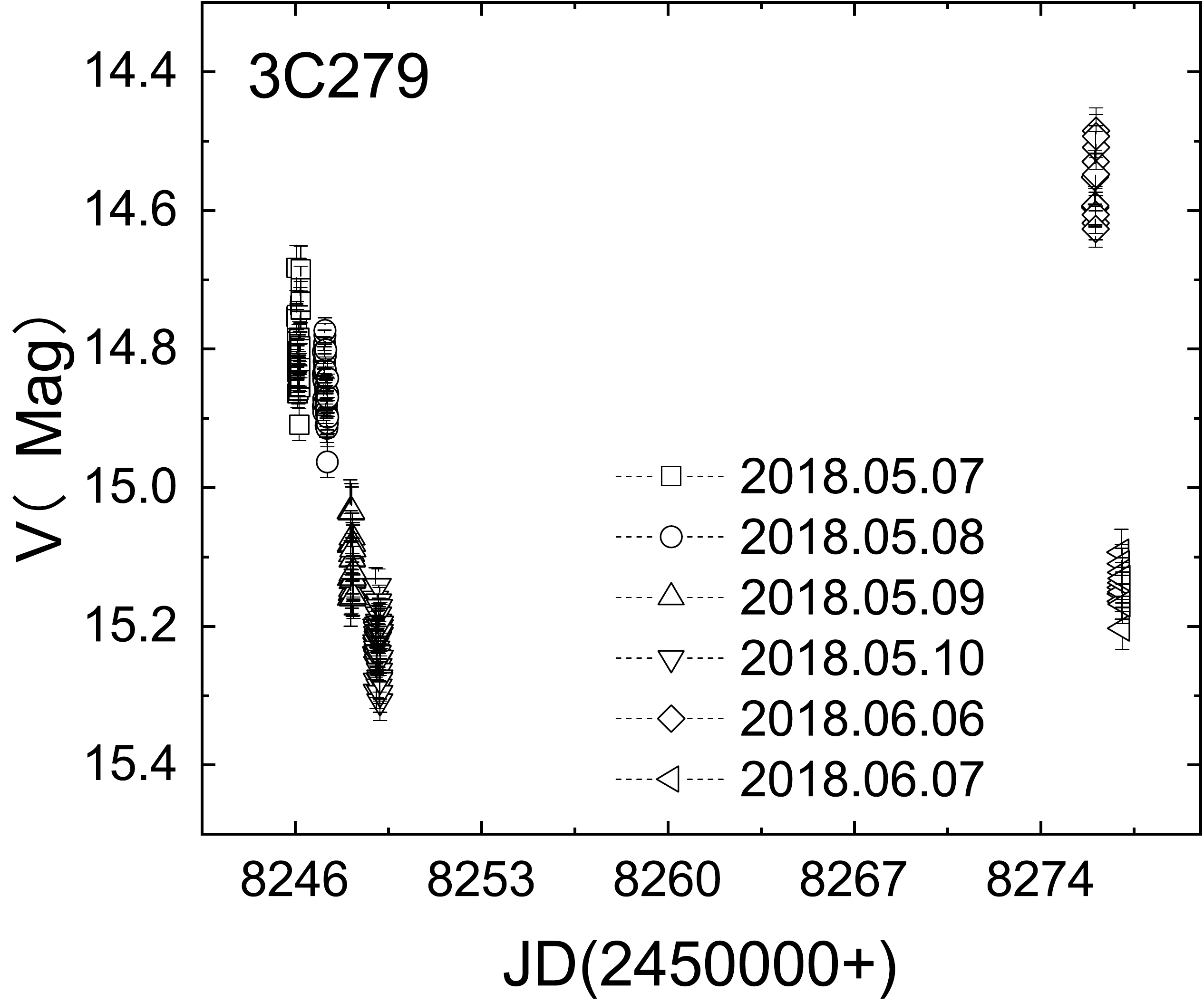}
\includegraphics[scale=.20]{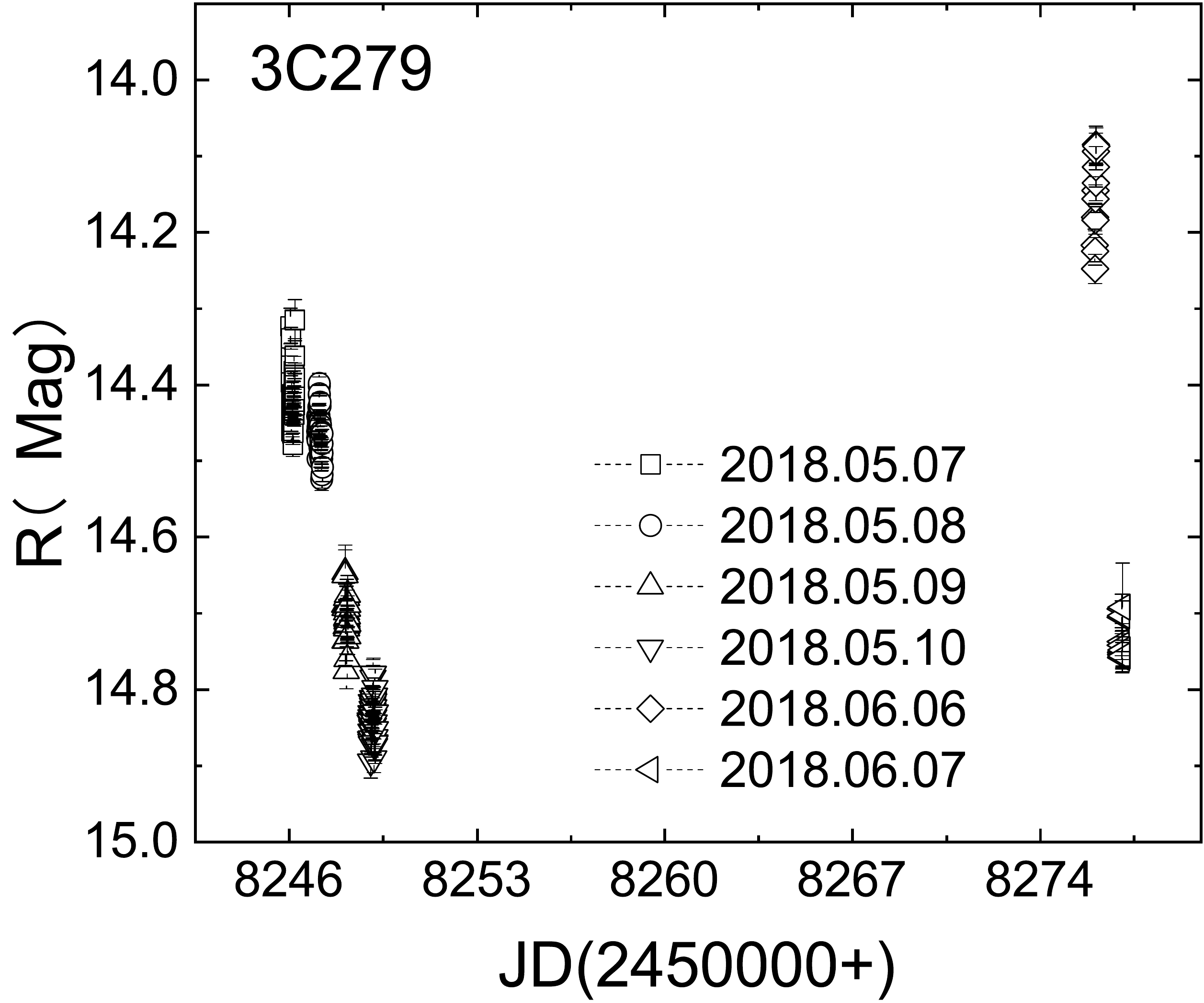}
\includegraphics[scale=.20]{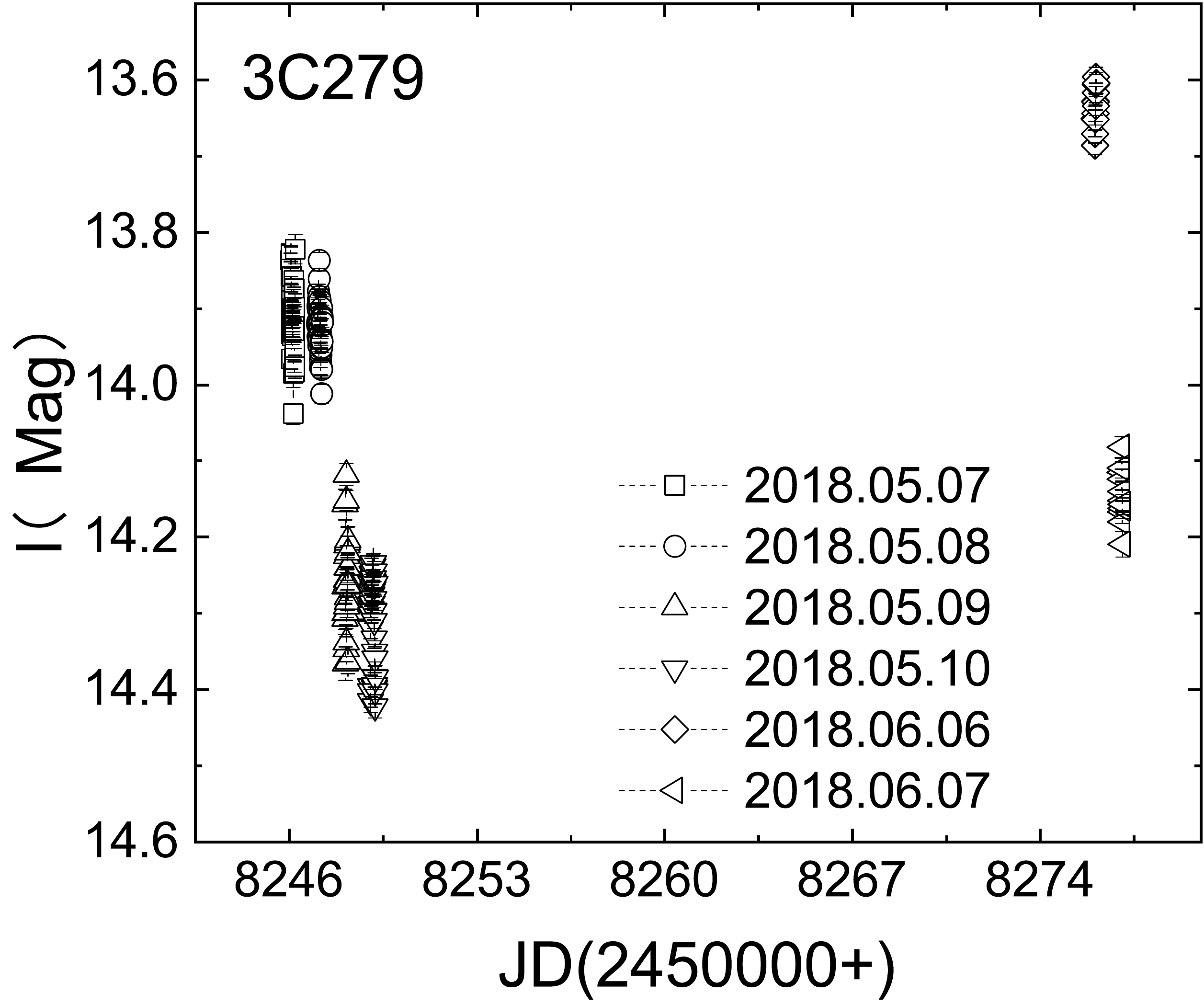}
\includegraphics[scale=.20]{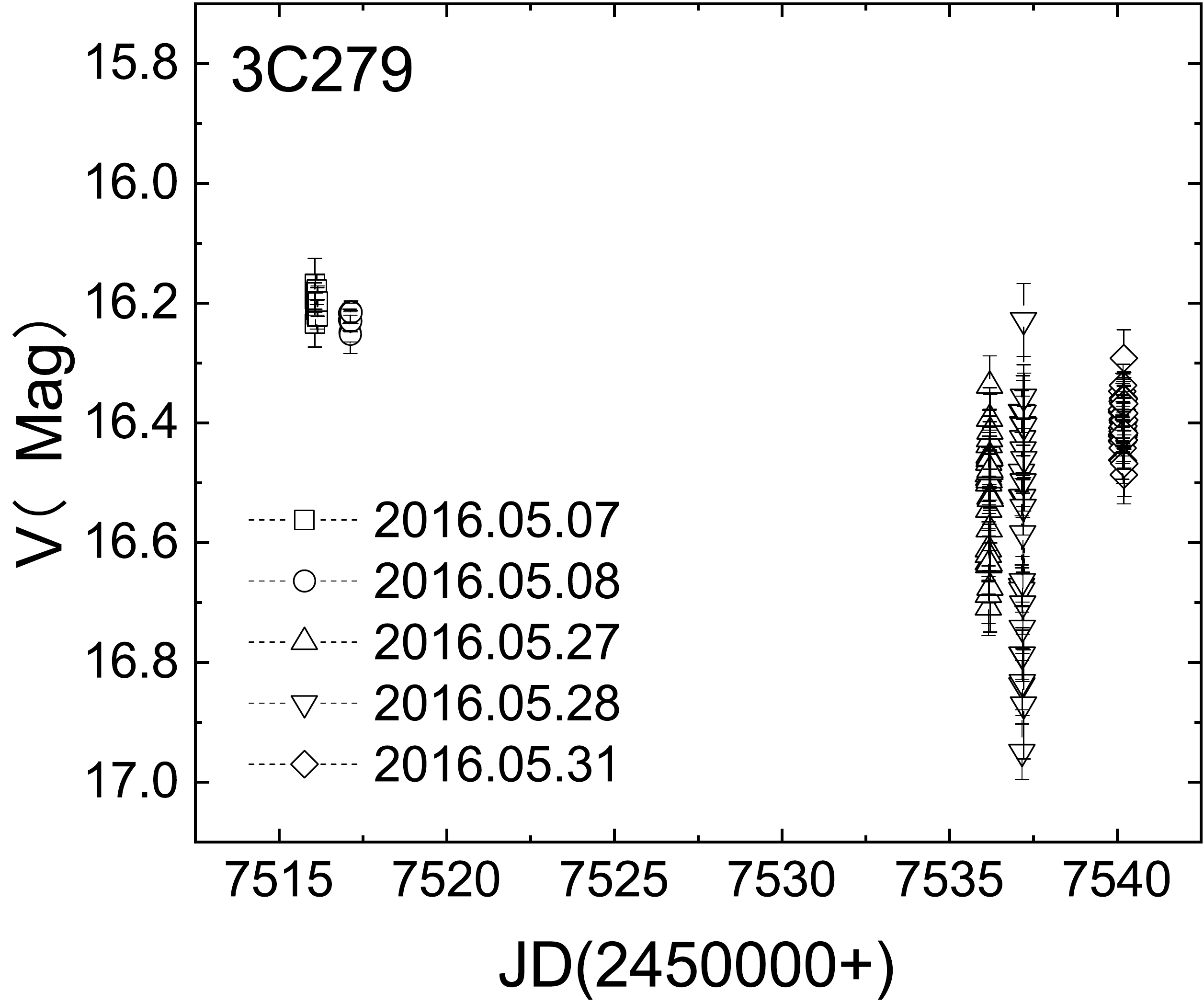}
\includegraphics[scale=.20]{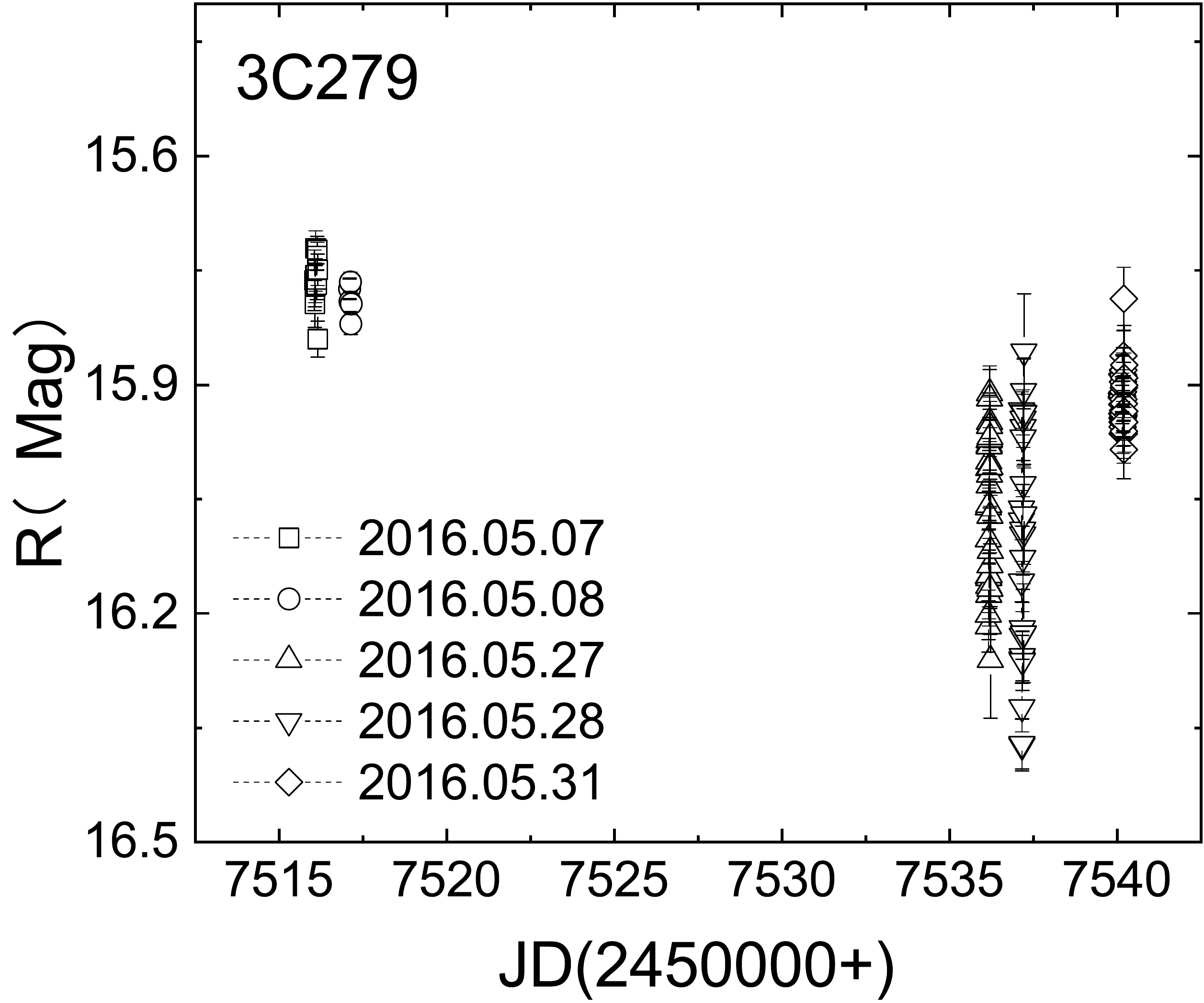}
\includegraphics[scale=.20]{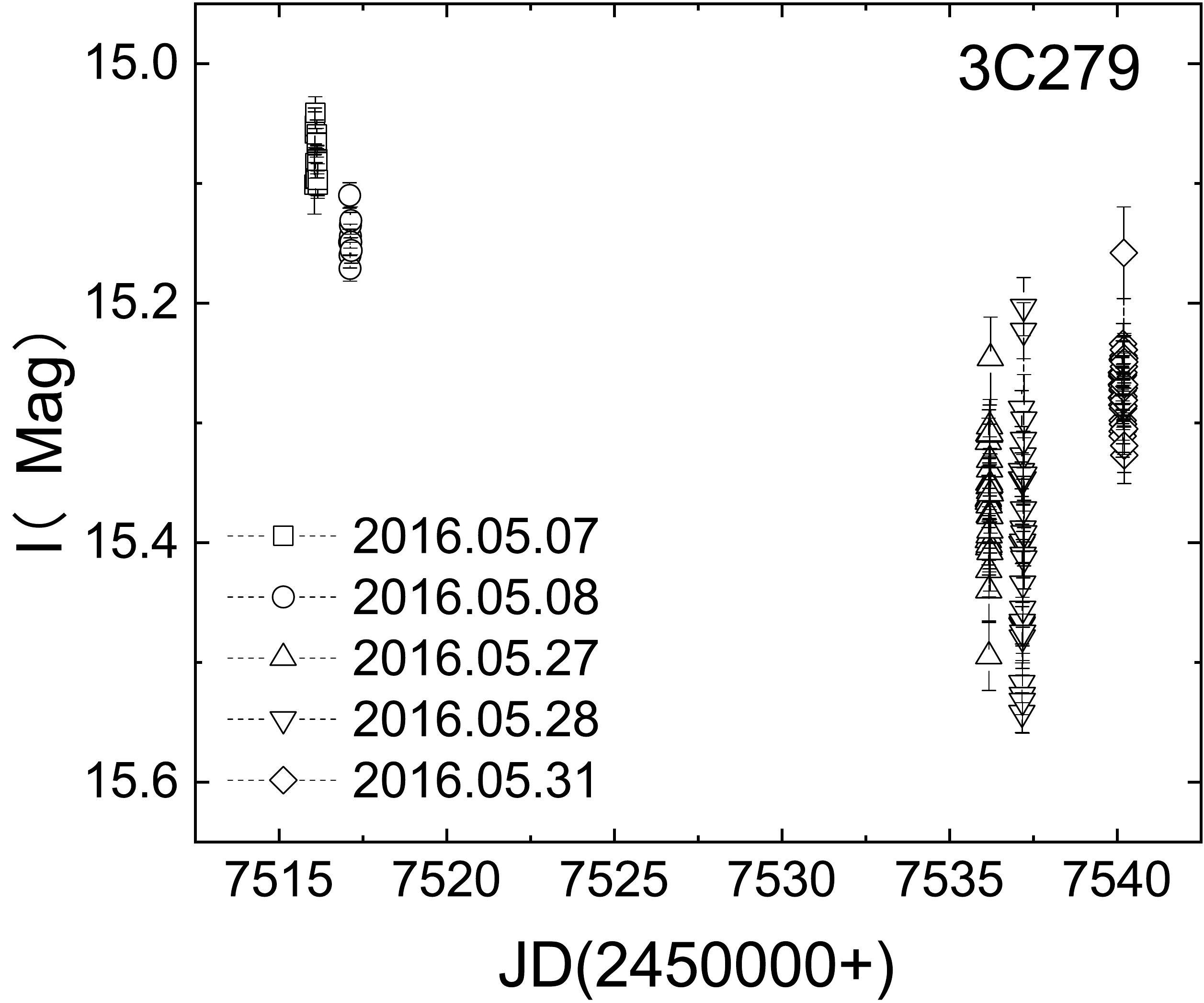}
\includegraphics[scale=.20]{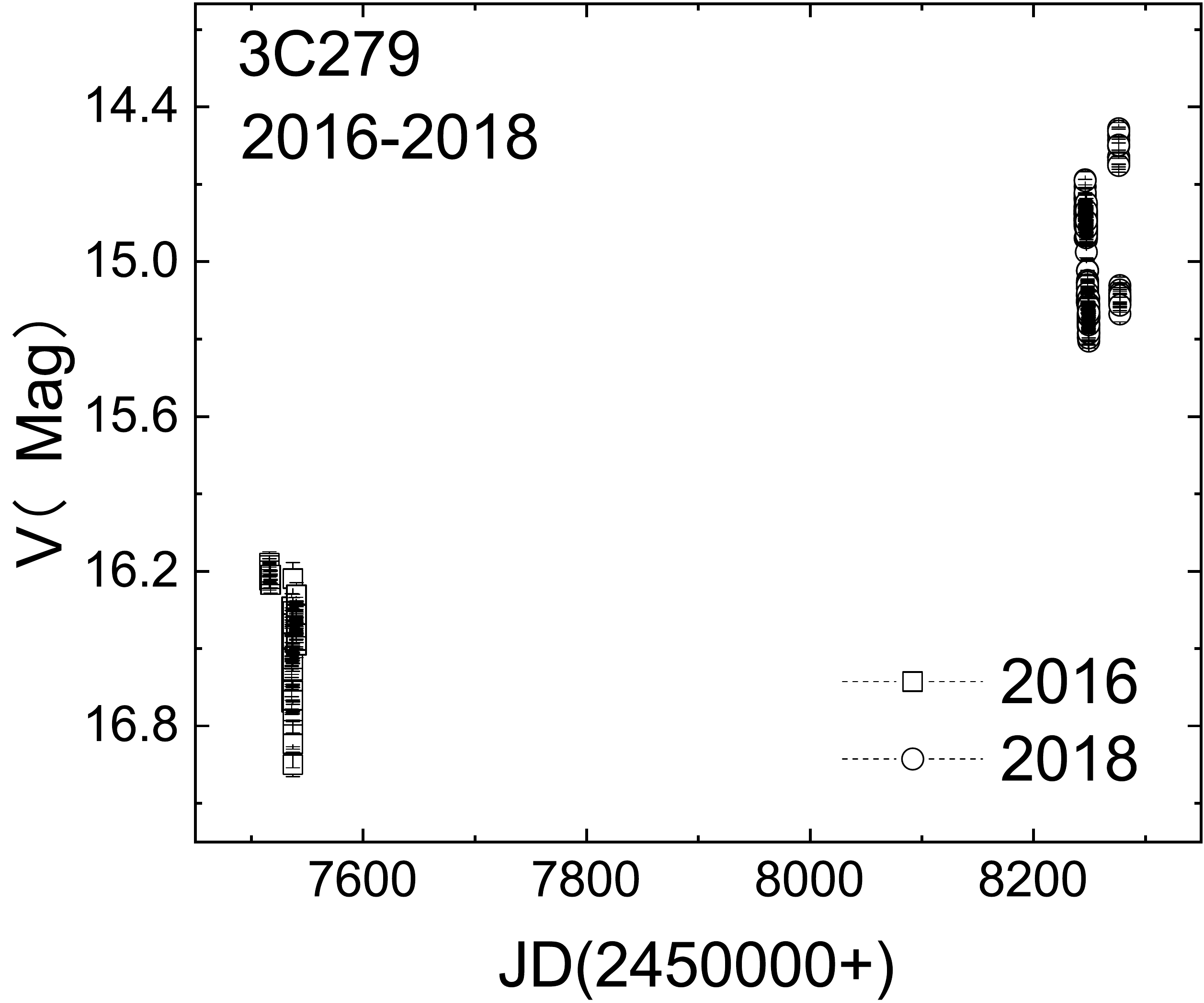}
\includegraphics[scale=.20]{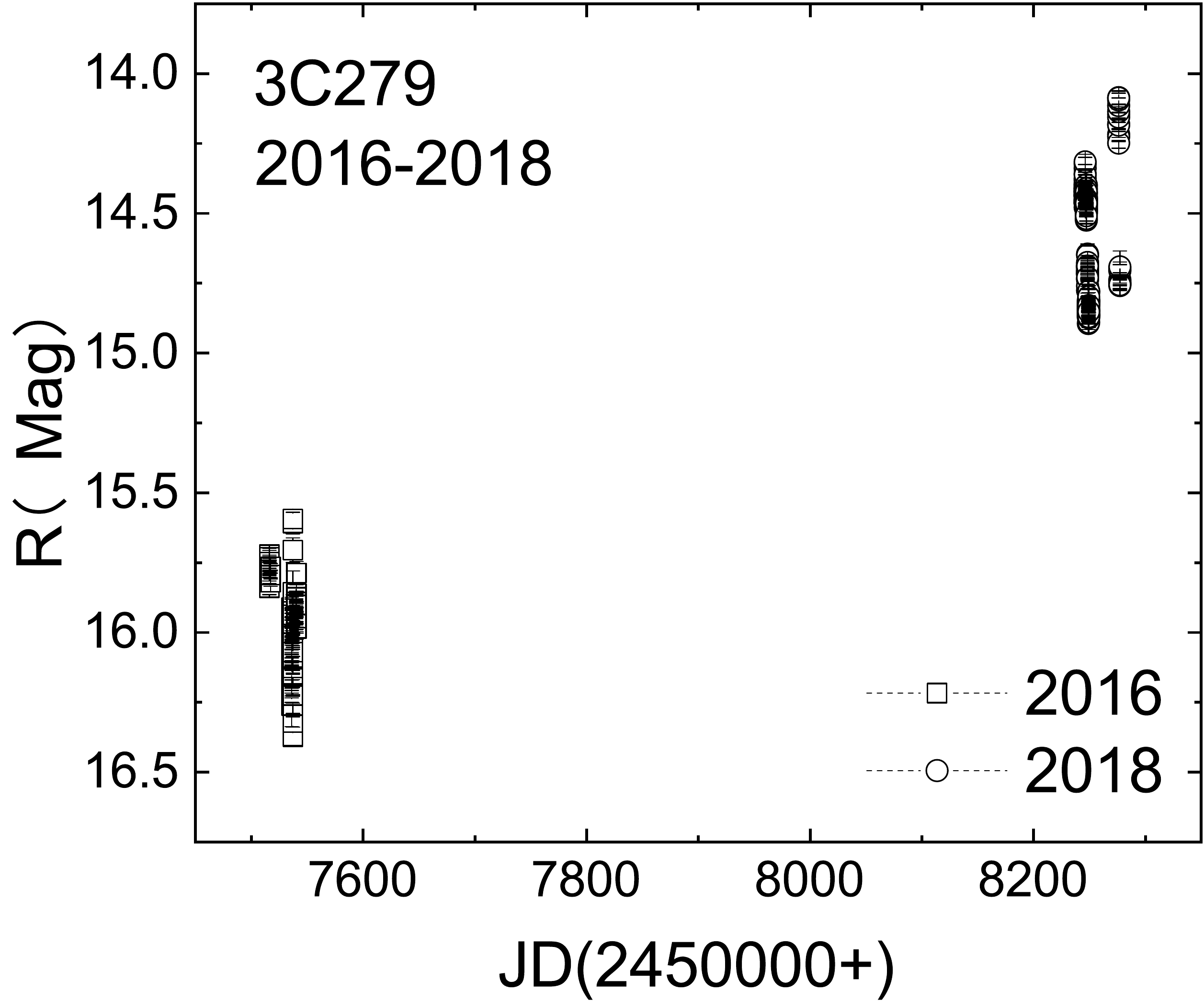}
\includegraphics[scale=.20]{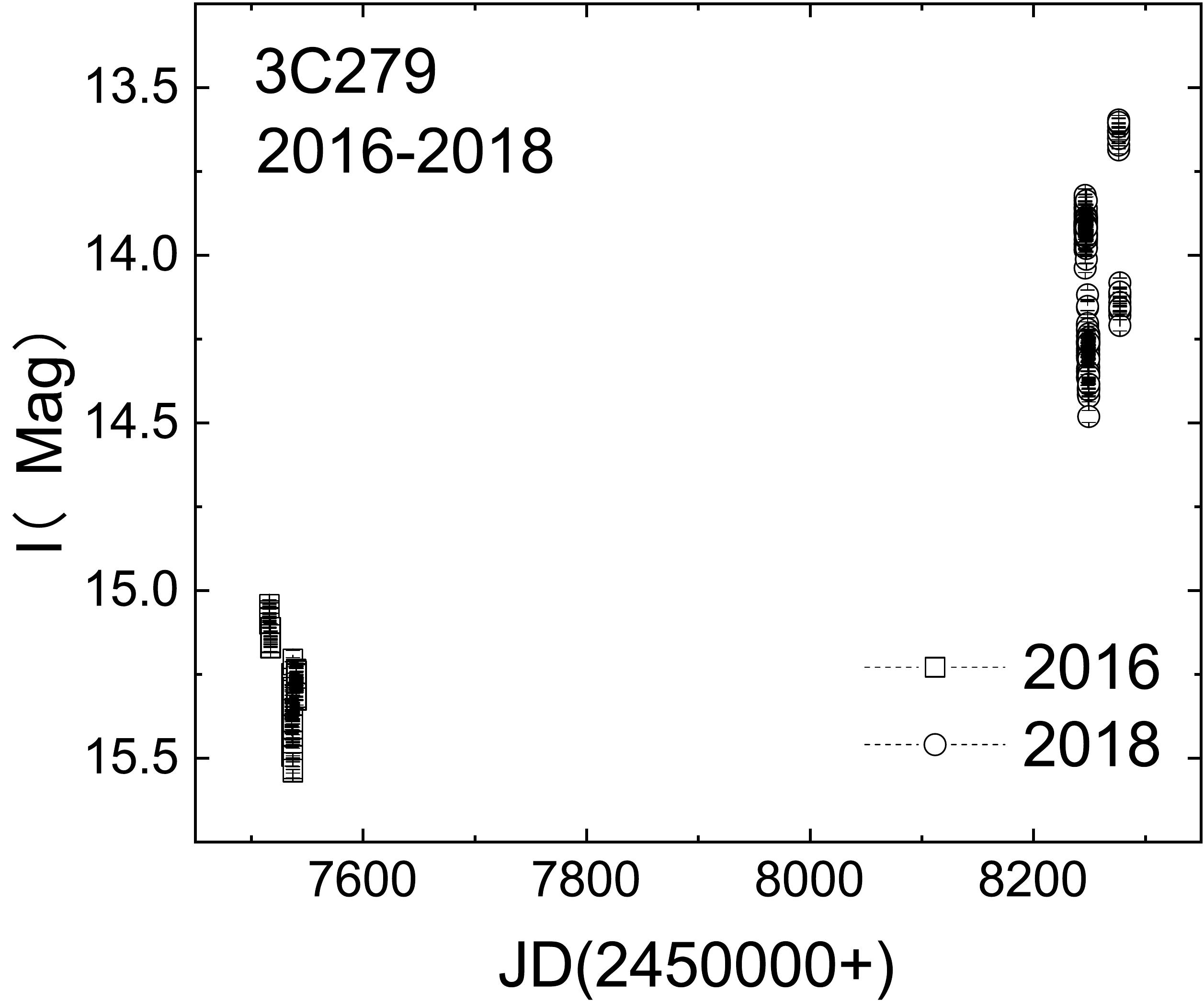}
\includegraphics[scale=.20]{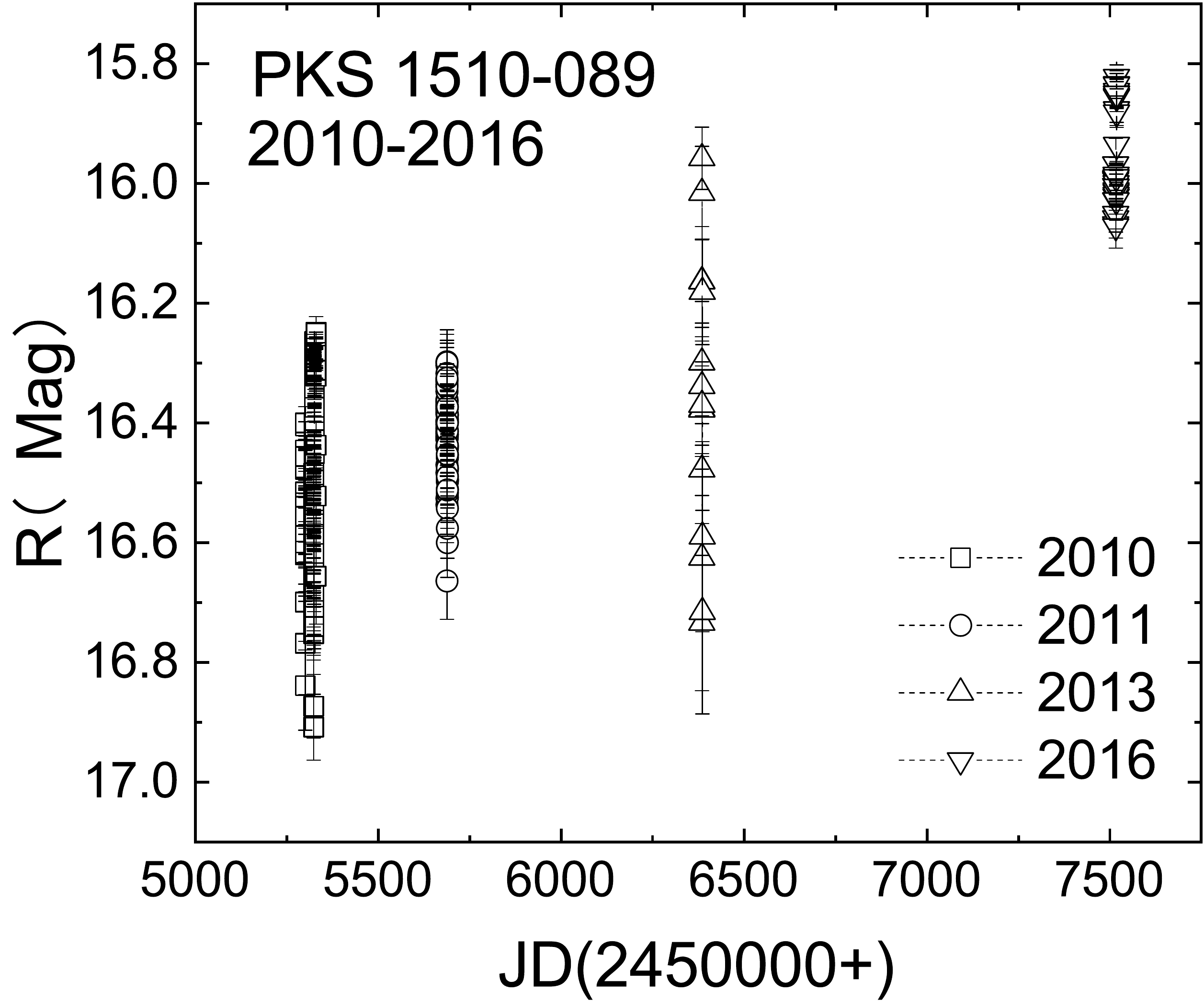}
\caption{Long-term light curves of 3C 279 and PKS 1510-089 for different periods and bands. \label{}}
\end{figure*}

\subsection{Auto-correlation Analysis and Variability Timescale}
We performed the auto-correlation function (ACF) analysis \citep{Alex97} to search for the characteristic timescale of variability. The ACF is written as
\begin{equation}
ACF(\tau)=\left\langle\left(m\left(t\right)-\left\langle m\right\rangle\right)\cdot\left(m\left(t+\tau\right)-\left\langle m\right\rangle\right)\right\rangle,
\label{eq:LebsequeIp10}
\end{equation}
where the brackets denote a time average. The ACF measures the correlation of the optical light curve with itself, shifted in time, as a function of the time lag $\tau$ \citep{Give99,Xion17}. The width of the ACF peak near zero time lag is proportional to the timescale if there is an underlying signal in the light curve with a typical variability timescale \citep{Give99,Liu08}. The zero-crossing time is the shortest time required for ACF to fall to zero \citep{Alex97}. Following \citet{Liu08}, \citet{Xion17} and \citet{Give99}, we choose the zero-crossing time of the ACF as the variability timescale. The variability timescale of the ACF is related to the characteristic size scale of the corresponding emission region \citep{Chat12}. The ACF was estimated by the code from \citet{Alex97}. Only nights detected with IDV are analyzed with ACF. The results of the ACF analysis that can determine the time delay are shown in Figure 4. Following \citet{Give99}, the fifth-order polynomial least-squares fitting was chosen to find the zero-crossing time, with the constraint that ACF ($\tau$ = 0) = 1. In order to obtain a better fitting effect on the light curves of 1ES 1426+42.8 (I band on April 13, 2010) and PKS 1510-089 (R band on May 06, 2010), we performed a ninth-order polynomial least-squares fitting for 1ES 1426+42.8, and an eighth-order polynomial least-squares fitting for PKS 1510-089. We adopted an error-weighted polynomial fitting method when performing polynomial fit. The fitting results show that the detected variability timescales of IDVs are 0.0322 day for the $I$ band of ON 231 on May 08, 2014, 0.01 day for the $I$ band of 1ES 1426+42.8 on April 13, 2010, and 0.0295, 0.0298 and 0.036 day for the $R$ band of PKS 1510-089 on May 06, 2010, May 08, 2010, and May 06, 2011. The minimum characteristic variability timescale of PKS 1510-089 is 0.0295 day.

\subsection{Black Holes Mass Estimation}
 The observed minimum timescale of variability are widely used to estimate the masses of the central black holes in blazars \citep[e.g.,][]{Abra82,Mill89,Liu15}. \citet{Liu15} proposed a new sophisticated model to limit the black hole mass $M$$_{\bullet}$ using the rapid variations for blazars as follows:
\begin{equation}
M_{\bullet} \lesssim \ 5.09 \times 10^4 \frac{\delta\Delta t^{\ ob}_{min}}{1+z} M_{\odot}  \qquad  (j \sim 1),
\label{eq:LebsequeIp3}
\end{equation}
\begin{equation}
M_{\bullet} \lesssim \ 1.70 \times 10 ^4 \frac{\delta\Delta t^{\ ob}_{min}}{1+z} M_{\odot}  \qquad (j = 0),
\label{eq:LebsequeIp4}
\end{equation}
where $\Delta t^{\ ob}_{min}$ is the minimum timescale in units of seconds, $z$ is the redshift, and $j = J/J_{max}$ is the dimensionless spin parameter of a black hole with the maximum possible angular momentum $J_{max} = GM^2_{\bullet}/c$ in which $G$ is the gravitational constant. Equations (6) is suitable for Kerr black holes. Equations (7) can be applied to Schwarzschild black holes.

We estimated the masses of black holes of ON 231, 1ES 1426+42.8 and PKS 1510-089. According to the results of ACF, the minimum timescales of optical IDV are $\tau$ = 0.0322 day, $\tau$  = 0.01 day, and $\tau$  = 0.0295 day, respectively. The Doppler factors are reported as $\delta$ = 1.56, $\delta = 27.3$ and $\delta$ = 13.18, respectively in the literatures \citep{Laht99,Gaur10}. Following Equations (6) and (7), the black hole mass of ON 231 is the $M_{\bullet} \lesssim 10^{8.30} M_{\odot}$ for the Kerr black hole and $M_{\bullet} \lesssim 10^{7.82} M_{\odot}$ for the Schwarzschild black hole. This is consistent with the $10^{8.38} M_{\odot}$ for the rapidly spinning case and $10^{7.58} M_{\odot}$ for the non-rotating case reported by \citet{Gaur10} derived as the relation between the period and the mass of SMBH. 1ES 1426+42.8 has $M_{\bullet} \lesssim 10^{9.03} M_{\odot}$ (Equation (6)) and $M_{\bullet} \lesssim 10^{8.55} M_{\odot}$ (Equation (7)). \citet{Woo02} obtained $M$$_{\bullet}$ = $10^{9.13} M_{\odot}$ of 1ES 1426+42.8 using stellar velocity dispersion which was estimated from an indirect method. \citet{Wu09} estimated the black hole mass $M_{\bullet} = 10^{8.51} M_{\odot}$ in 1ES 1426+42.8 using the $R$ band magnitudes of host galaxy. Thus, our estimates of black hole mass for 1ES 1426+42.8 are consistent with the black hole mass in \citet{Wu09}. In the case of PKS 1510-089, $M_{\bullet} \lesssim 10^{9.01} M_{\odot}$ is derived with Equation(6) and $M_{\bullet} \lesssim 10^{8.62} M_{\odot}$ is derived with Equation (7). \citet{Woo02} obtained $M$$_{\bullet}$ = $10^{8.65} M_{\odot}$ of PKS 1510-089 using the BLR size-luminosity relation, which is consistent with our result of the Kerr black hole.

\begin{figure*}
\centering
\includegraphics[scale=.24]{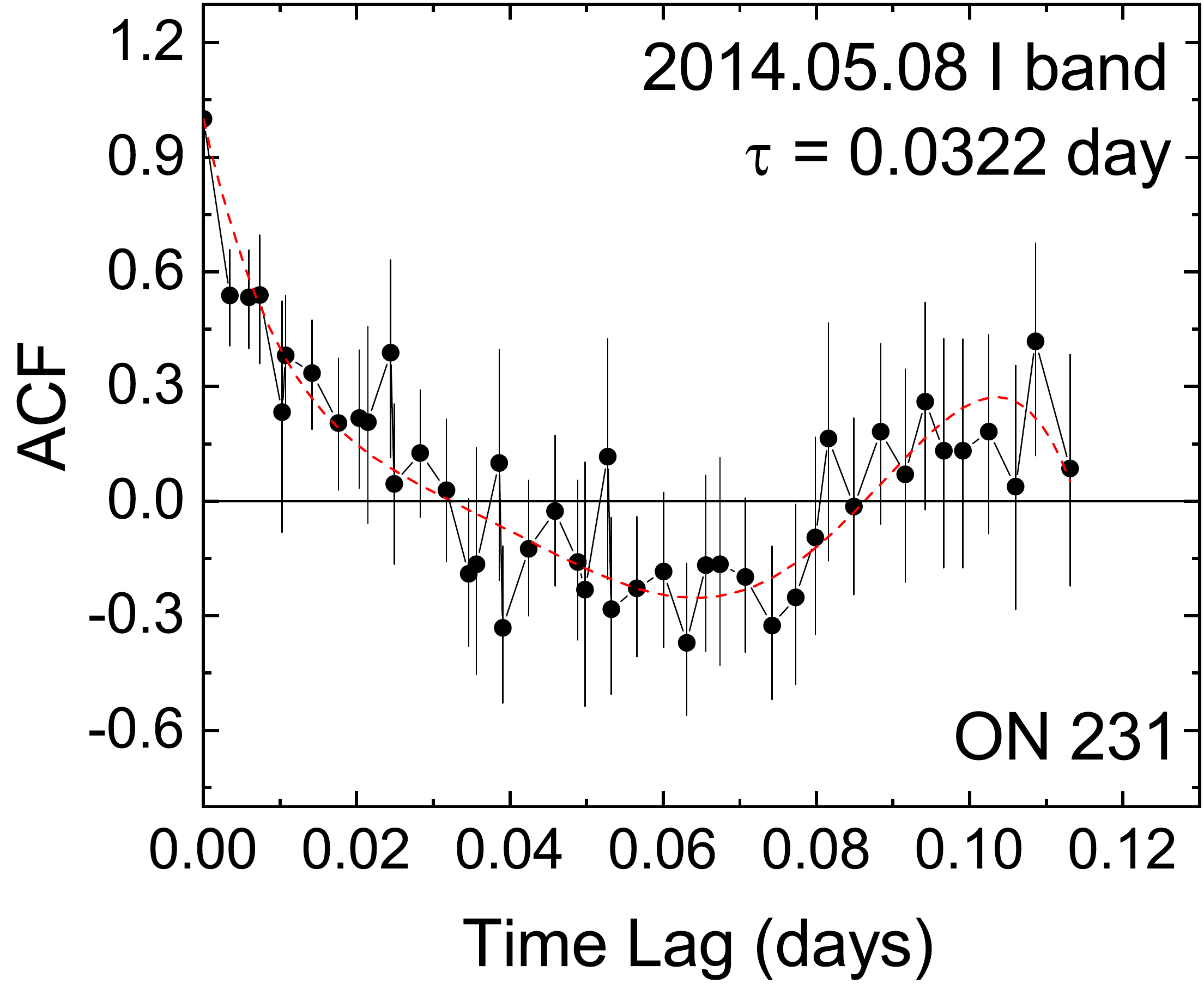}
\includegraphics[scale=.24]{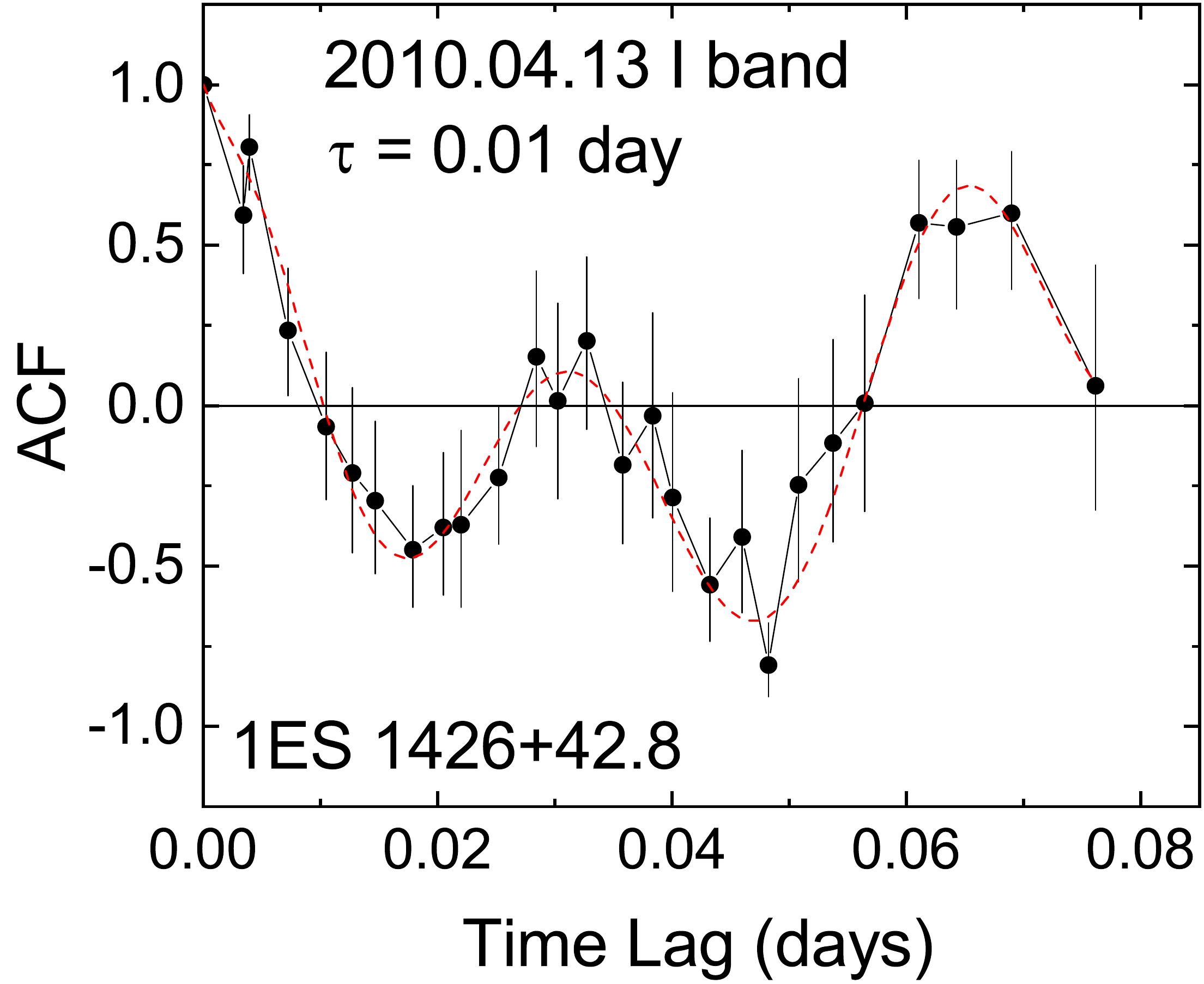}
\includegraphics[scale=.24]{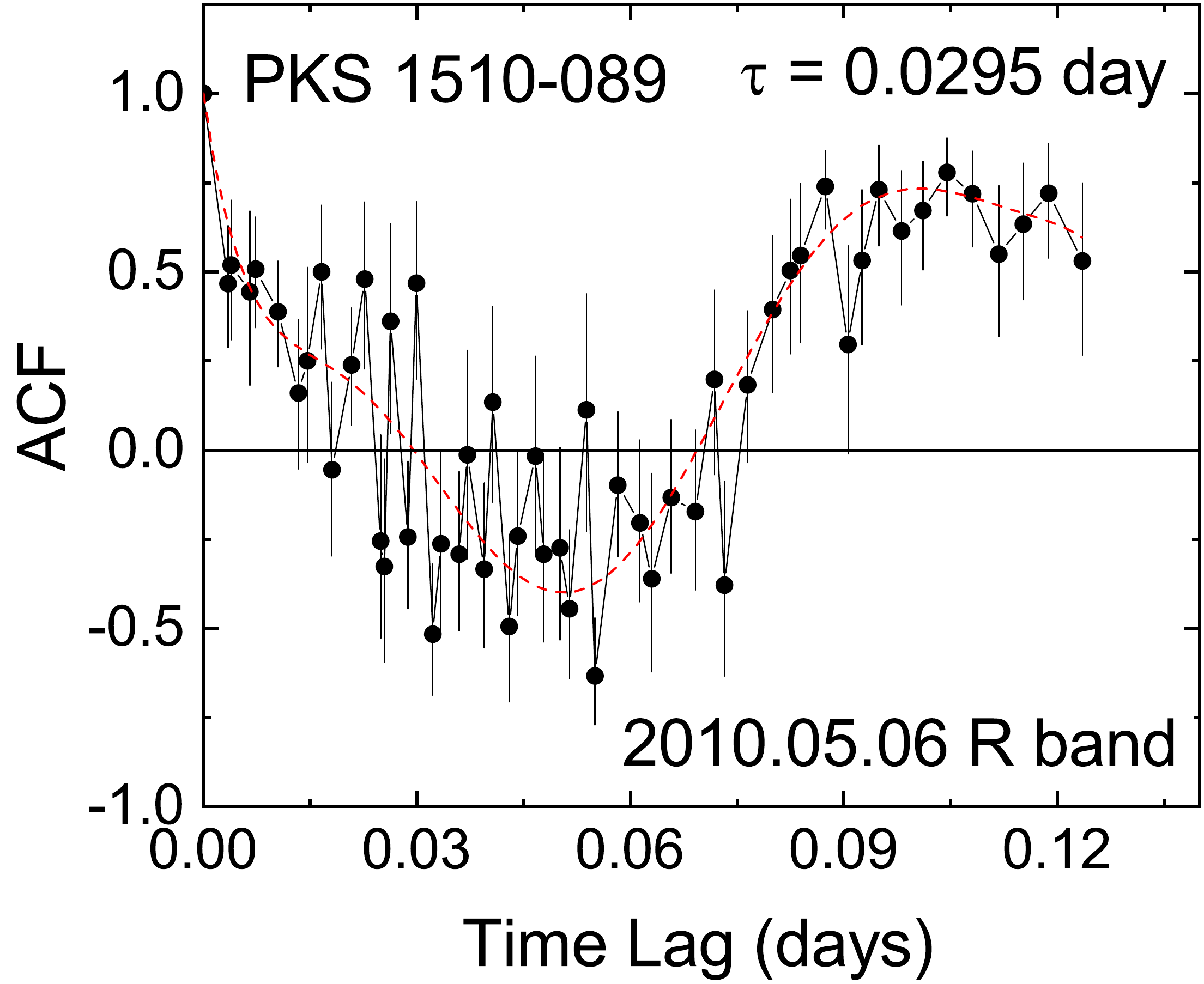}
\includegraphics[scale=.24]{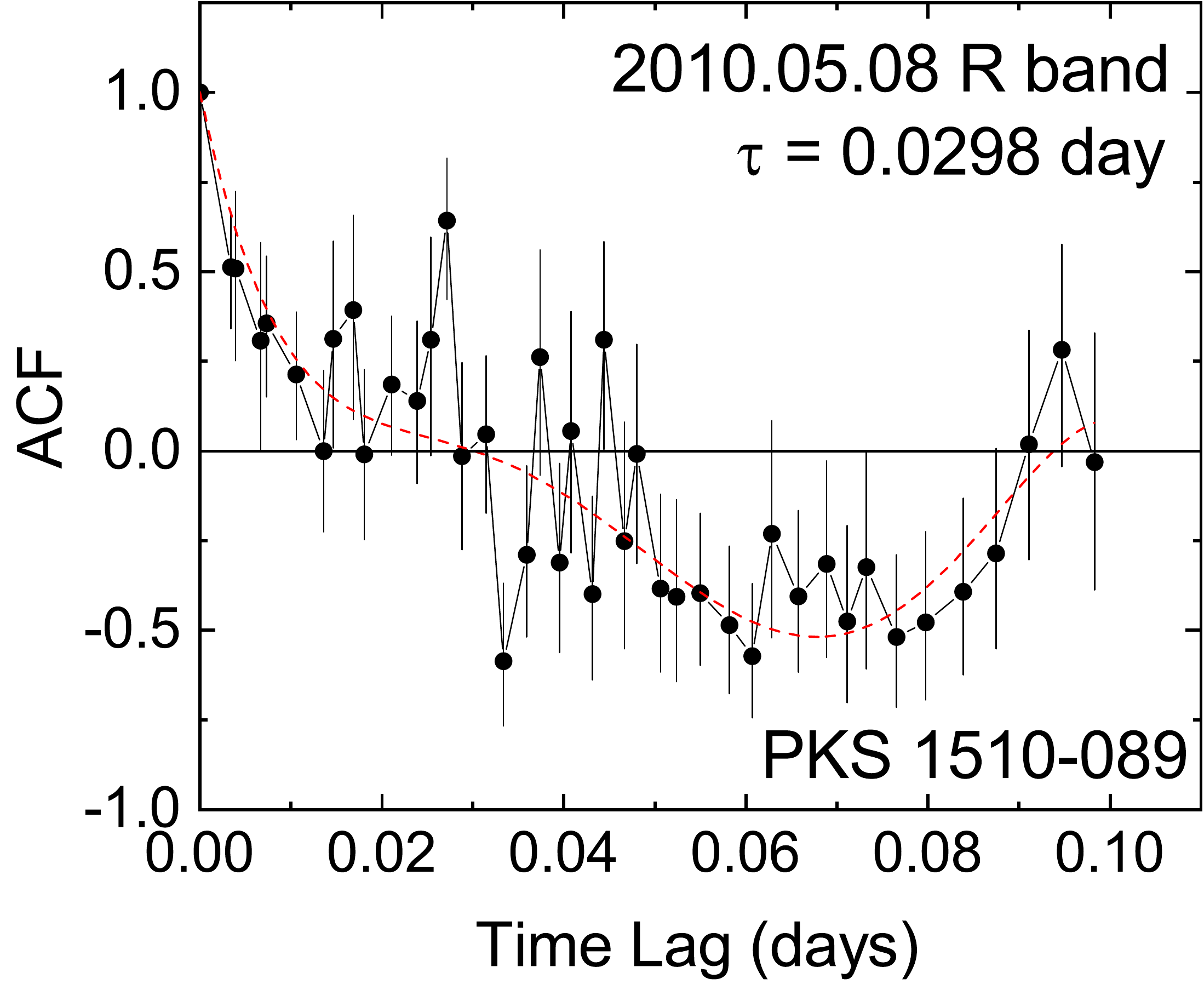}
\includegraphics[scale=.24]{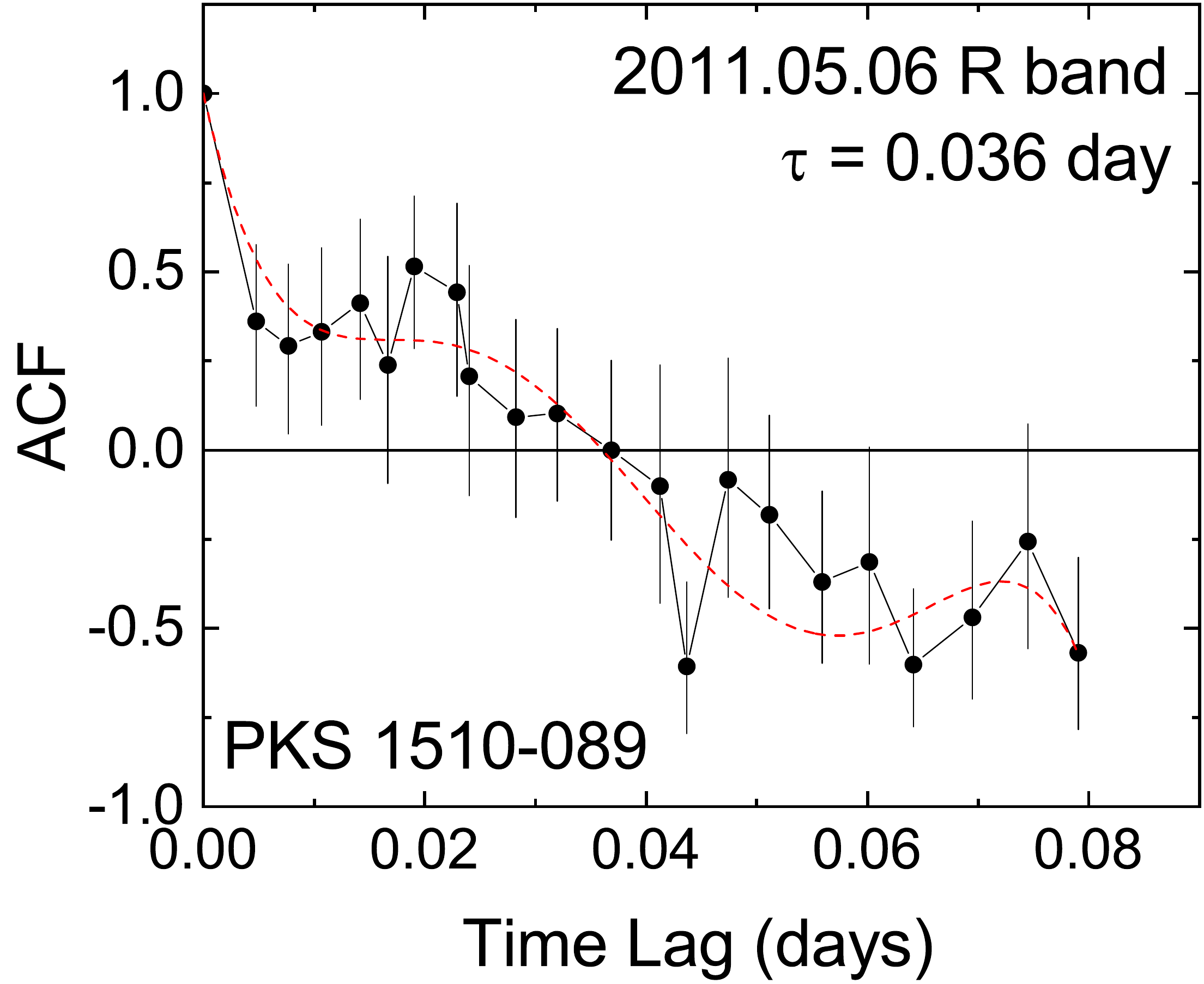}
\caption{Results of the ACF analysis for the three blazars. The red dashed line is a polynomial least-squares fit. \label{}}
\end{figure*}

\subsection{The Correlation between Magnitudes and Colours}

In order to explore the optical spectral behaviors, we analyzed the correlations between the magnitudes and colours on intraday timescales, short-term timescales, and long-term timescales. In this Section, we focus on the correlation between $V - I$ index and $V$ magnitude, which is widely studied. Only data obtained by quasi-simultaneous measurements are analyzed. We use the correlation coefficient from error-weighted linear regression analysis to indicate the intensity of the BWB trend. The results of Spearman correlation coefficient analysis and the corresponding $p$-values are given in Table 4. In Table 4, $r$ is the Spearman correlation coefficient, and $p$ is the chance probability. Generally, if $r$ is positive, it means positive correlation, and if $r$ is negative, it means negative correlation. The absolute value of $r$ of 0-0.1 means no correlation, 0.1-0.3 means weak correlations, 0.3-0.5 means moderate correlations, and 0.5-1.0 means strong correlations \citep{Cohe88}. The $p$-value is a parameter that indicates whether the correlation is significant or not. If the $p$-value is less than 0.001, it indicates that there is a significant correlation; if the $p$-value is greater than 0.05, it indicates that there is no correlation \citep{Cohe88}.

Figure 5 shows the correlations between the $V - I$ index and $V$ magnitude on intraday timescales. Figure 6 shows the correlations between the $V - I$ index and $V$ magnitude on short-term timescales and long-term timescale, respectively. Table 4 and Figure 5 show that the BWB trend appears in twenty-one nights on the intraday timescales, of which twenty nights have strong correlations ($p< 0.001$) and one night has moderate correlation (3C 279 on May 10, 2018). On the short-term timescales, four objects (ON 231, 3C 279, 1E 1458.8+2249, BL Lacertae) have strong correlations, two objects (OJ 287, OT 546) have no correlations, and one object (3C 279 in May-June 2018) has a weak correlation ($r = 0.208$, $p = 0.004$). On the long-term timescales, 3C 279 shows a strong correlation from 2016 to 2018 ($r = 0.835$, $p < 0.0001$). An insignificant RWB trend is detected on short-term timescale for PKS 1510-089 ($r = -0.431$, $p = 0.017$; see Figure 6). Therefore, the BWB chromatic trend is dominant for our objects on intraday timescales. Except for three sources, the other four targets all display the BWB trend on the short timescales. There is one target with BWB trend on the long timescales. A FSRQ with RWB trend is found.

\subsection{Search for Periodicity in 1ES 1426+42.8}
The study of blazars' QPO is of great value for exploring the physical mechanism and radiation process in blazars \citep{Gupt09}. In our observations, most of the light curves of IDV are monotonous. The light curve of 1ES 1426+42.8 on April 13, 2010 is the only one that had three obvious peaks. It was likely periodic, so we performed a periodic analysis for the 1ES 1426+42.8. At present, Lomb-Scargle periodogram (LSP) \citep{Lomb76, Scar82} and weighted wavelet Z-transform (WWZ) \citep{Fost96,Bhat17} are commonly used for investigating the periodicity of variability in the light curves. We used these two methods to explore whether there is QPO in the $I$ band light curve of 1ES 1426+42.8.

LSP is a method widely used in the search of QPOs, which can be used to eliminate the aliasing problem caused by unevenly sampled data \citep{Lomb76, Scar82, Pres92}. The LSP method extracts the average value of the signal and employs a phase shift of the basis functions. The resulting normalized periodogram does not need to interpolate the missing data. Therefore, false artificial peaks are avoided \citep{Wang14}. The periodogram is defined as \citep{Li16}:
\begin{equation}
\begin{aligned}
P_X(f)=\frac{1}{2}\left\{\frac{\left[\sum^N_{i=1}X(t_i)\cos[2\pi{f}(t_i-\tau)]\right]^2}{\sum^N_{i=1}\cos^2[2\pi{f}(t_i-\tau)]}\right. \\ \left.+\frac{\left[\sum^N_{i=1}X(t_i)\sin[2\pi{f}(t_i-\tau)]\right]^2}{\sum^N_{i=1}\sin^2[2\pi{f}(t_i-\tau)]}\right\},
\label{eq:LebsequeIp6}
\end{aligned}
\end{equation}
where $X(t_i)(i=1,2,3,...,N)$ is a time series, $f$ is the test frequency, and $\tau$ is the time offset, which can be calculated by the formula:
\begin{equation}
\tau=\frac1{2\omega}\tan^{-1}\left[\frac{\sum^N_{i=1}\sin2\omega t_i}{\sum^N_{i=1}\cos2\omega t_i}\right],
\label{eq:LebsequeIp7}
\end{equation}
where $\omega = 2\pi{f}$. For the real signal $X(t_i)$, the power in $P_X(f)$ would present a peak, and the most likely period corresponds to the peak of the maximum power $P_{peak}(f)$ \citep{Wang14}.

The WWZ is a periodicity analysis method in both the time and frequency domains \citep{Li21b}, which can be used to cross-validate the possible QPO of 1ES 1426+42.8. The WWZ method can effectively handle irregularly sampled variability data by introducing the weighted projection of the light curves on trial functions \citep{Wang14}. Following \citet{Fost96} and \citet{Bhat17}, we constructed the WWZ spectra using the Morlet mother function for each artificial light curve. WWZ projects the Morlet wavelet transforms onto three trial functions: $\varphi_1(t)=1(t)$, $\varphi_2(t)=\cos\left[\omega\left(t-\tau\right)\right]$, $\varphi_3(t)=\sin\left[\omega\left(t-\tau\right)\right]$, and also includes statistical weights $\omega_\alpha=exp\left[-c\omega^2{(t_\alpha-\tau)}^2\right],\;(\alpha=1,2,3)$ on the projection, where $c$ is an adjustable parameter \citep{Li21b}. The WWZ power estimates the confidence level of a detected periodicity with frequency $\omega$ and time shift $\tau$ as follows:
\begin{equation}
WWZ=\frac{(N_{eff}-3)V_y}{2(V_x-V_y)},
\label{eq:LebsequeIp8}
\end{equation}
where $N_{eff}$ is the effective number of data points contributing to the signal, and $V_x$ is the weighted variations of the uneven data $x$, while $V_y$ is the model function $y$, respectively. These factors are defined as follows:
\begin{equation}
\begin{aligned}
&N_{eff}=\frac{{(\sum\omega_\alpha)}^2}{\sum\omega_\alpha^2},\\
&V_x=\frac{\sum_\alpha\omega_\alpha x^2(t_\alpha)}{\sum_\lambda\omega_\lambda}-\left[\frac{\sum_\alpha\omega_\alpha x(t_\alpha)}{\sum_\lambda\omega_\lambda}\right]^2,\\
&V_y=\frac{\sum_\alpha\omega_\alpha y^2(t_\alpha)}{\sum_\lambda\omega_\lambda}-\left[\frac{\sum_\alpha\omega_\alpha y(t_\alpha)}{\sum_\lambda\omega_\lambda}\right]^2,
\label{eq:LebsequeIp9}
\end{aligned}
\end{equation}
where $\lambda$ is the number of test frequencies. Then, the WWZ periodogram can be obtained by decomposing the data into observing epoch and time/frequency domain, and the peaks of WWZ power can be used to determine the probable period \citep{Li21b}.

In addition, frequency-dependent red-noise needs to be considered for the periodicity analysis of blazars, because the periodicity of blazars generally exhibits red-noise behavior at lower frequencies \citep{Vaug05,Bhat17,Fan14,Sand16}. This can be solved by using the power response method, which characterizes the power spectral density (PSD) \citep{Uttl02,Ren21}. The random fluctuations of blazars are generally approximated as a power-law PSD: $P(f) \propto f^{{-}\alpha}$, where $P(f)$ is the power at temporal frequency $f$, and $\alpha$ is the spectral slope. Following \citet{Vaug05}, we estimated the power spectral slope $\alpha$ by fitting a linear function to the log-periodogram. The best-fit PSD is shown in the upper right panel of Figure 7. The PSD result show that the slope is $\alpha = 1.85\pm0.06$. Then, we assessed the confidence level of the QPO of 1ES 1426+42.8 by modeling the optical variability as red-noise with the spectral slope ($\alpha = 1.85$). We simulate a large number of light curves, using the Monte Carlo method described by \citet{Timm95} to establish the red-noise background. Once 10000 light curves were simulated by using even sampling intervals, the LSP was computed \citep[see][]{Yang20}. Consequently, using the power spectral distribution of the simulated light curves, local 95\%, 99\%, and 99.7\% ($3\sigma$) confidence contour lines were evaluated. We also simulated 10000 light curves and re-sampled them according to the source light curves to evaluate the confidence level of the WWZ \citep{Timm95}. Then, the 95\%, 99\%, and 99.7\% ($3\sigma$) confidence contour lines were evaluated, respectively.

As shown in the middle left panel of Figure 7, there is an obvious peak in the periodogram of LSP, which hints a possible QPO with $48.67 \pm 13.90$ minutes ($> 99\%$ confidence level) for 1ES 1426+42.8 in I band on April 13, 2010. We use the FWHM of the peak as the uncertainty of period value. The middle right panel of Figure 7 shows that there is a strong peak in the periodogram of WWZ, which hints a possible QPO with $47.23 \pm 11.21$ minutes ($> 3\sigma$). The periodicity of WWZ method is very close to the $48.67 \pm 13.90$ minutes obtained by the LSP method, and the result indicates that 1ES 1426+42.8 has periodicity in IDV.

The REDFIT method also can be used to evaluate the QPO and red-noise. \citet{Schu02} presented a program (REDFIT\footnote{http://www.geo.uni-bremen.de/geomod/staff/mschulz/}), which can be used to test if peaks in the spectrum of a time series are significant against the red-noise background by fitting a first-order autoregressive (AR1) process. The results of REDFIT are shown in the bottom panel of Figure 7. From Figure 7, we can see that the peak (0.0289 day $\sim$ 41.62 minutes) in the spectrum of a time series is significant ($> 3\sigma$) against the red-noise background. The three methods support the possible QPO ($48.67 \pm 13.90$ minutes) within the error range.

We performed new observations of 1ES 1426+42.8 in March 2021 to further confirm the reliability of this QPO. For PSD on March 16, 2021 of 1ES 1426+42.8, we fit the part where frequency $<$ 200 $\rm day^{-1}$, because the part where frequency $>$ 200 $\rm day^{-1}$ is dominated by white-noise. The PSD result (upper right panel of Figure 8) shows that the slope is $1.64\pm0.08$. Then, the confidence level is assessed with the spectral slope of $ \alpha = 1.64$. The results show that the close QPOs ($>3\sigma$) are confirmed by the LSP method, WWZ method, and REDFIT method. The corresponding QPOs are $30.70\pm6.55$ minutes, $30.74\pm5.22$ minutes, and $28.88\pm8.66$ minutes, respectively. This new result further confirms the existence of QPO in 1ES 1426+42.8 after considering the errors of QPQ. The light curve of 1ES 1426+42.8 on March 16, 2021 and the corresponding results of different methods are shown in Figure 8.

\begin{figure*}
\centering
\includegraphics[scale=.14]{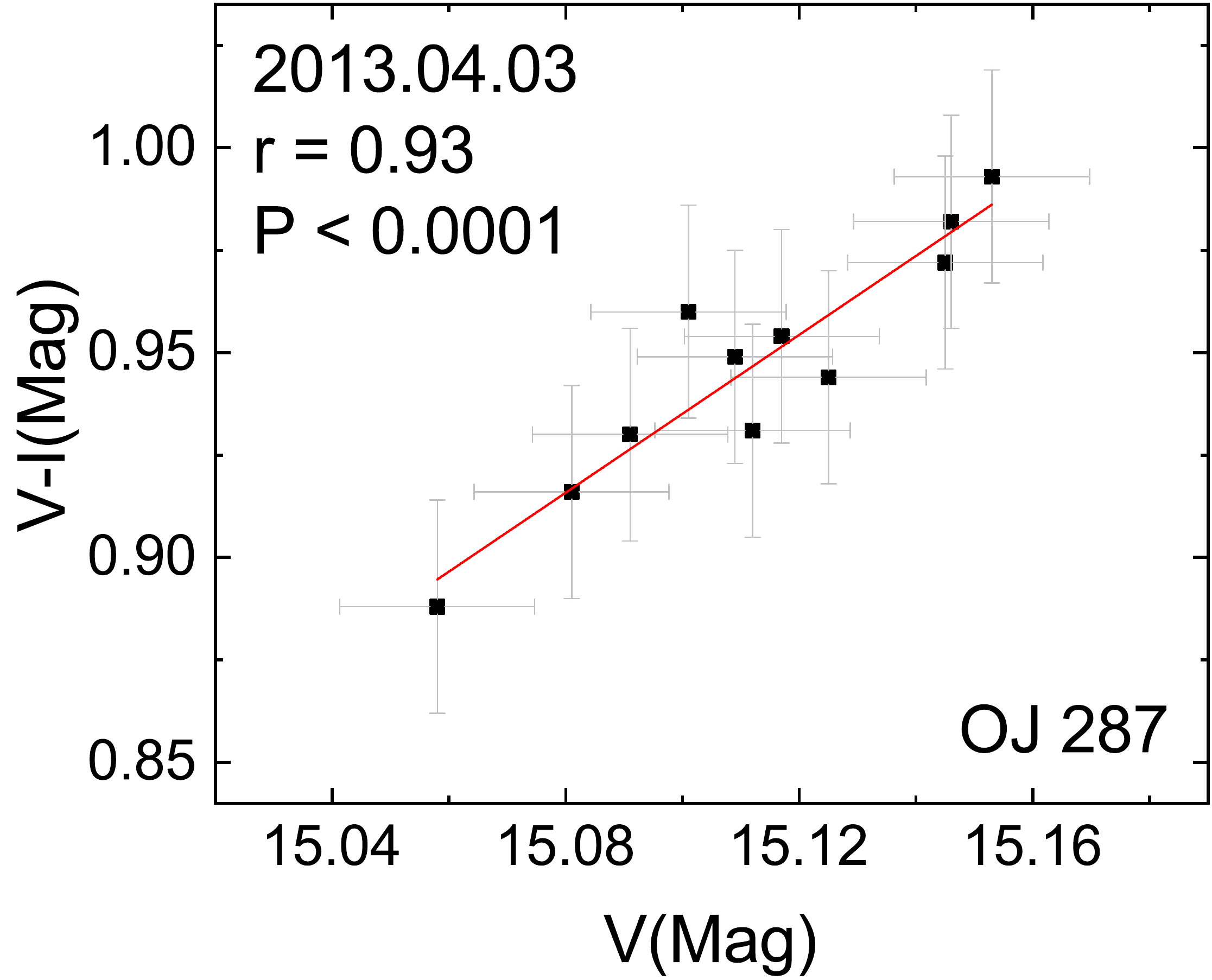}
\includegraphics[scale=.14]{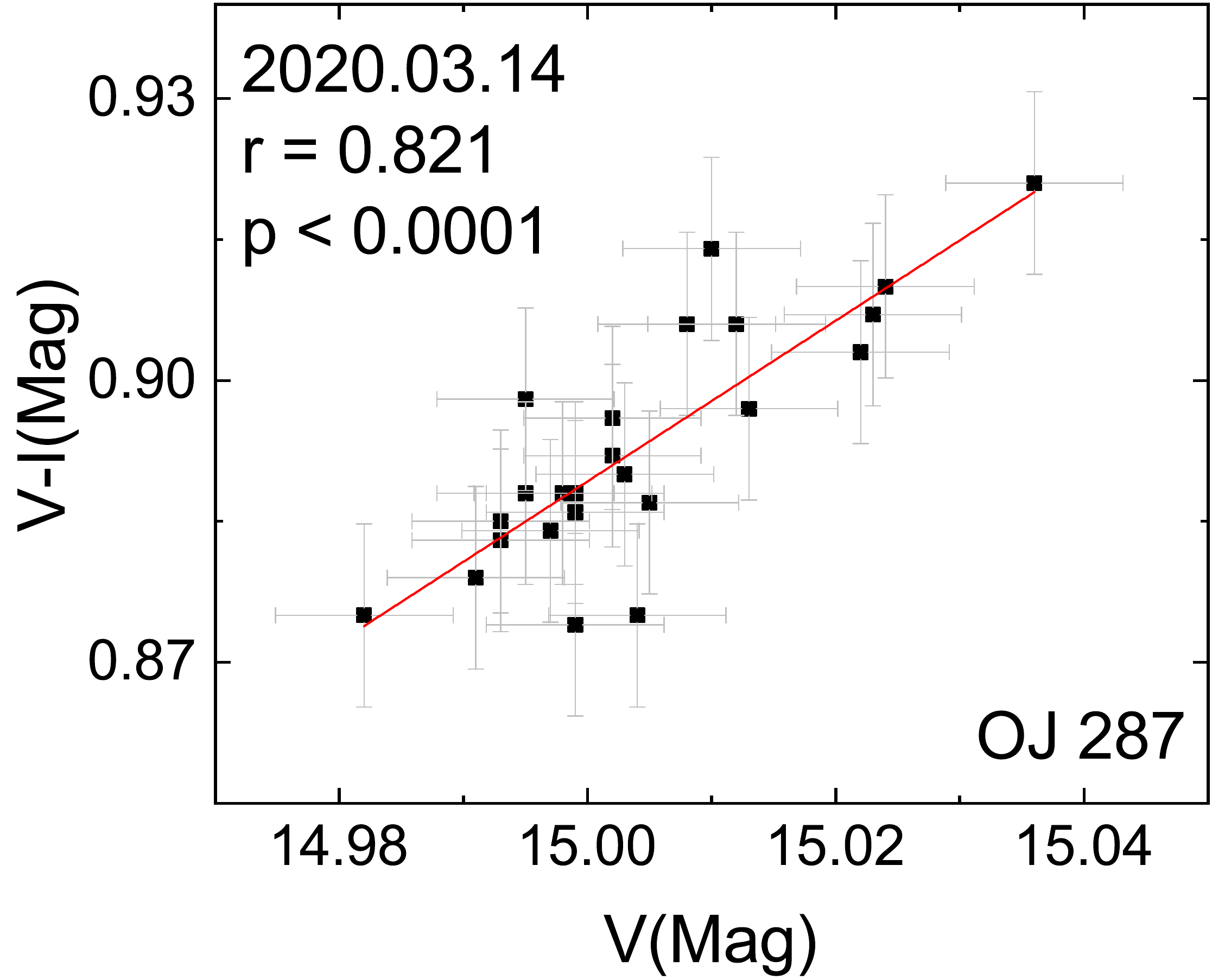}
\includegraphics[scale=.14]{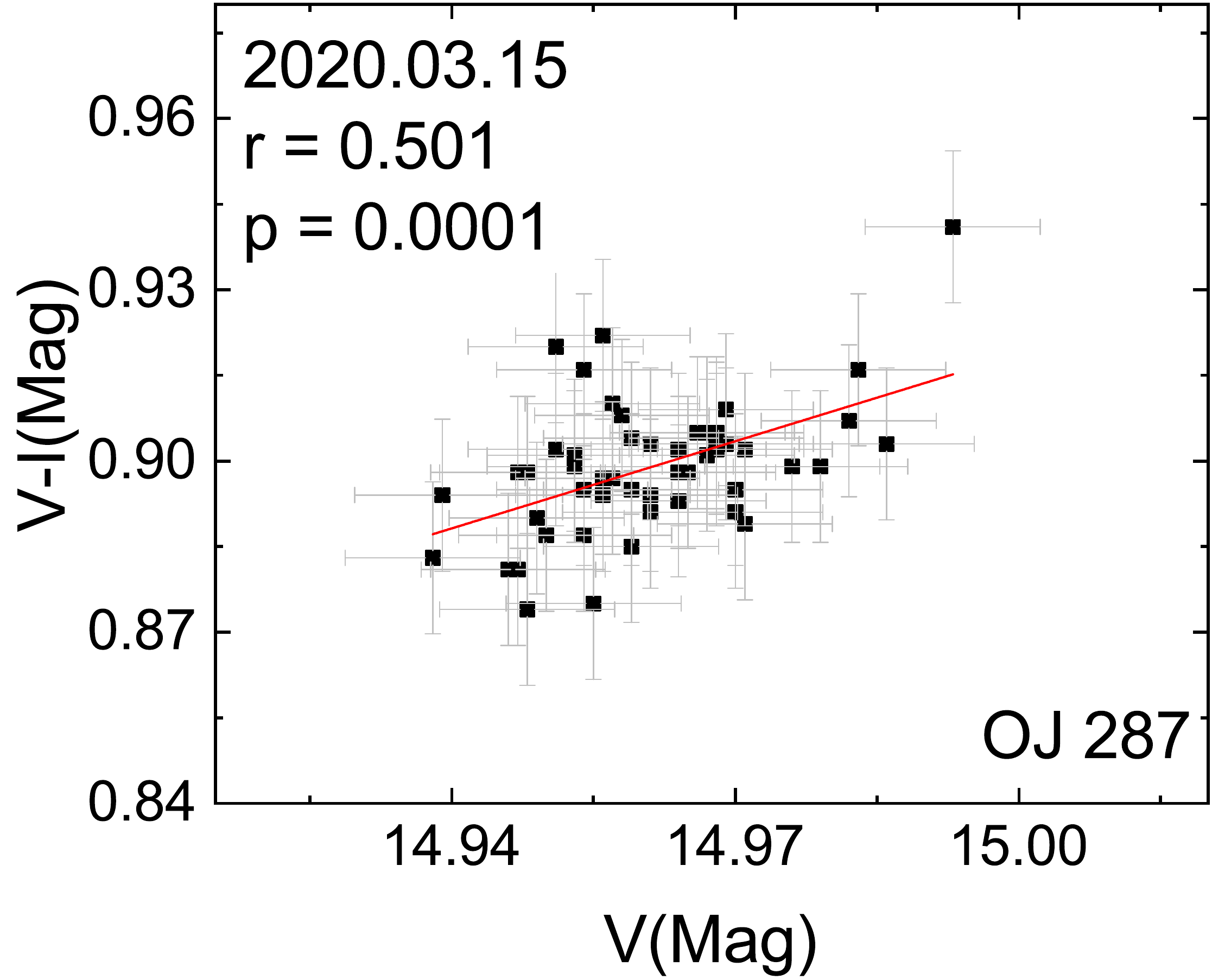}
\includegraphics[scale=.14]{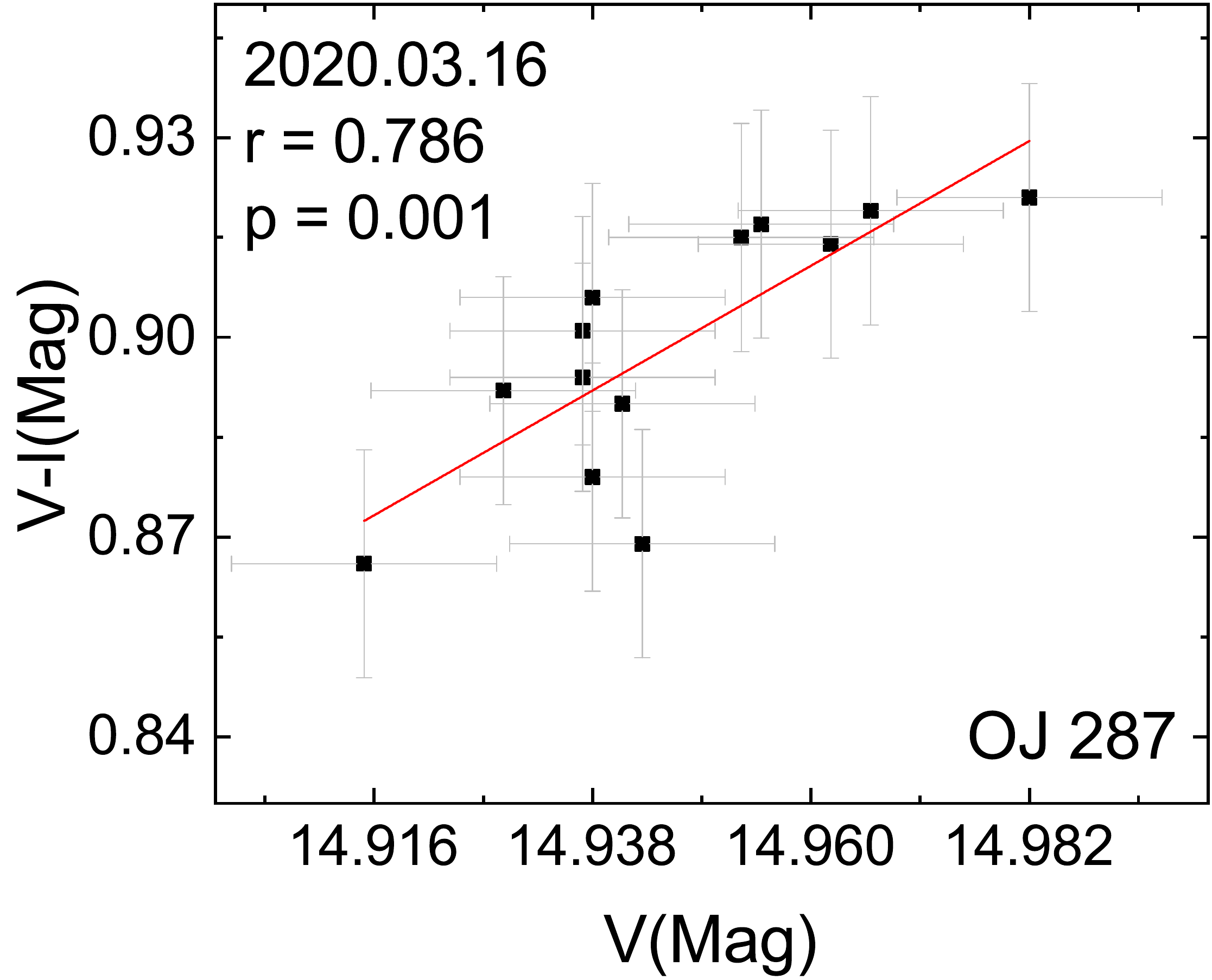}
\includegraphics[scale=.14]{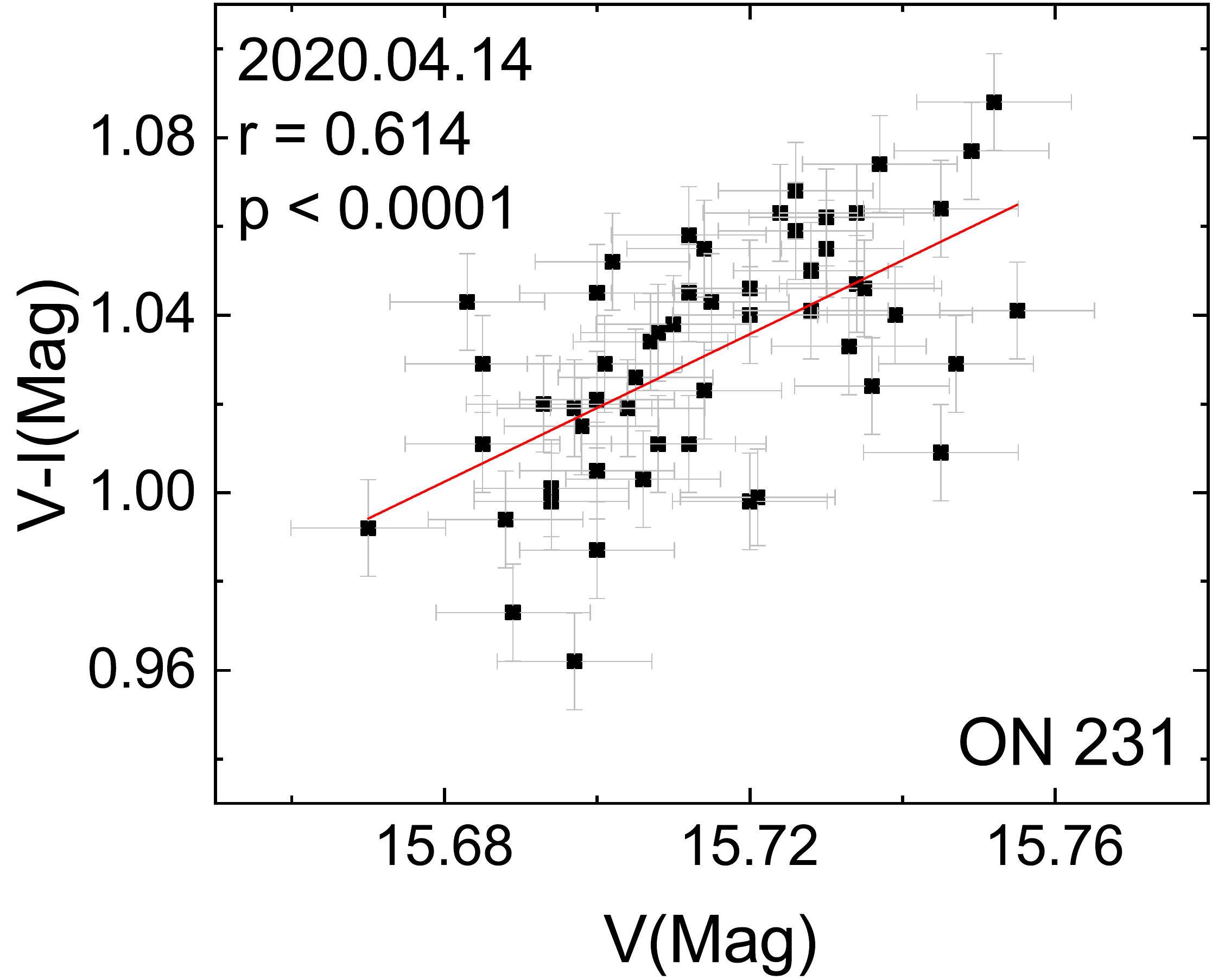}
\includegraphics[scale=.14]{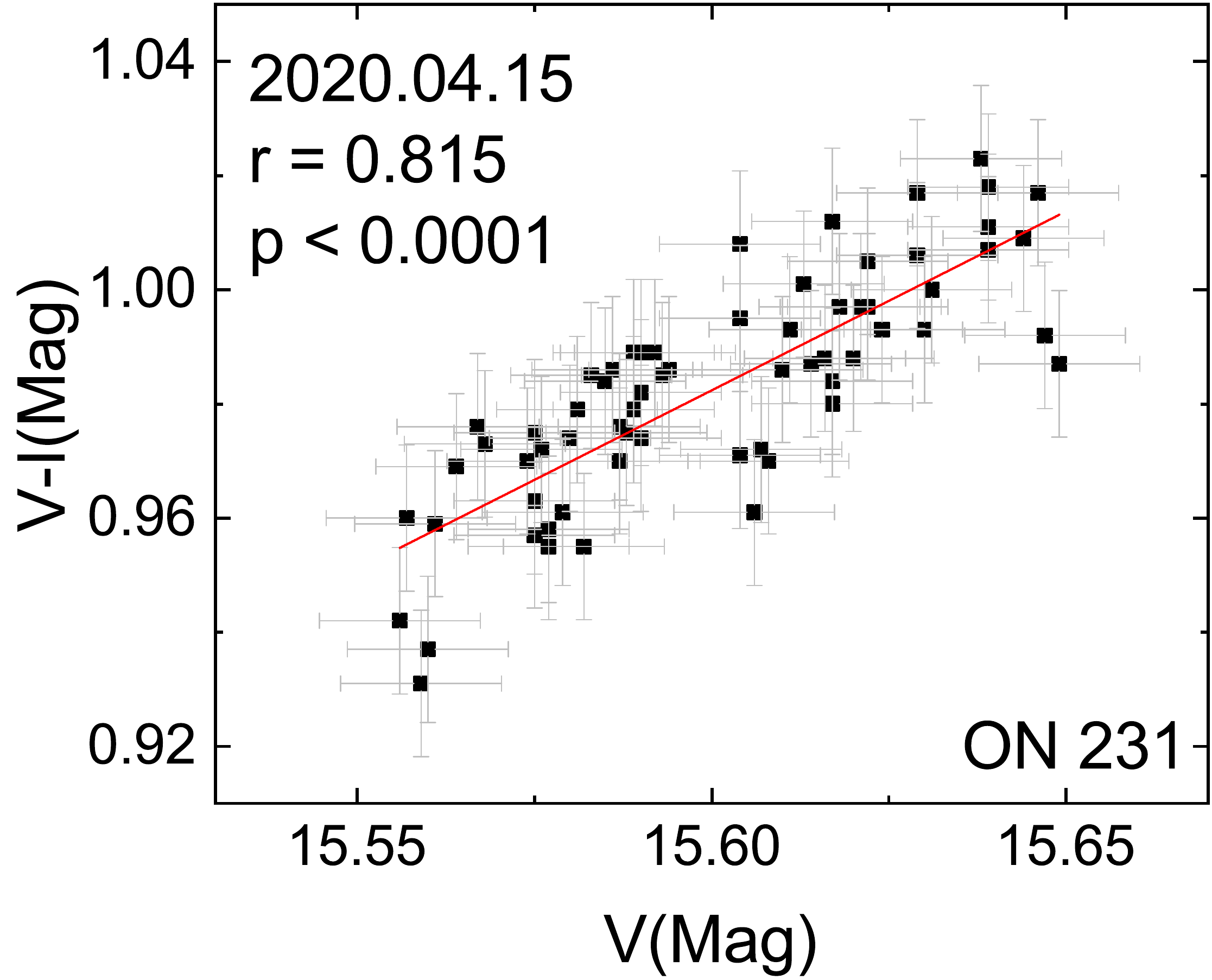}
\includegraphics[scale=.14]{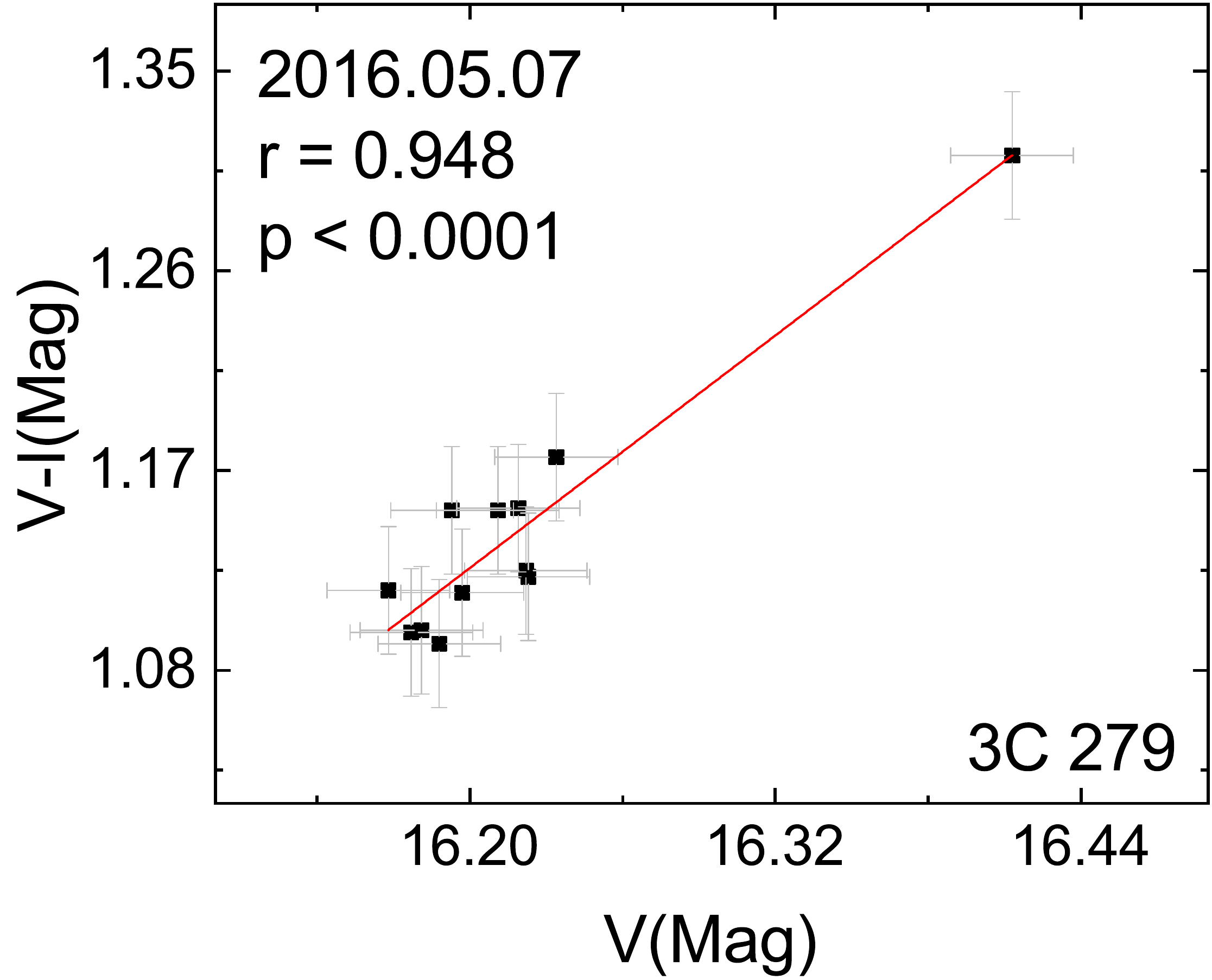}
\includegraphics[scale=.14]{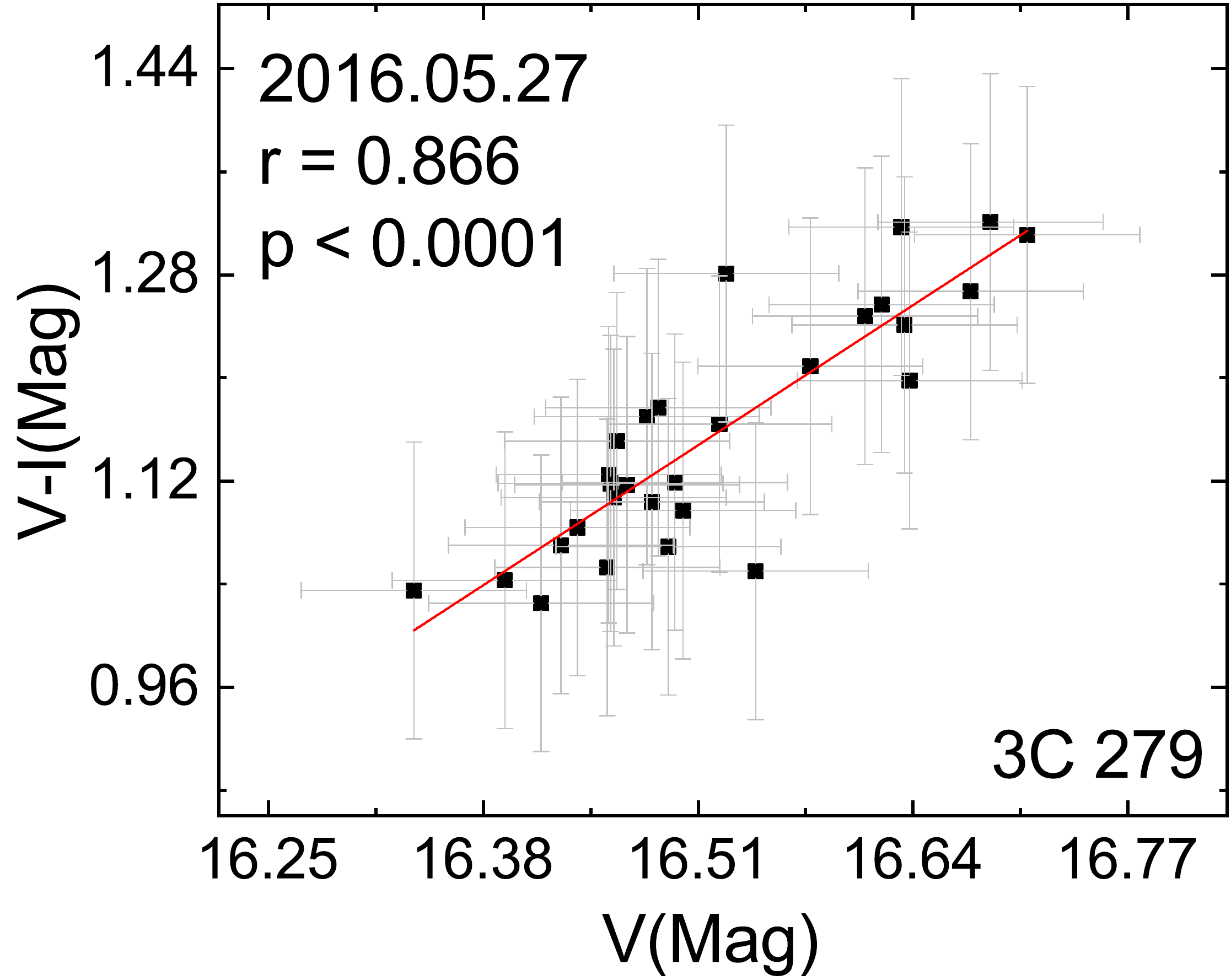}
\includegraphics[scale=.14]{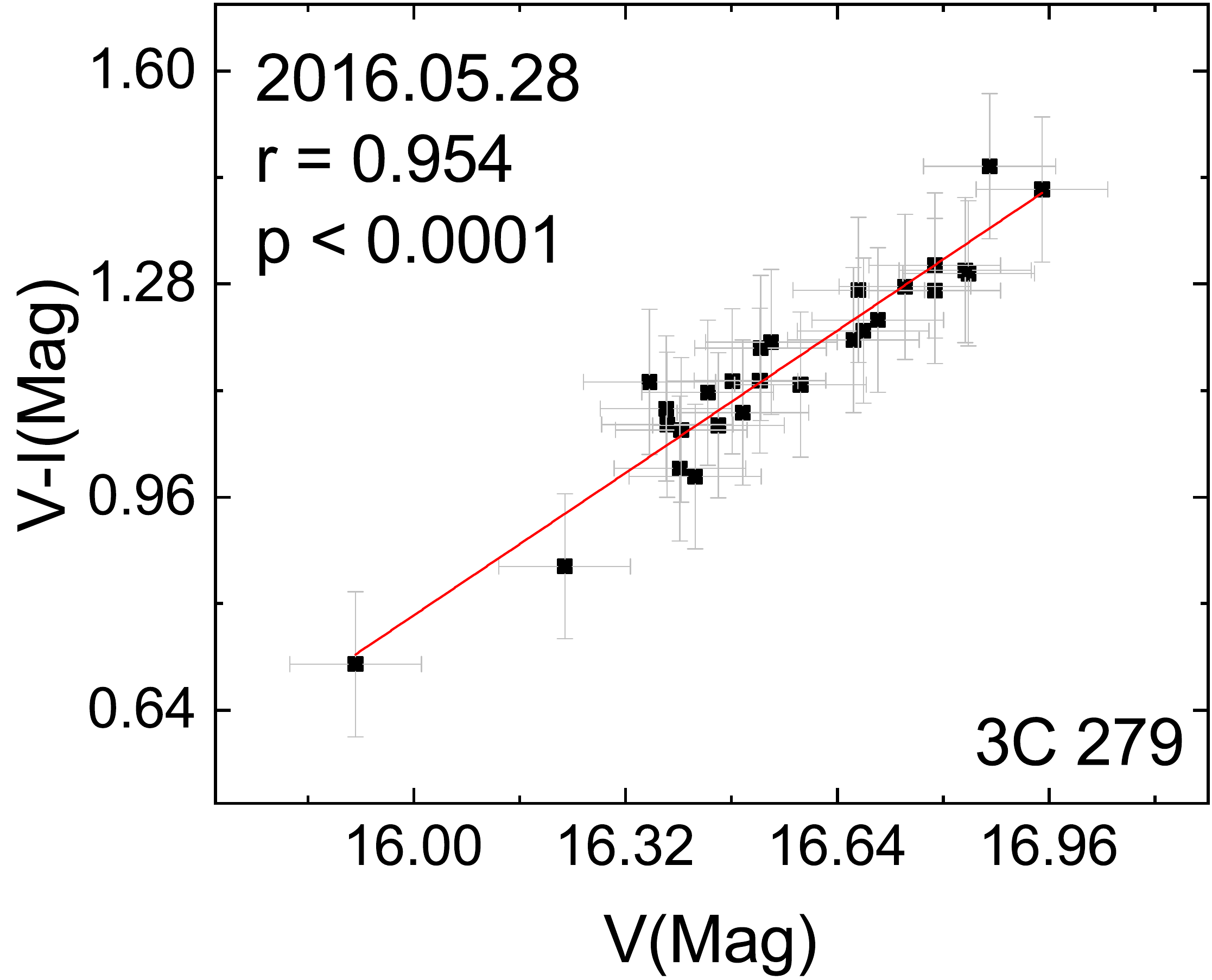}
\includegraphics[scale=.14]{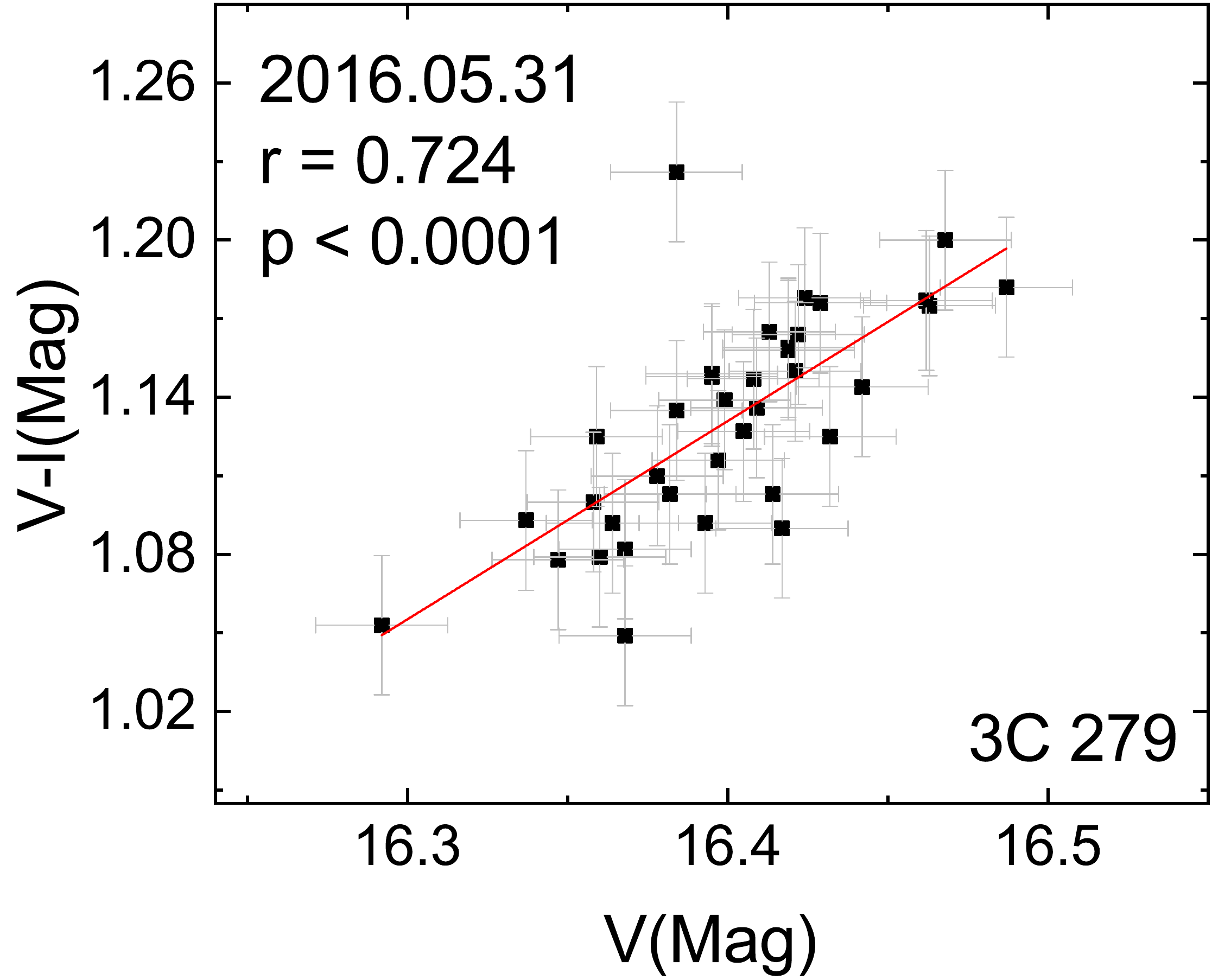}
\includegraphics[scale=.14]{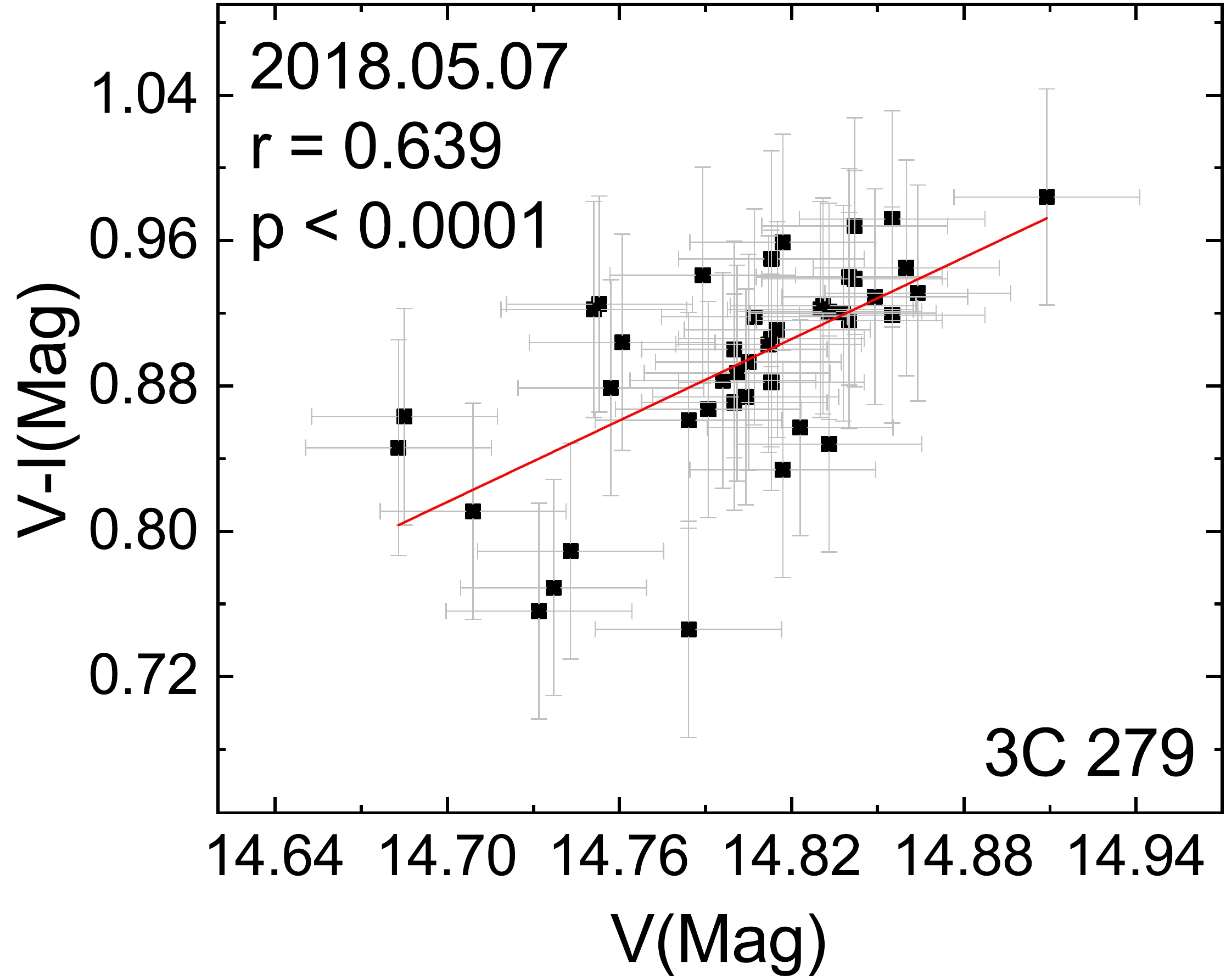}
\includegraphics[scale=.14]{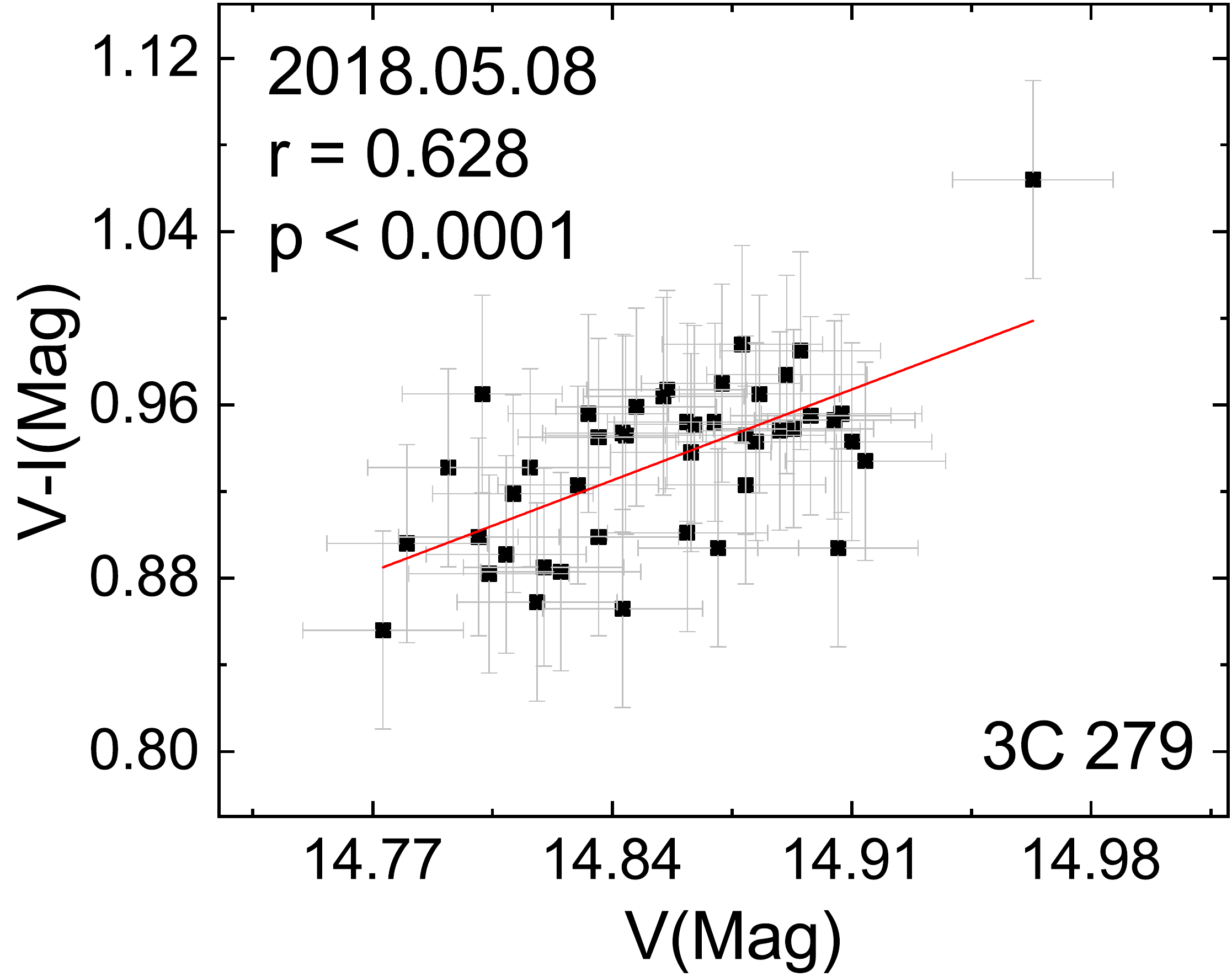}
\includegraphics[scale=.14]{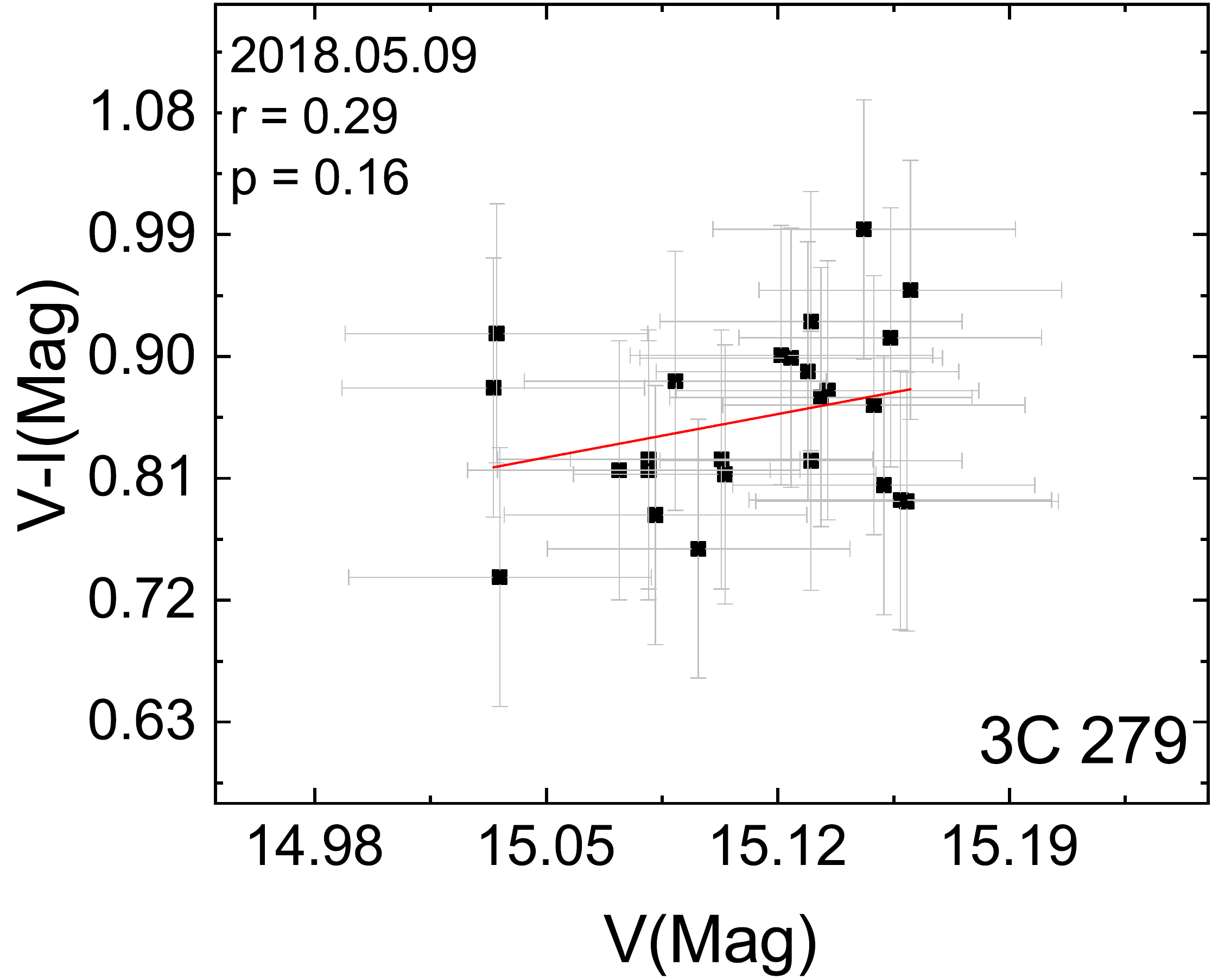}
\includegraphics[scale=.14]{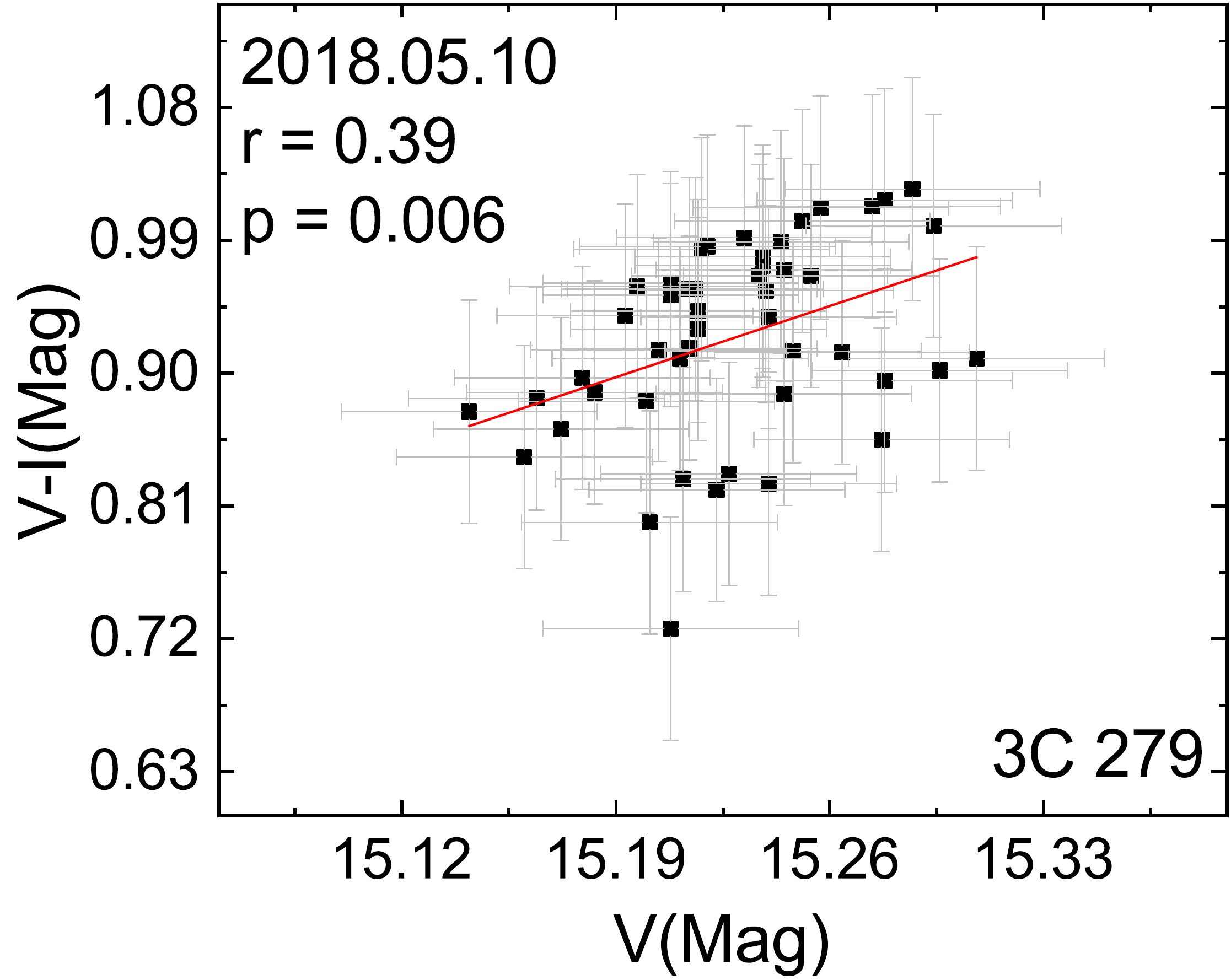}
\includegraphics[scale=.14]{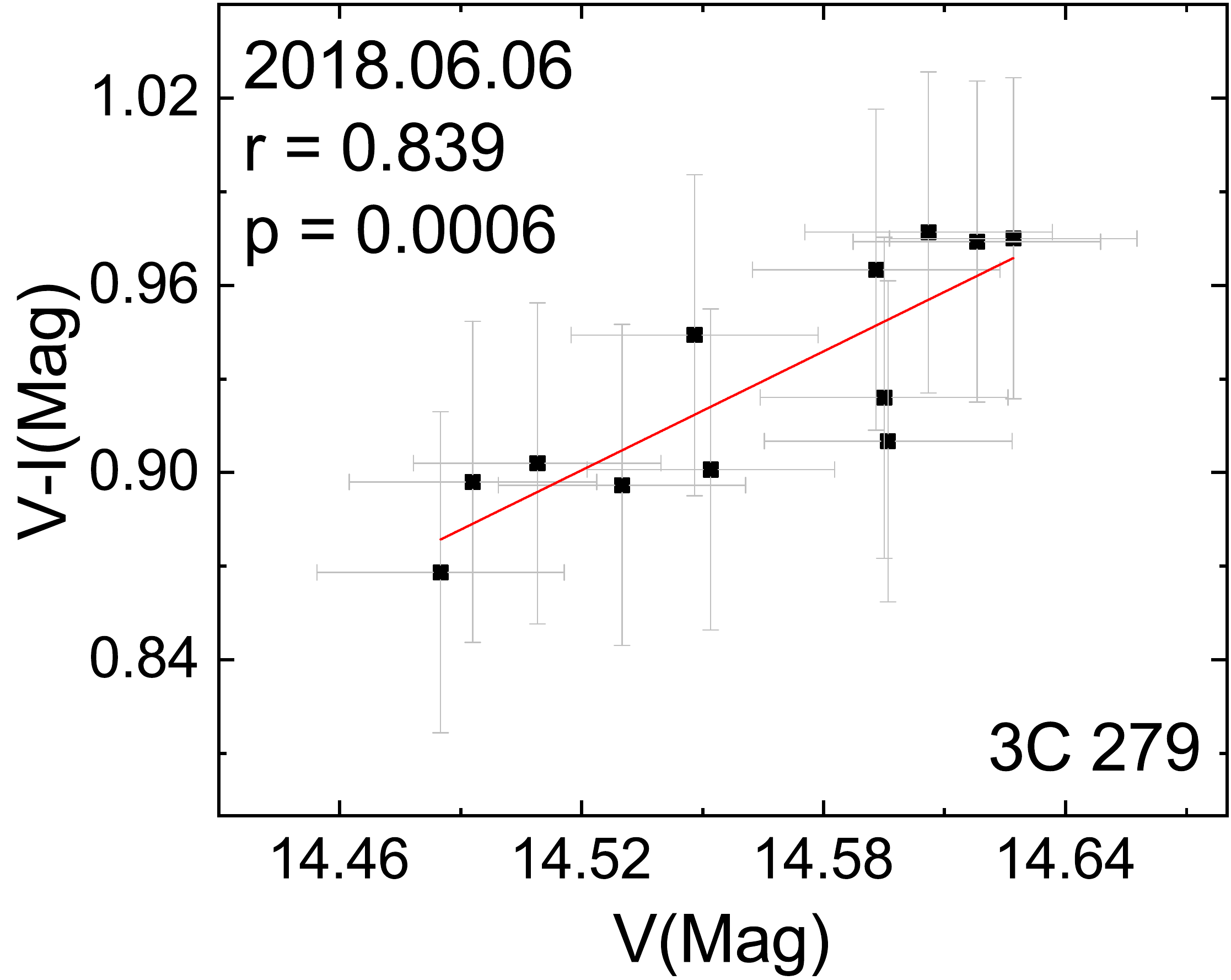}
\includegraphics[scale=.14]{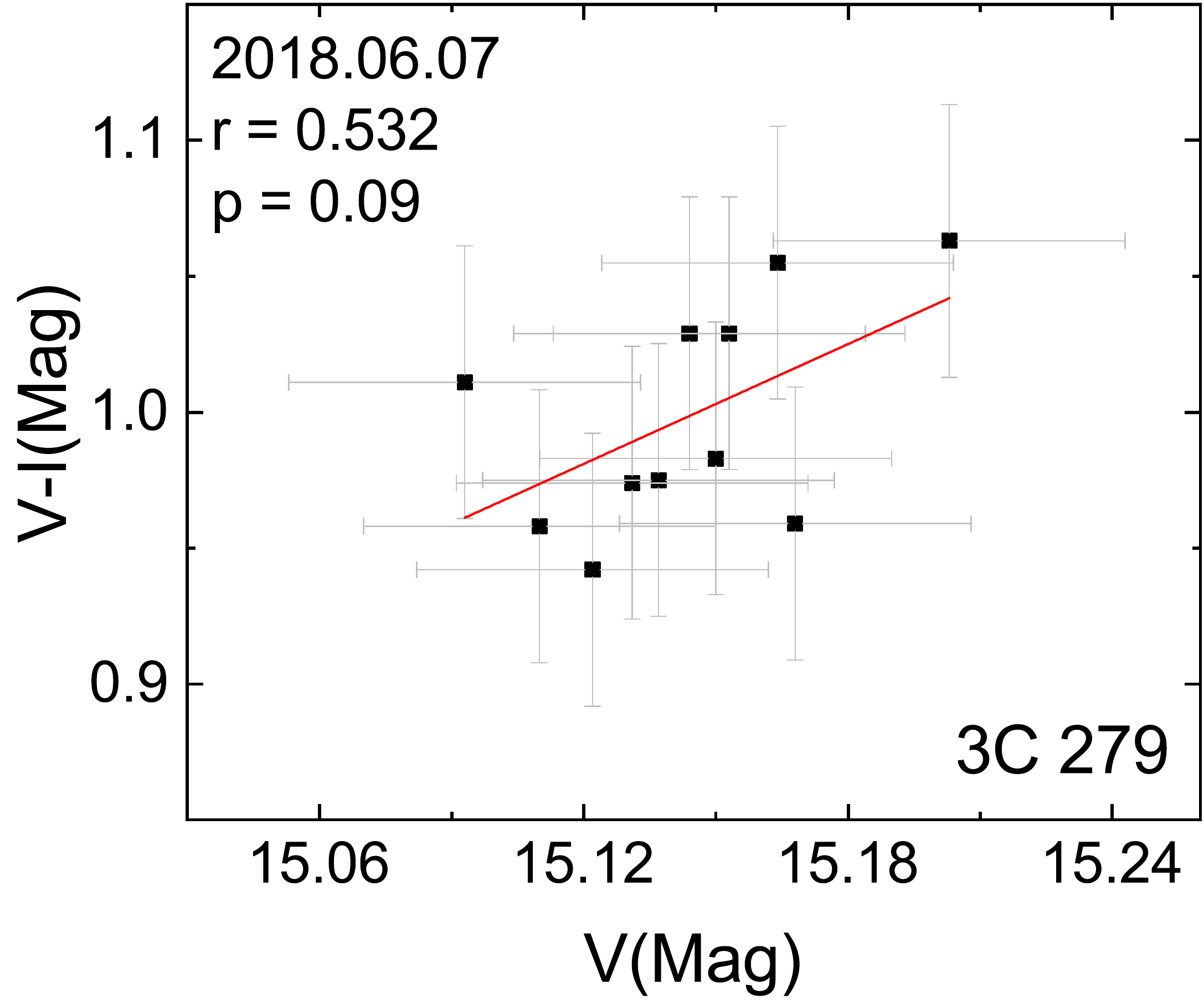}
\includegraphics[scale=.14]{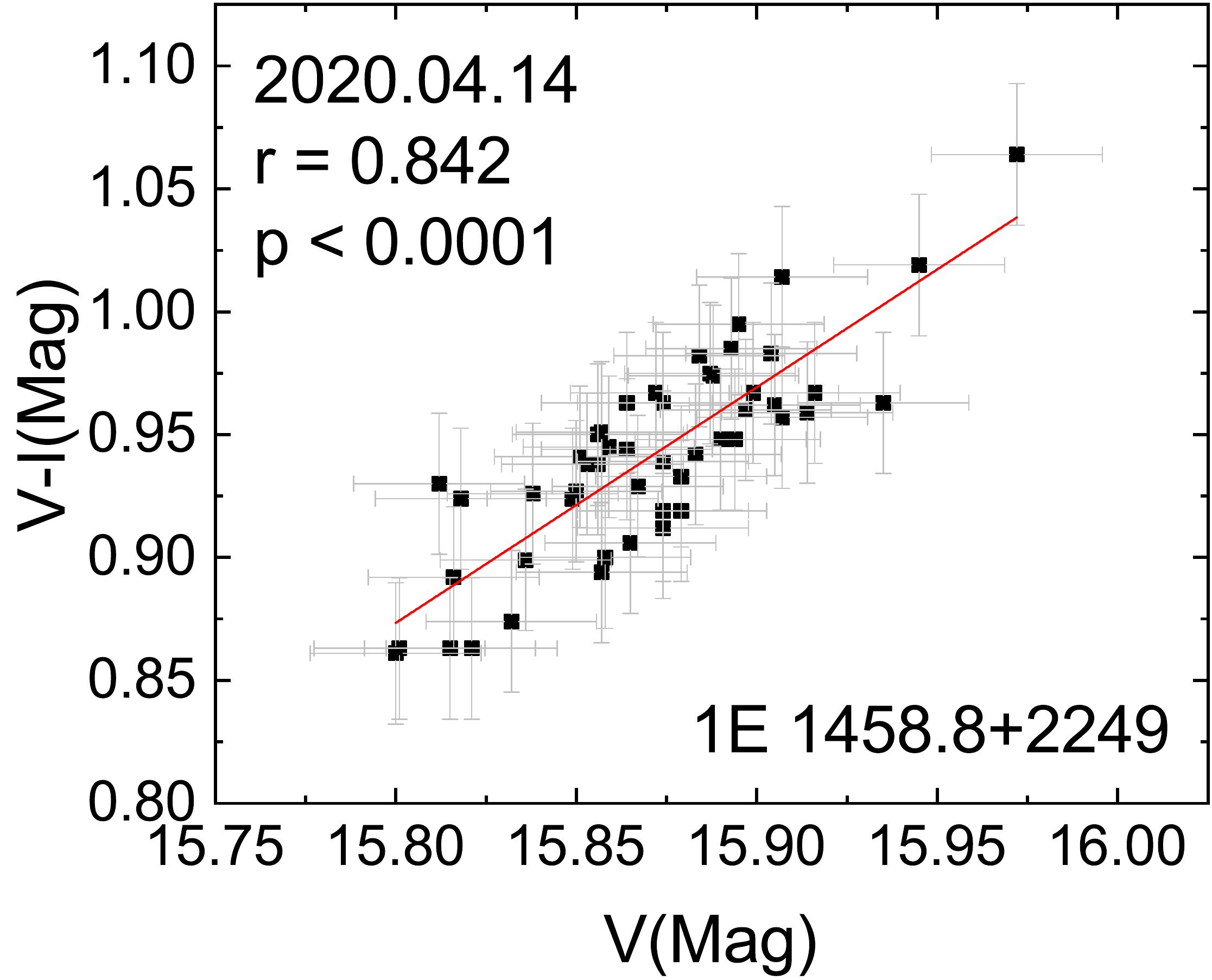}
\includegraphics[scale=.14]{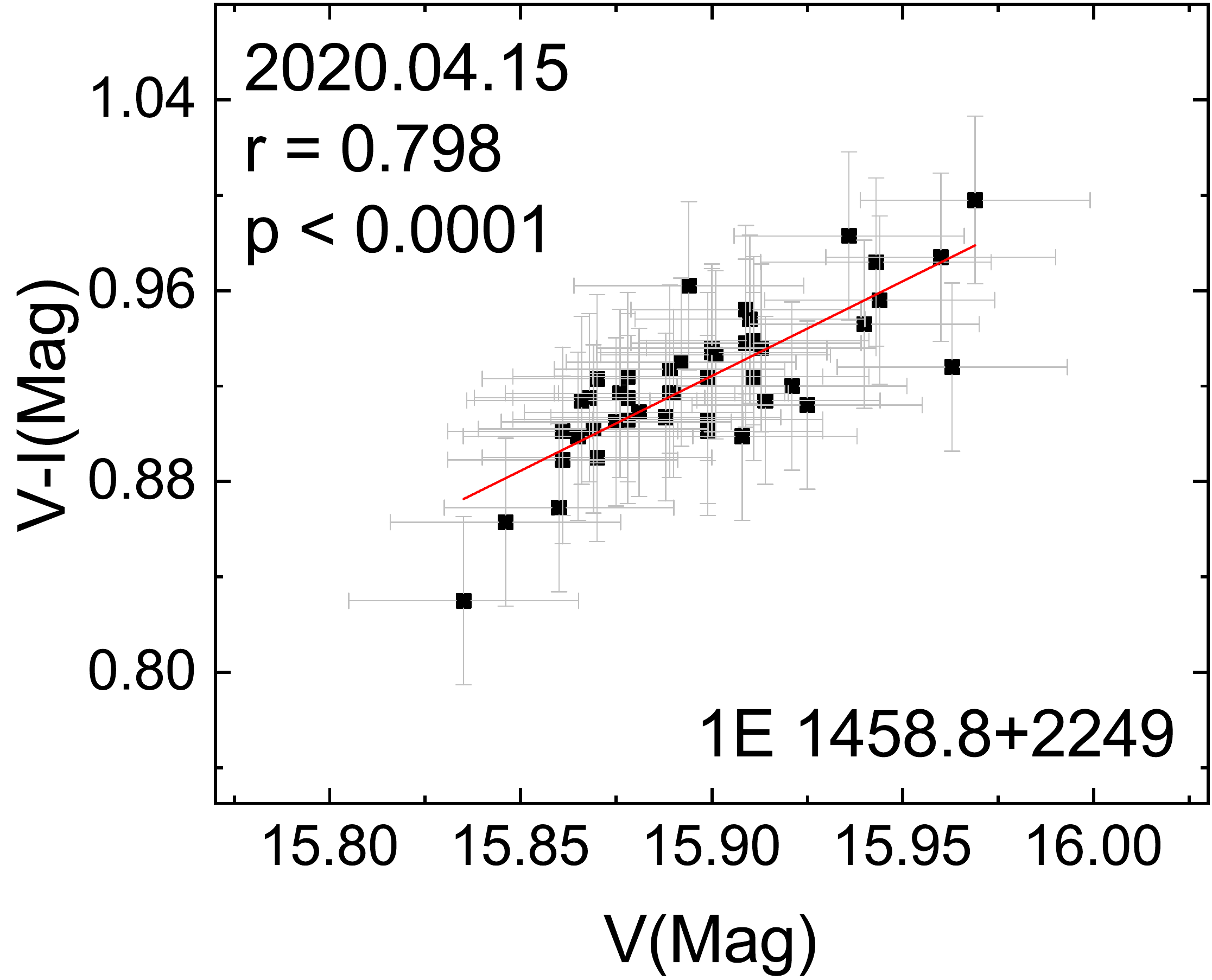}
\includegraphics[scale=.14]{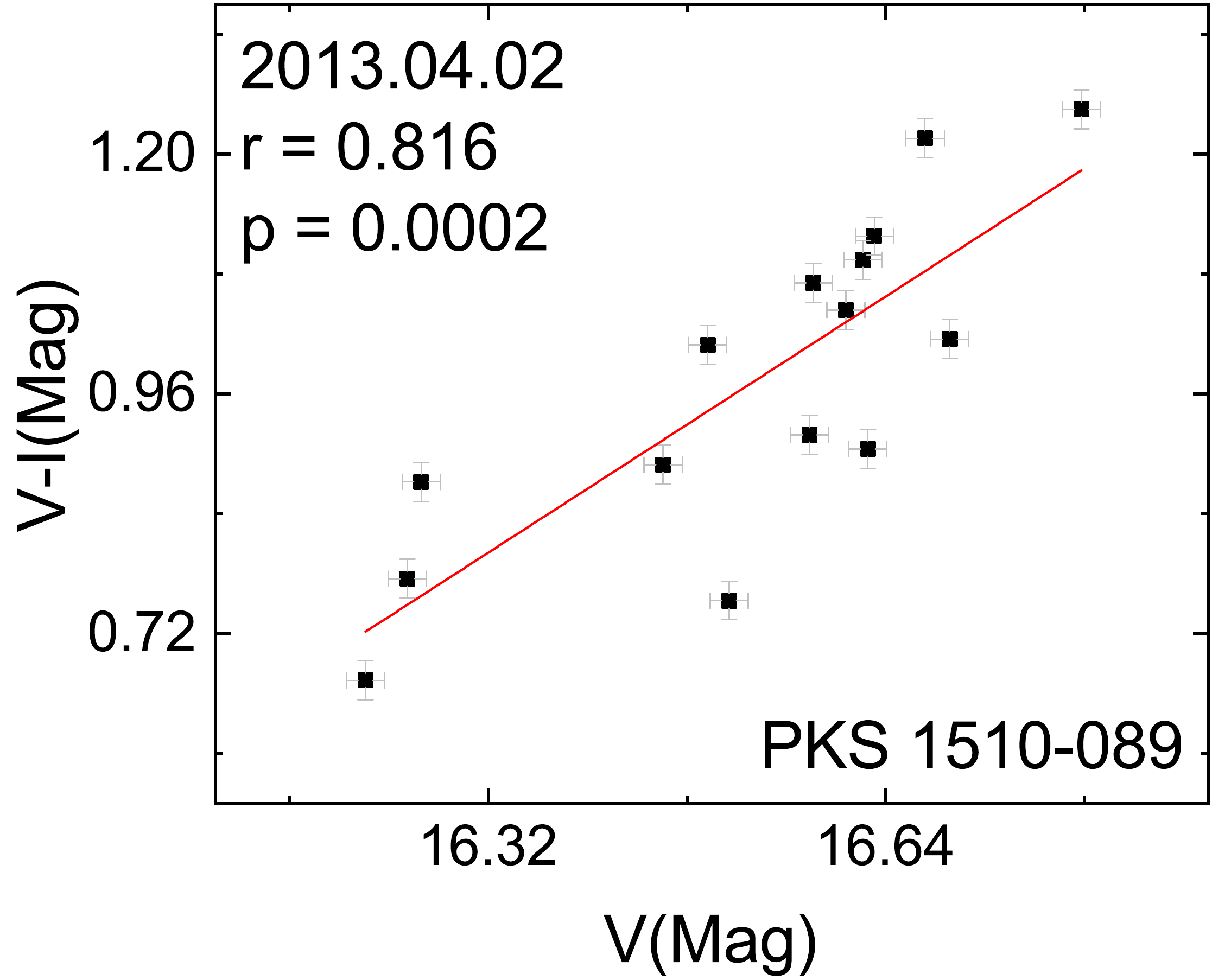}
\includegraphics[scale=.14]{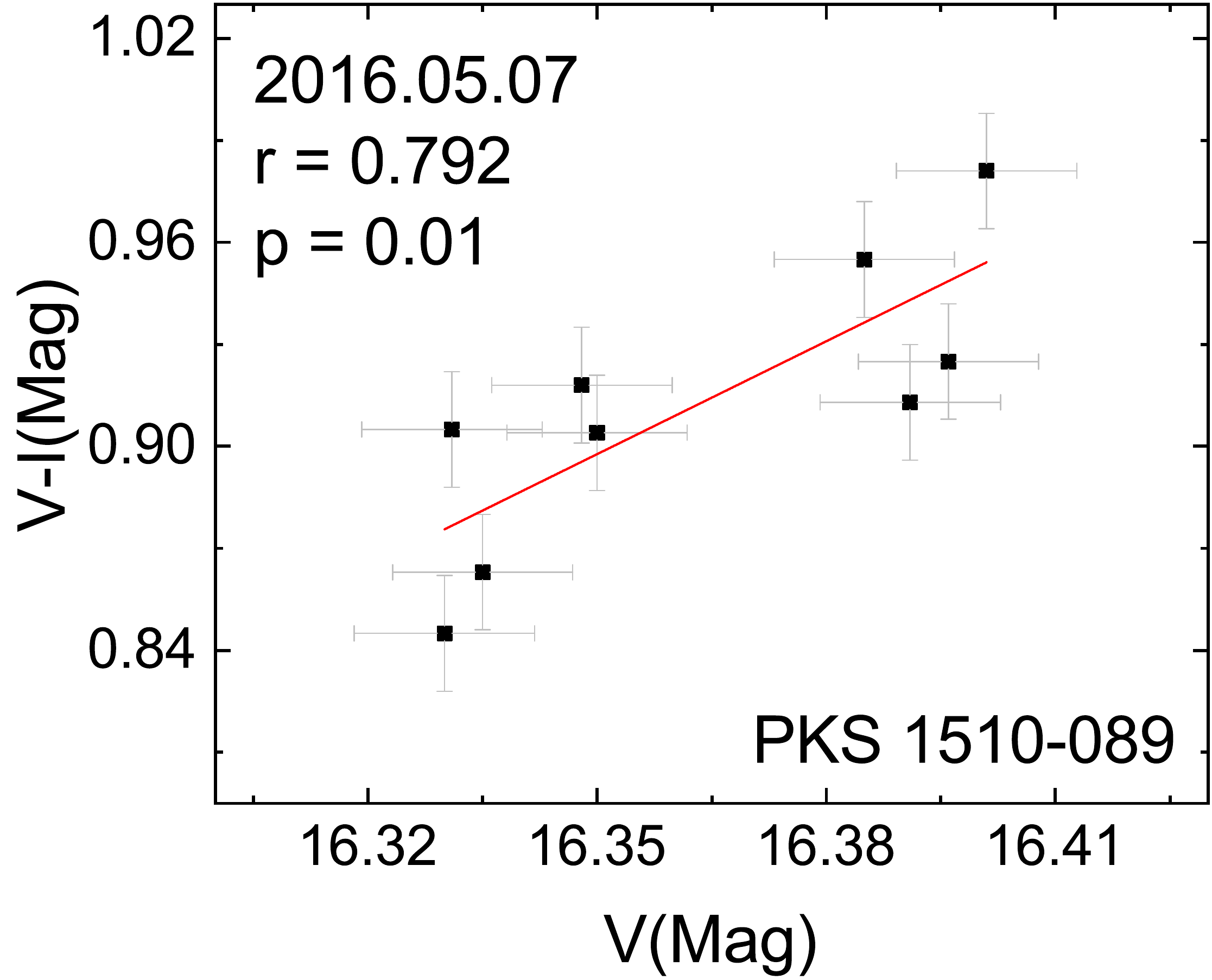}
\includegraphics[scale=.14]{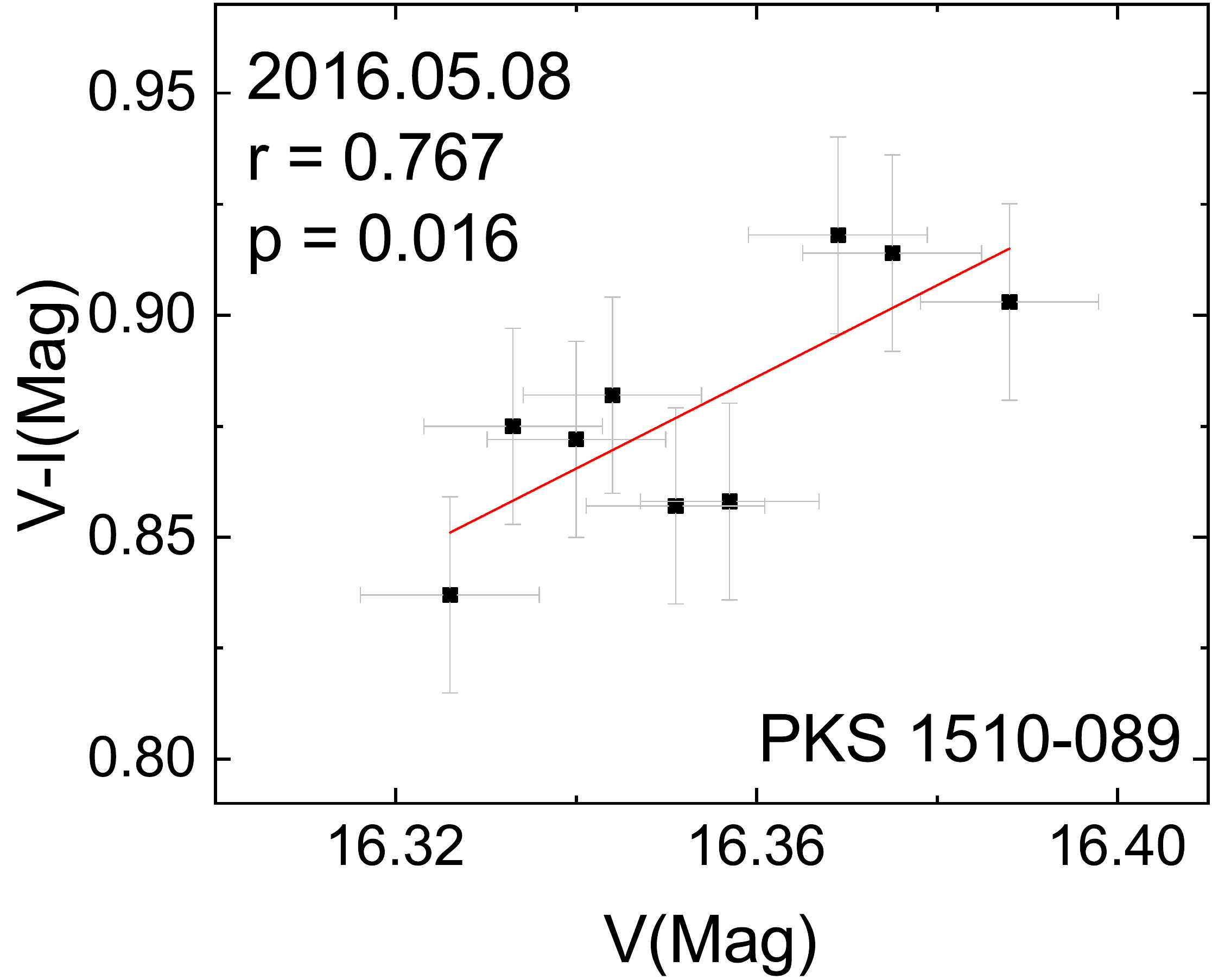}
\includegraphics[scale=.14]{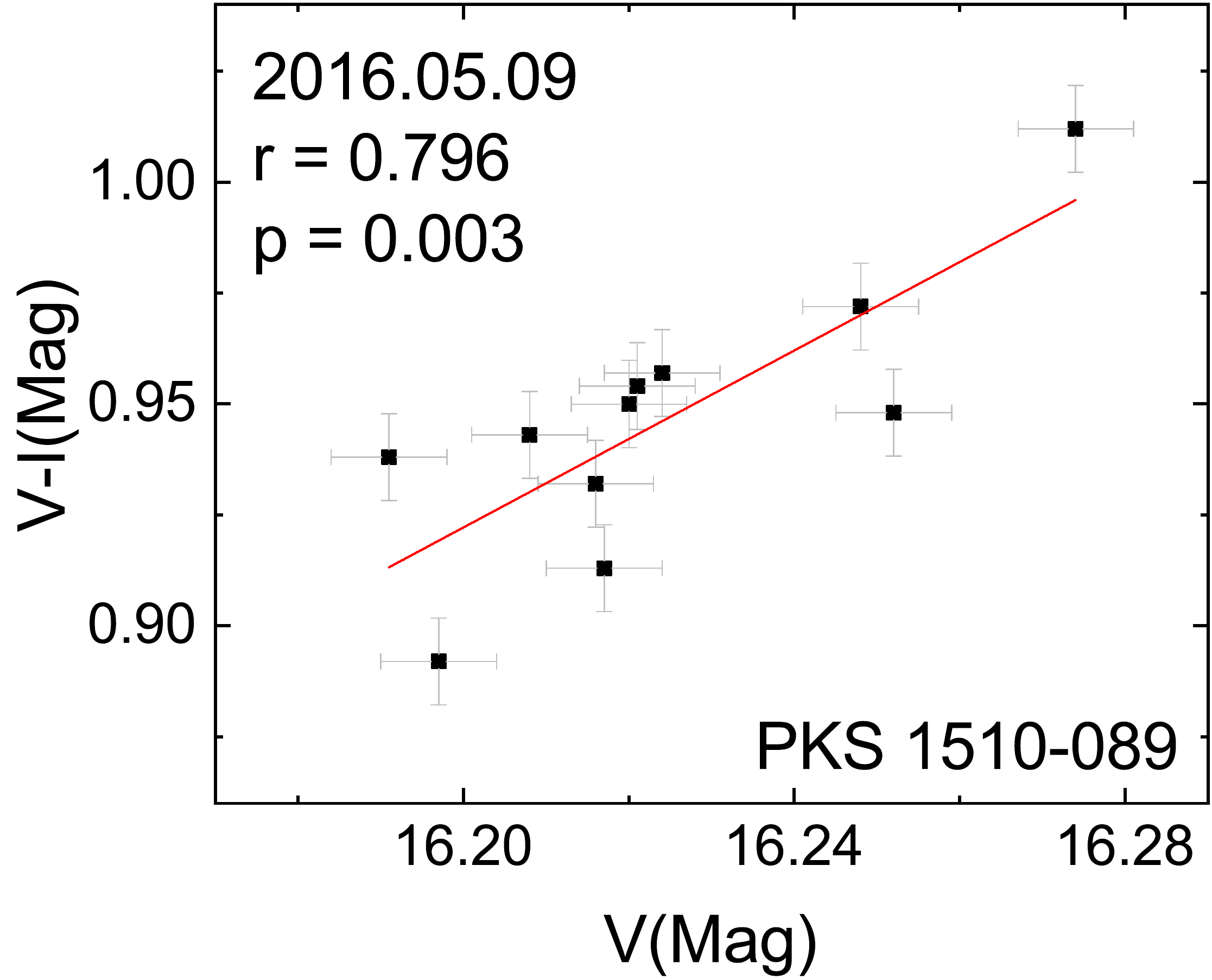}
\includegraphics[scale=.14]{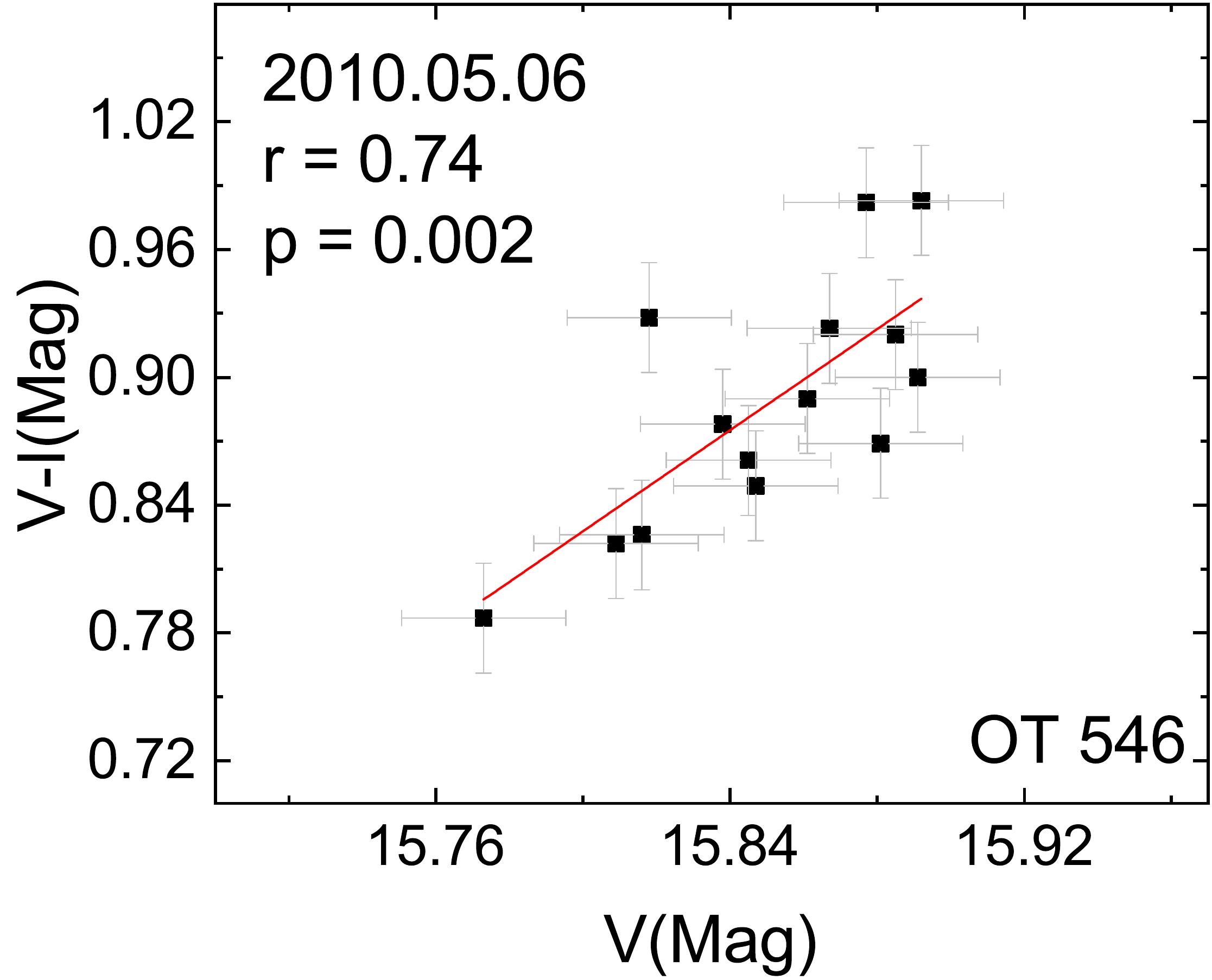}
\includegraphics[scale=.14]{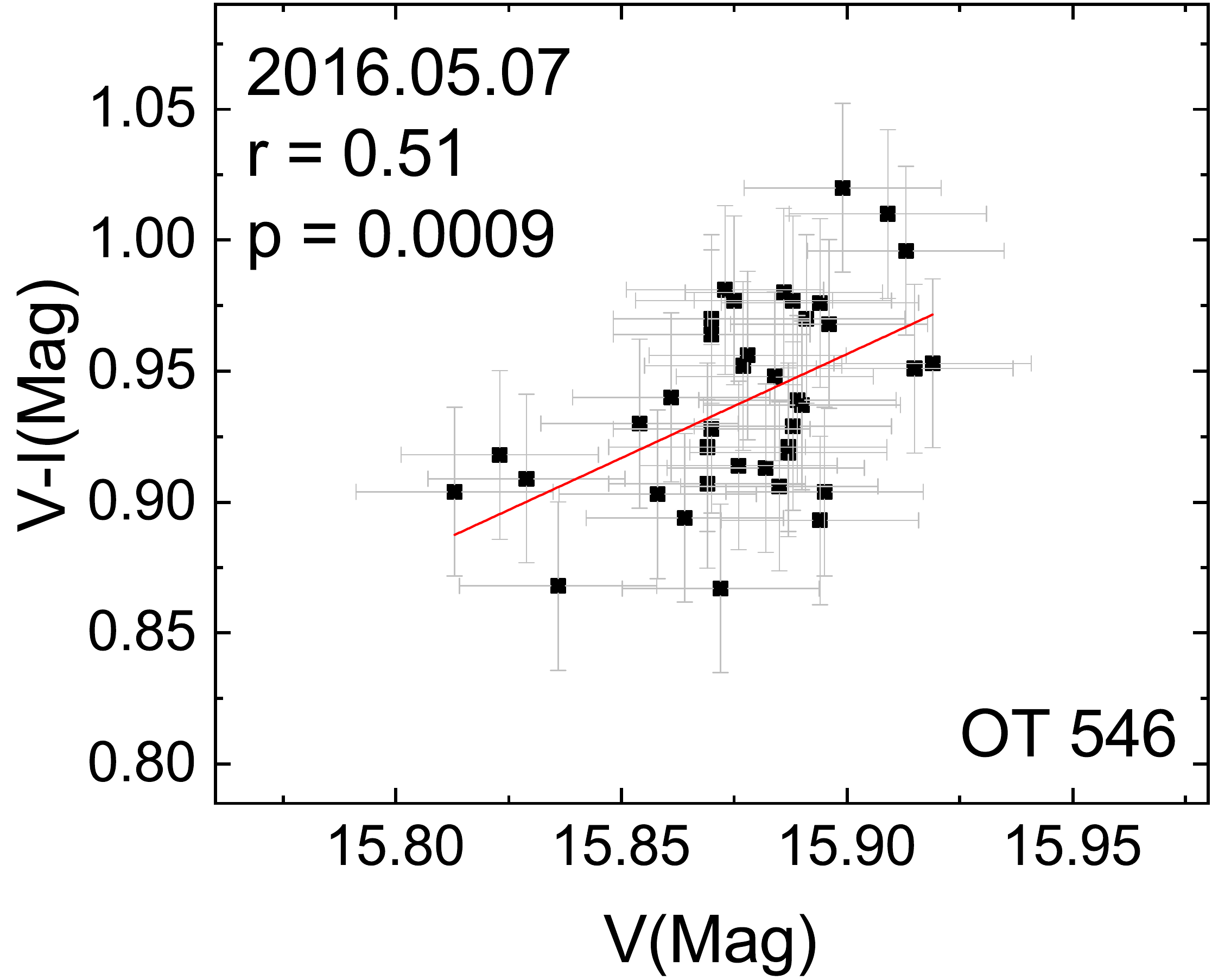}
\includegraphics[scale=.14]{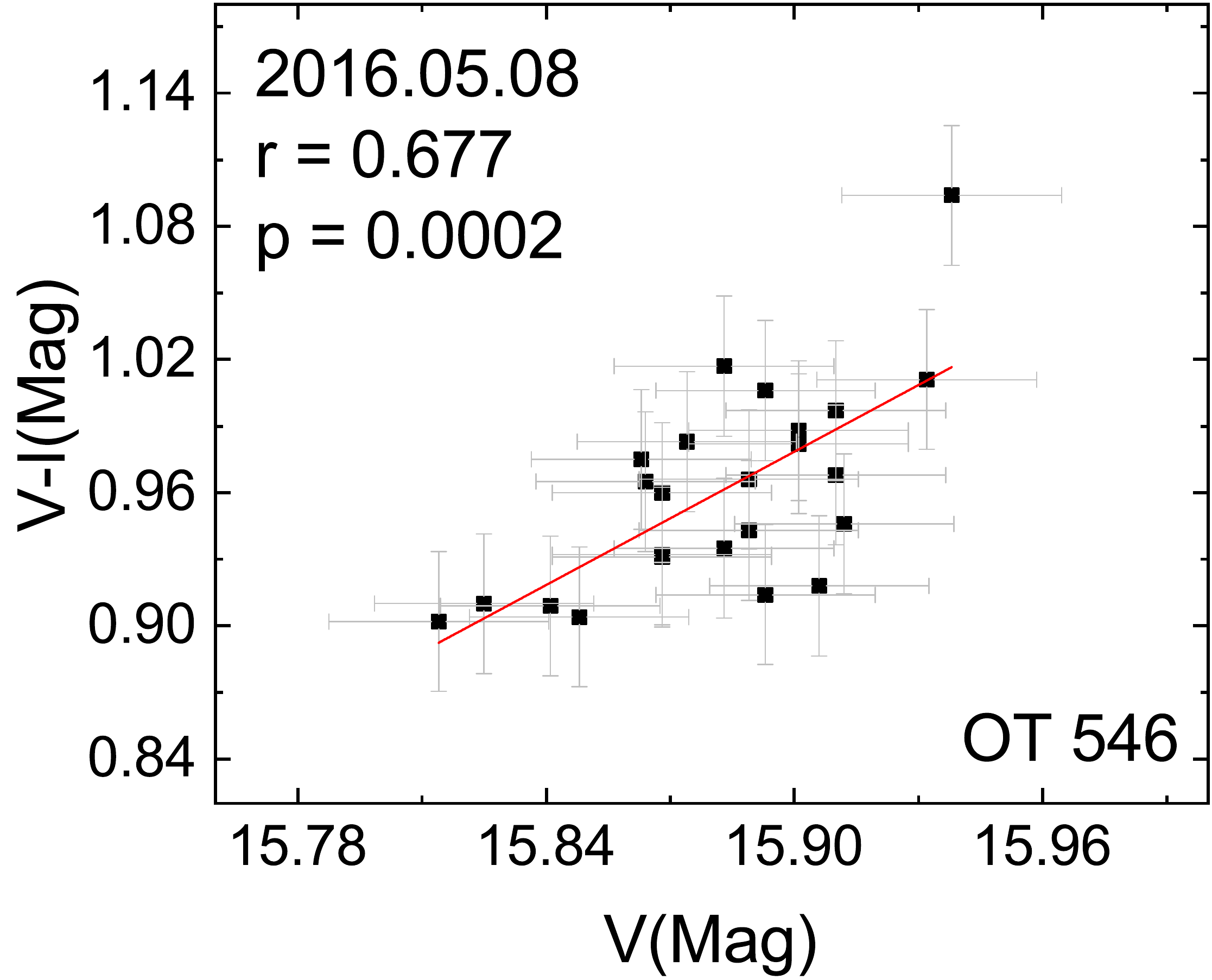}
\includegraphics[scale=.14]{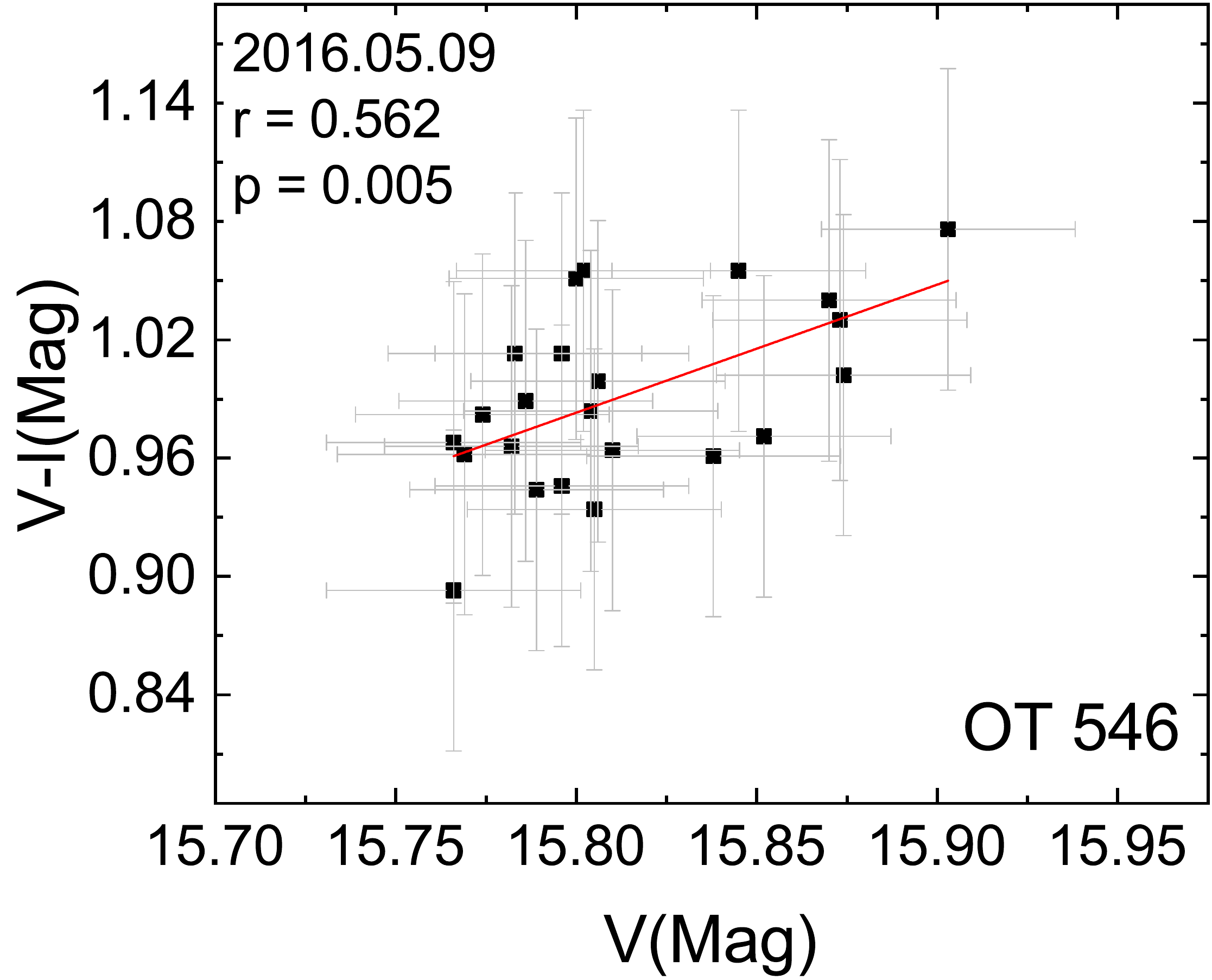}
\includegraphics[scale=.14]{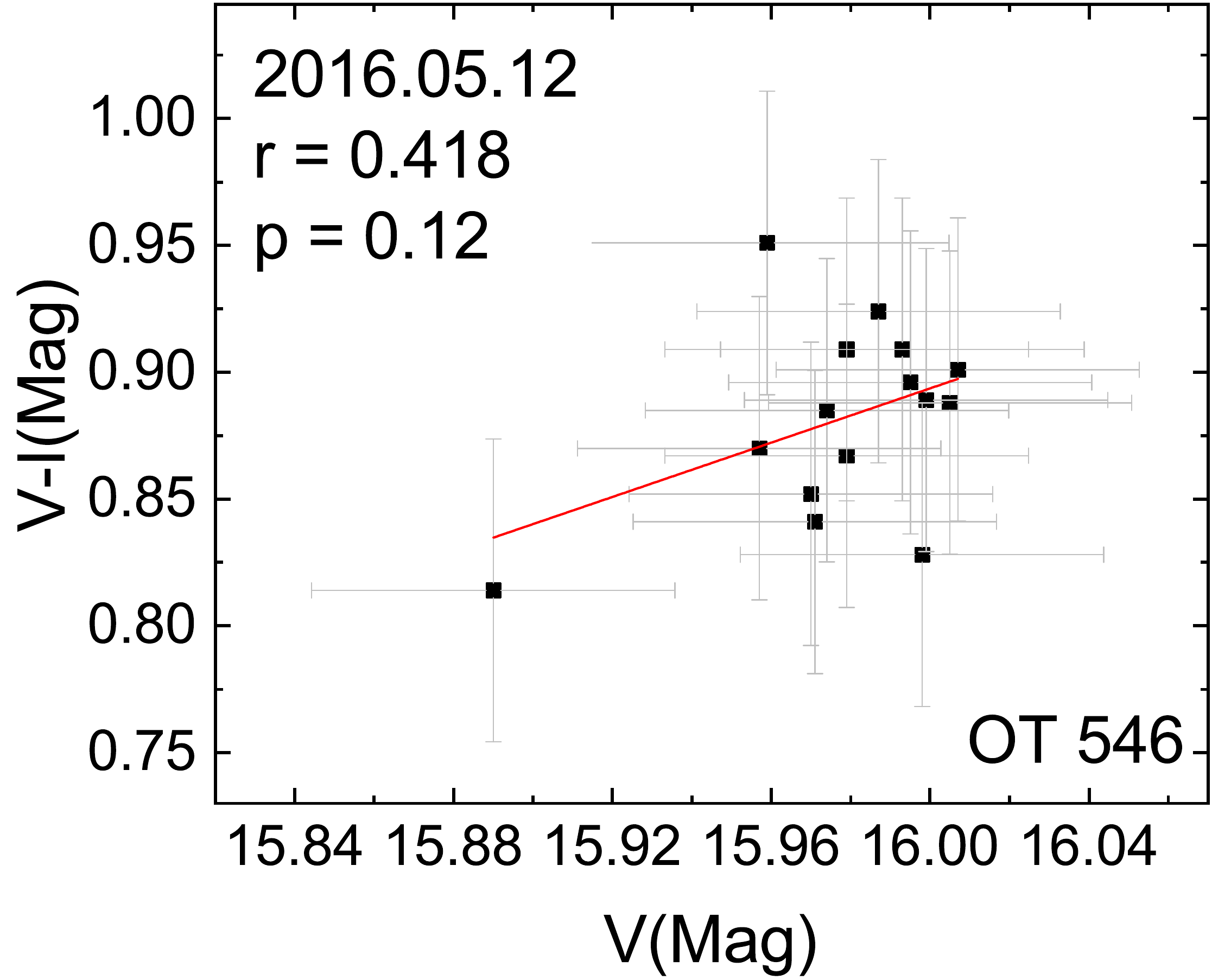}
\includegraphics[scale=.14]{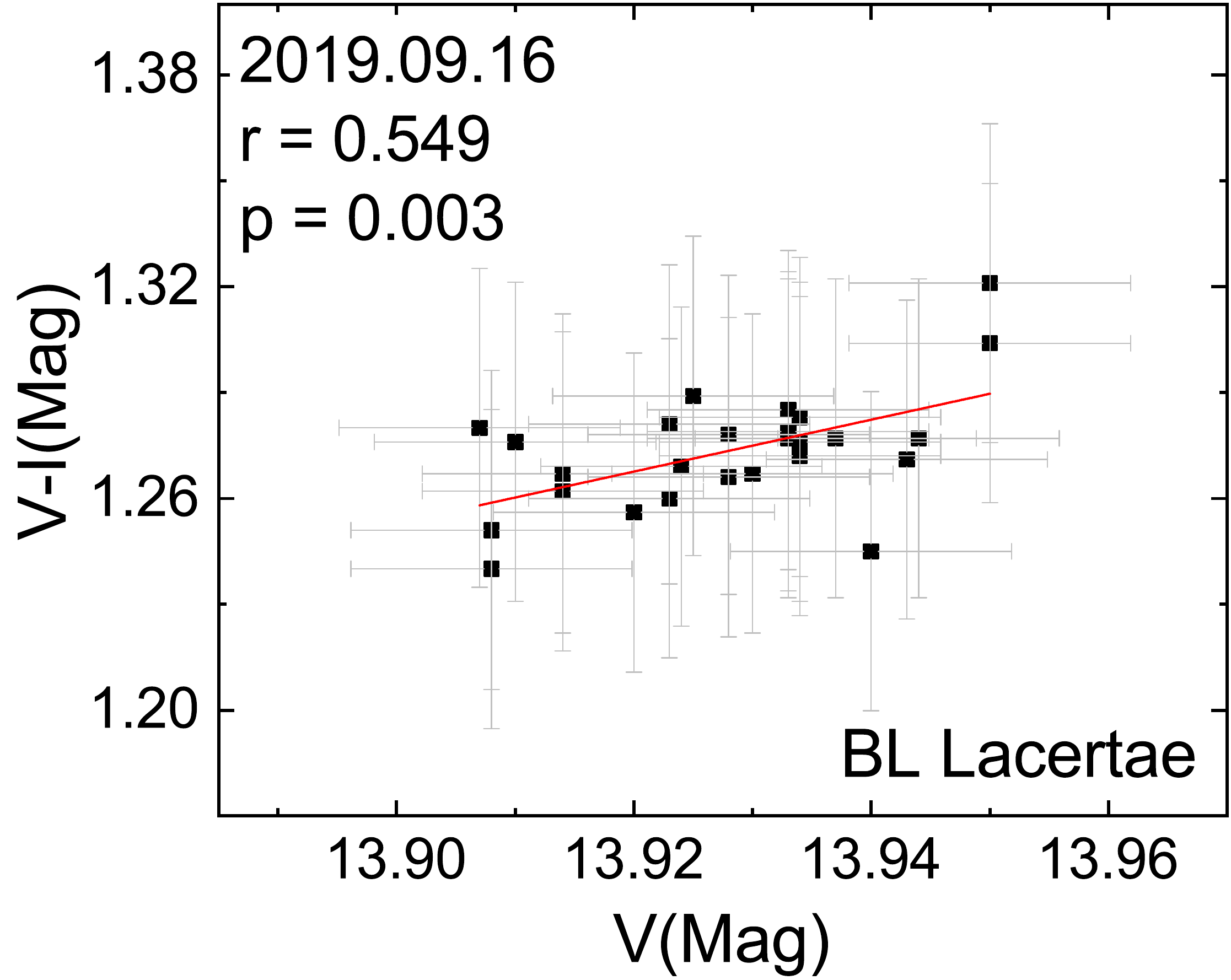}
\includegraphics[scale=.14]{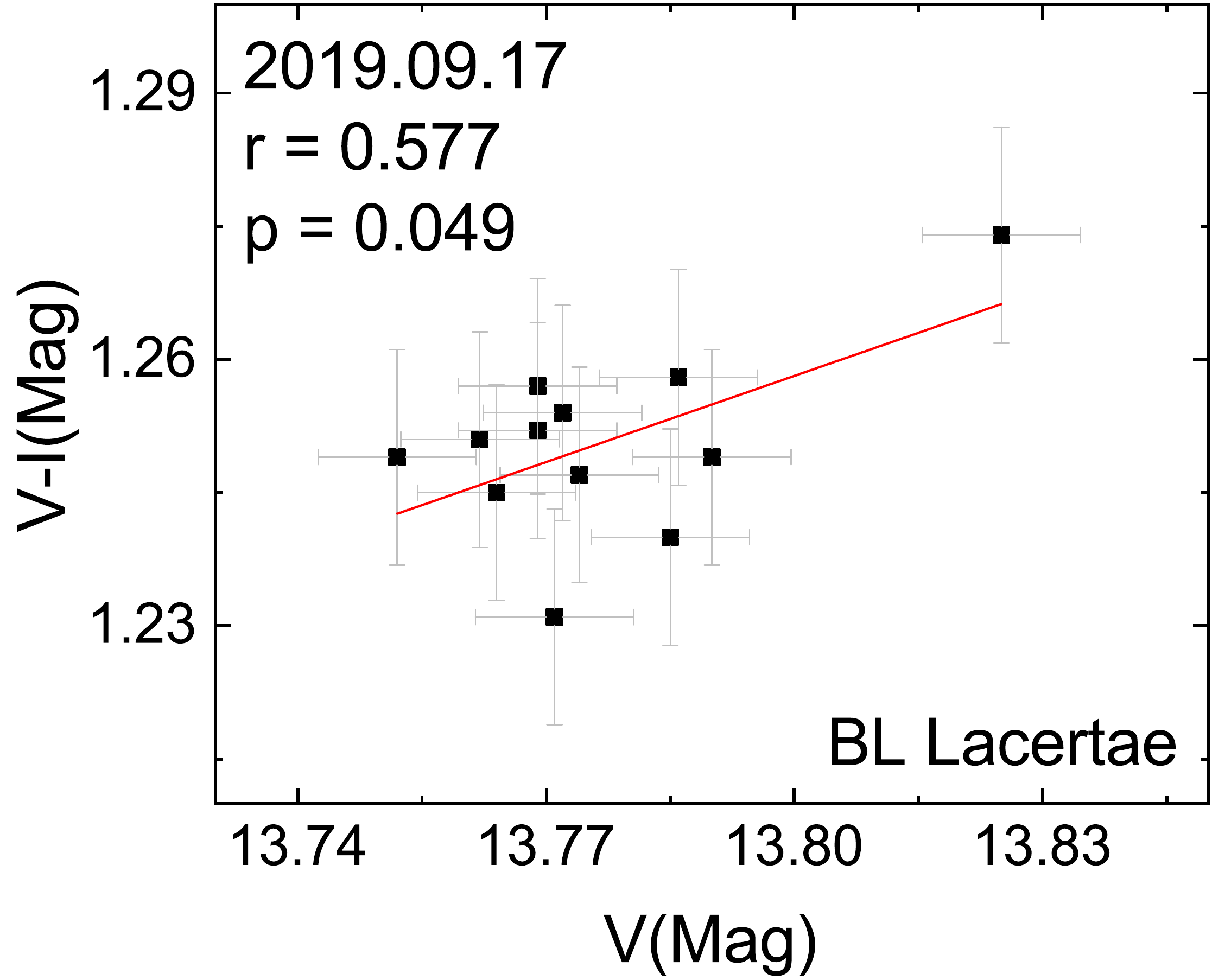}
\includegraphics[scale=.14]{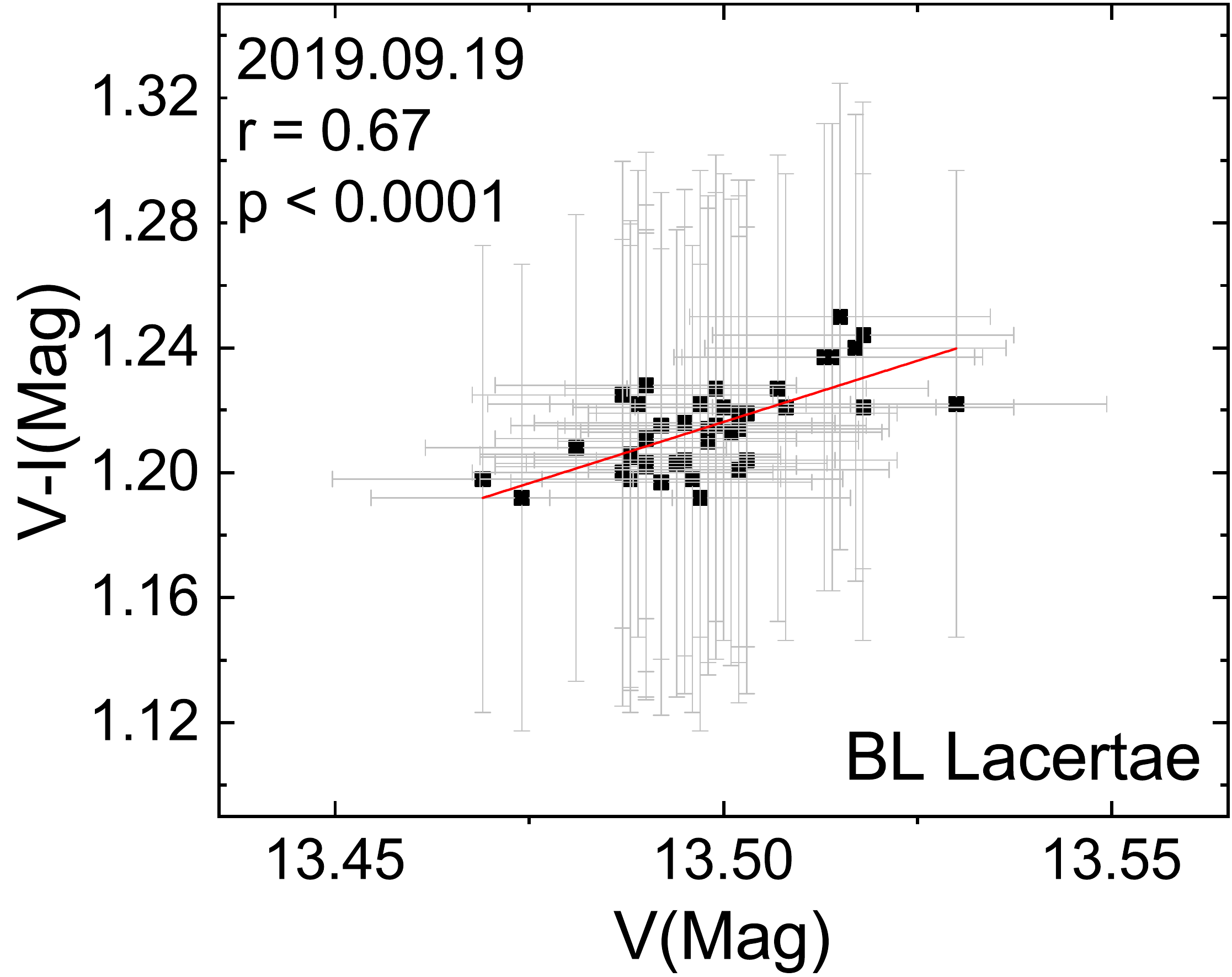}
\caption{Correlations between the $V - I$ index and $V$ magnitude for intraday. The red solid lines are the results of linear regression analysis. $r$ is the correlation coefficient; $p$ is the chance probability. \label{}}
\end{figure*}

\begin{figure*}
\centering
\includegraphics[scale=.15]{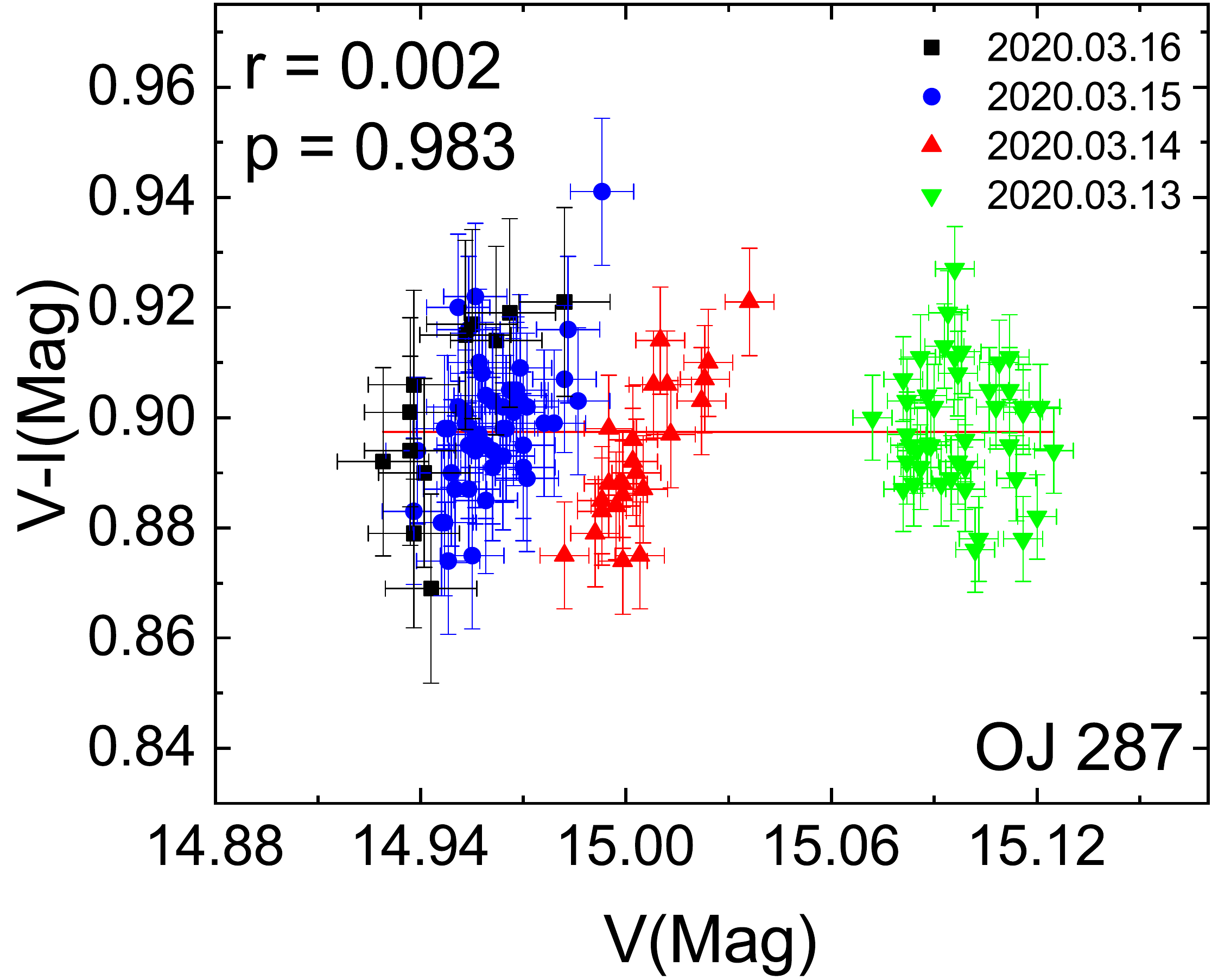}
\includegraphics[scale=.15]{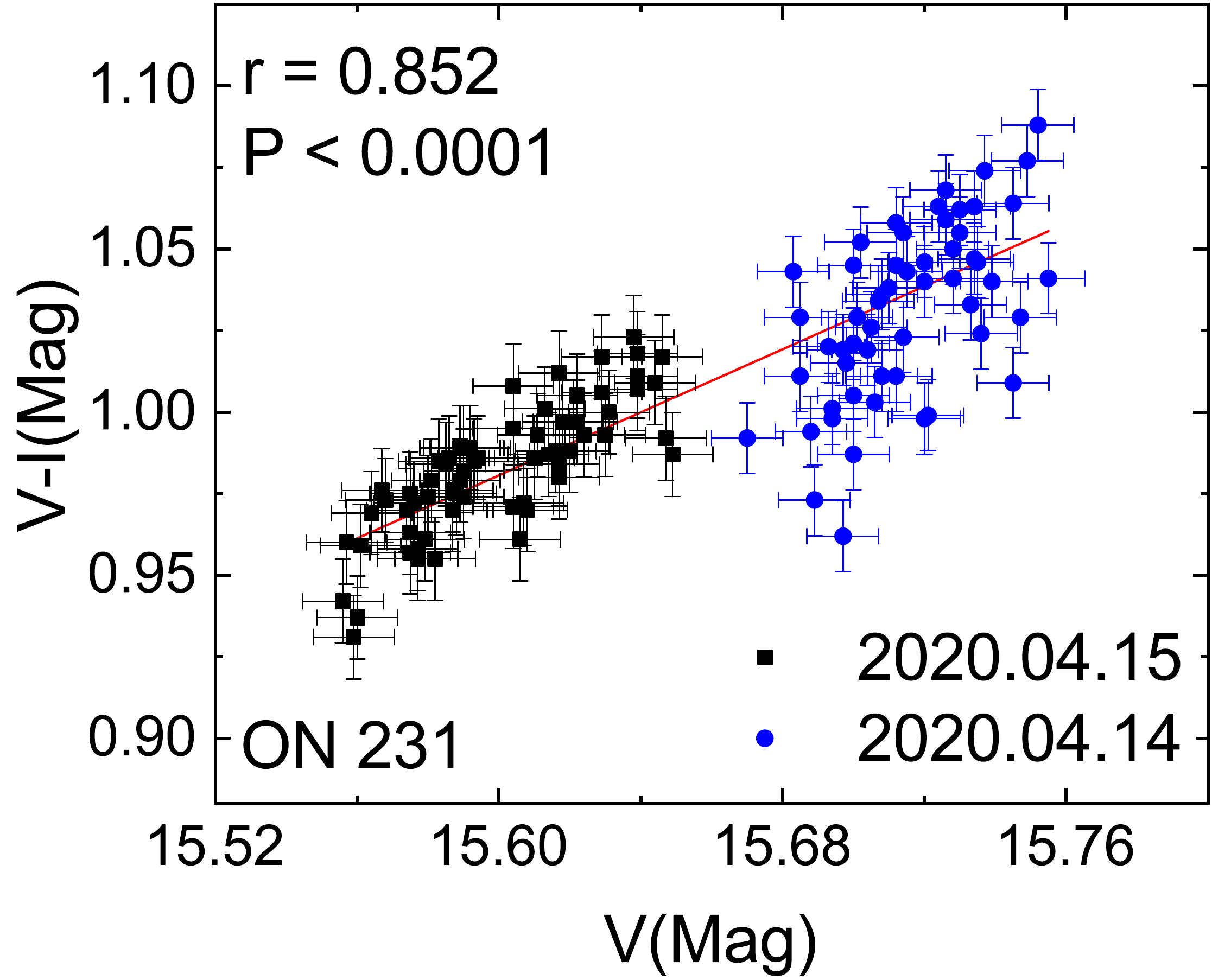}
\includegraphics[scale=.15]{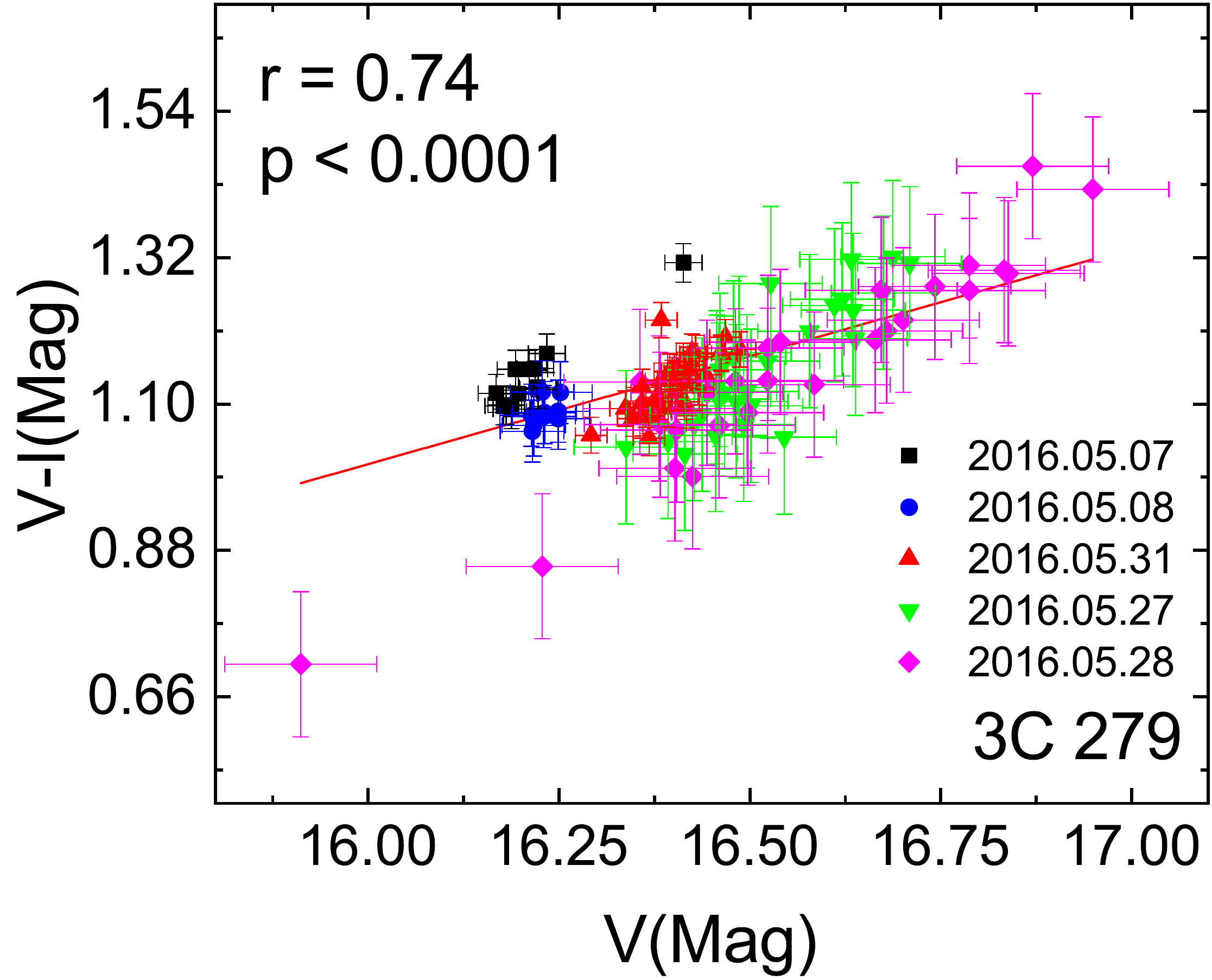}
\includegraphics[scale=.15]{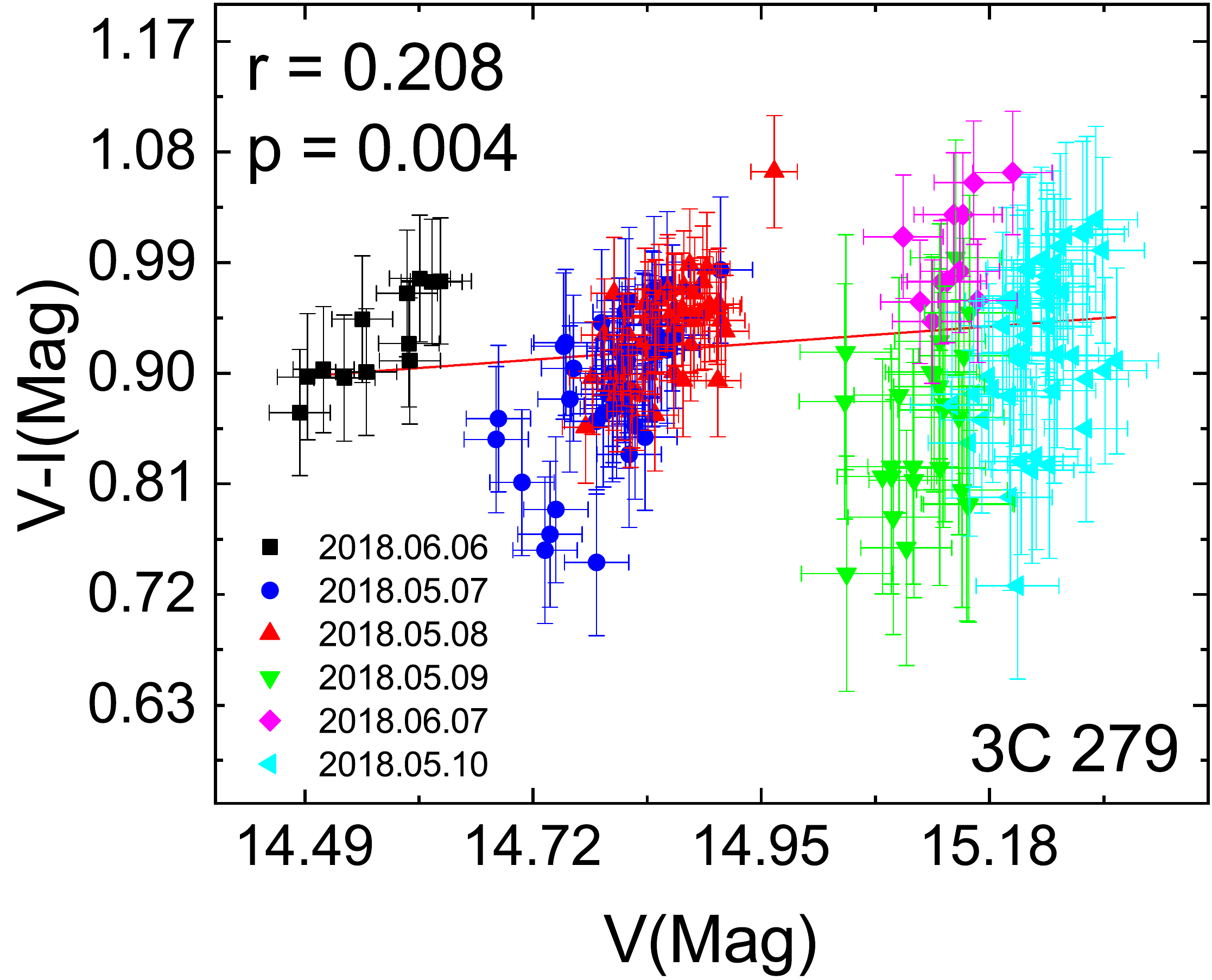}
\includegraphics[scale=.15]{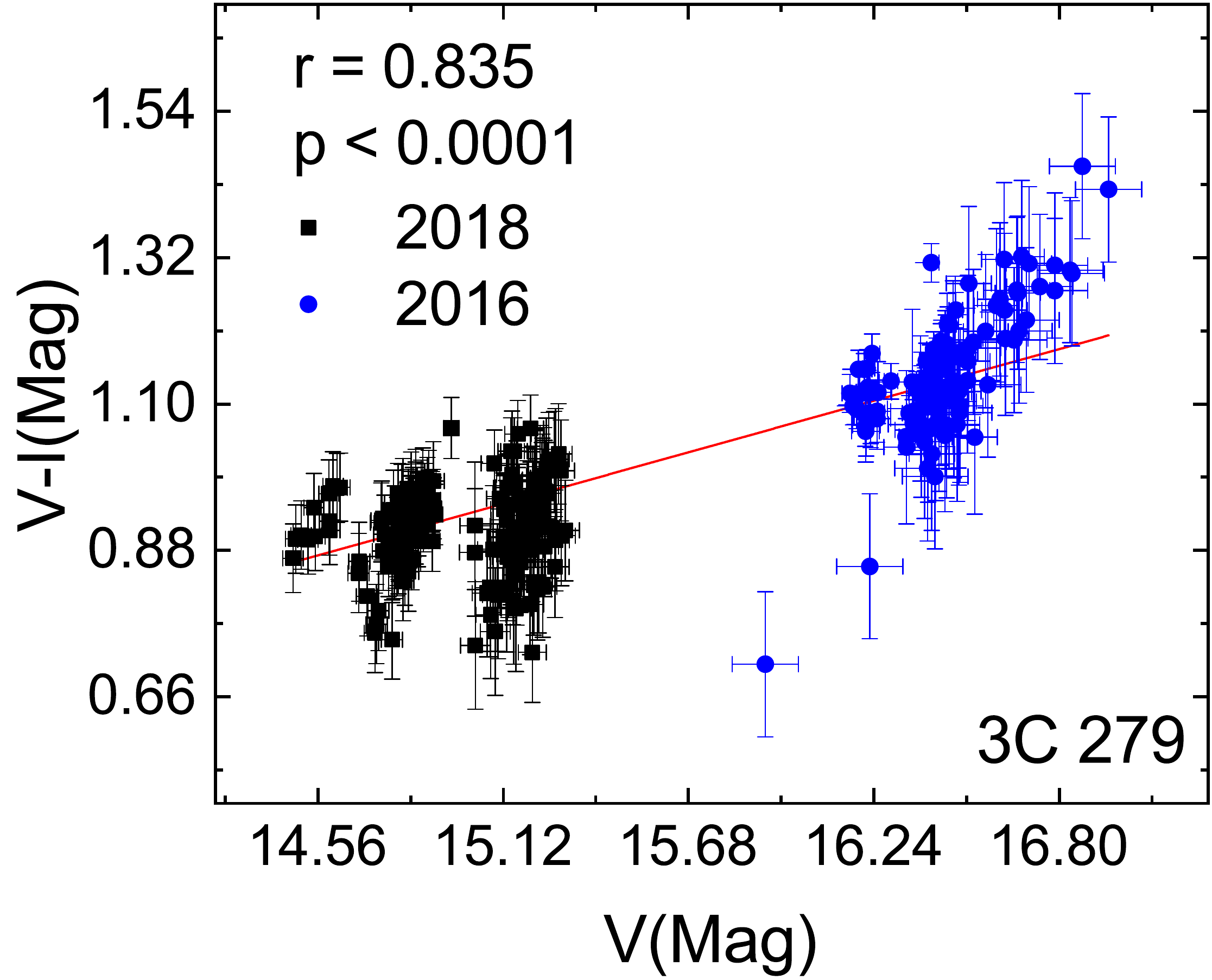}
\includegraphics[scale=.15]{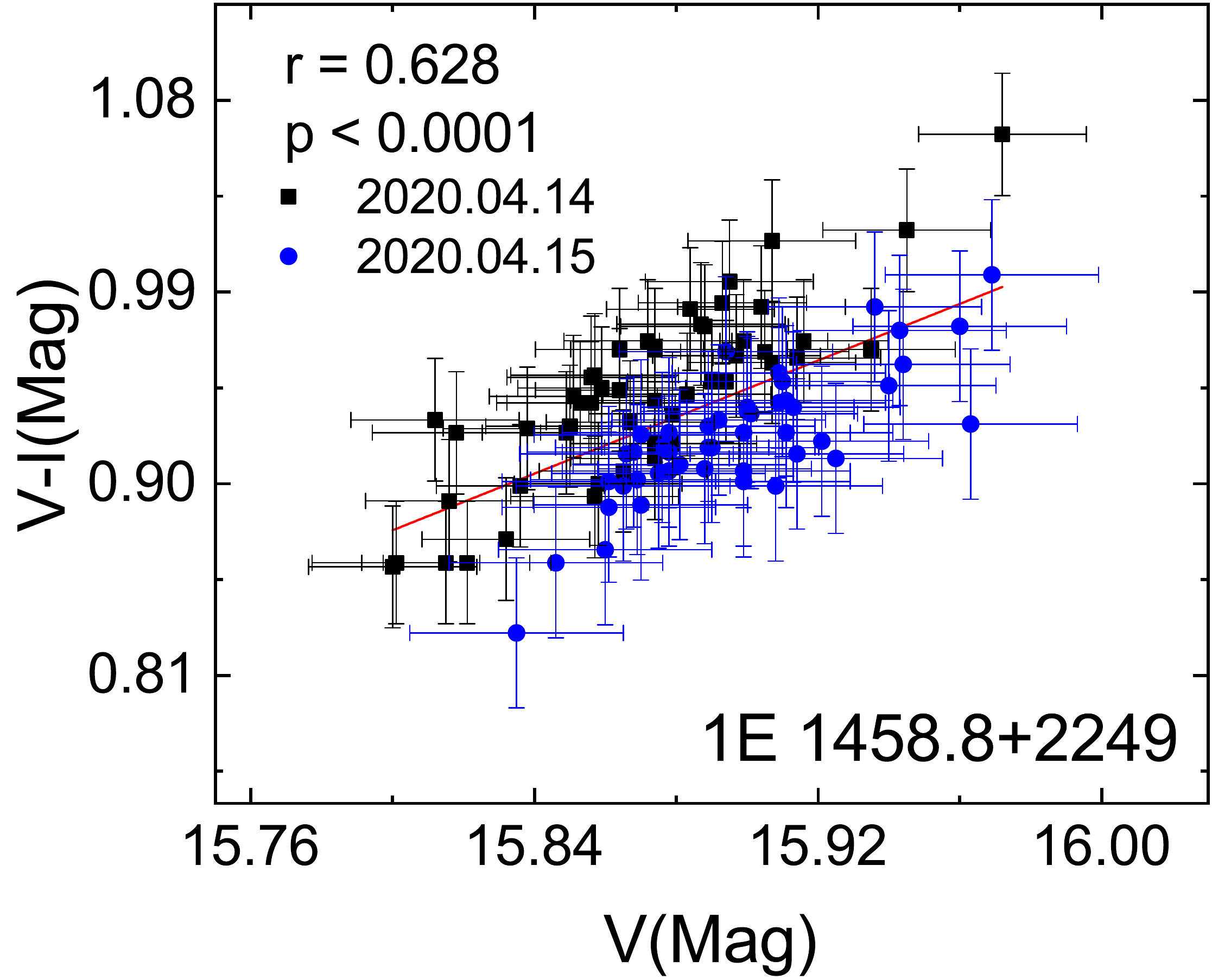}
\includegraphics[scale=.15]{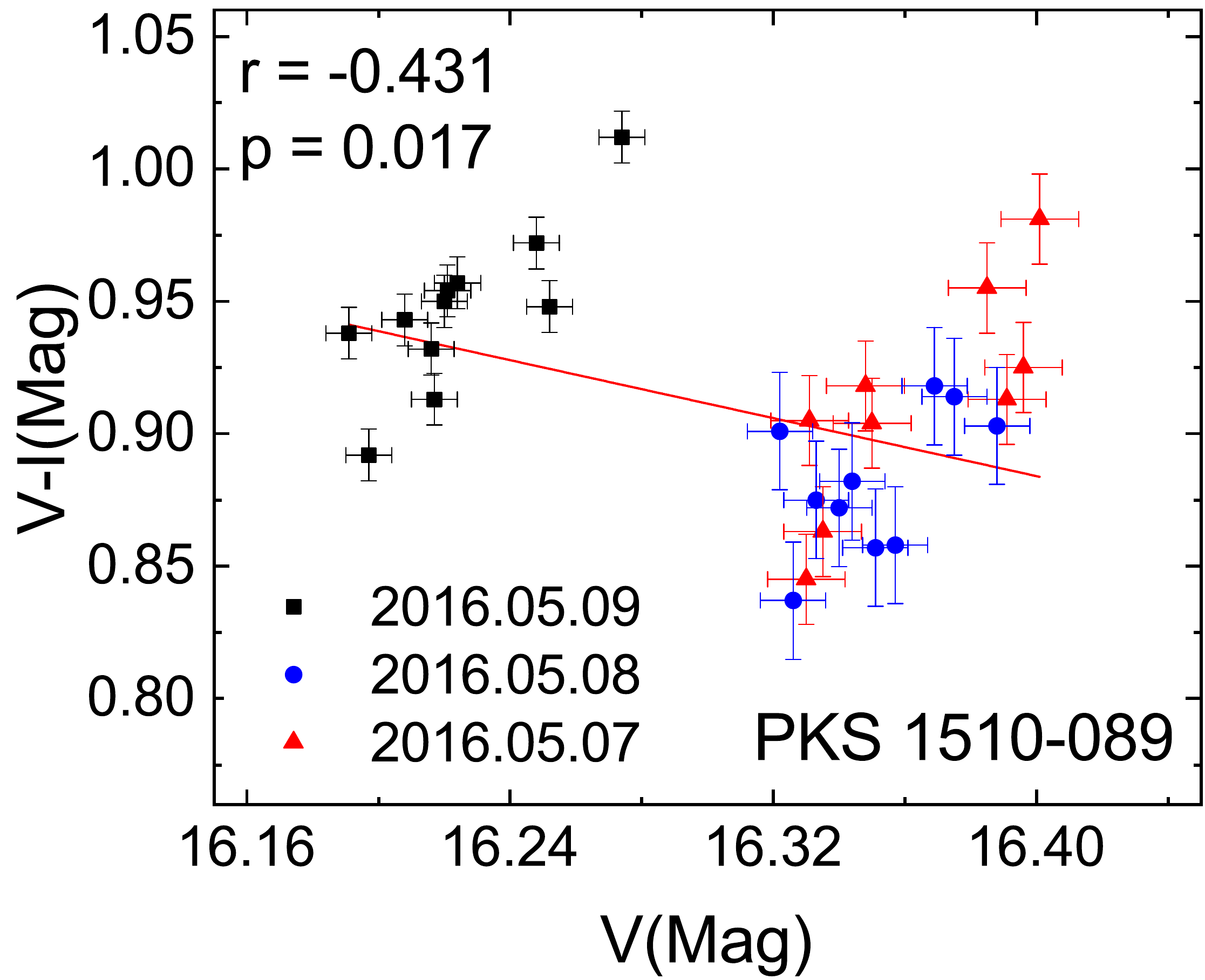}
\includegraphics[scale=.15]{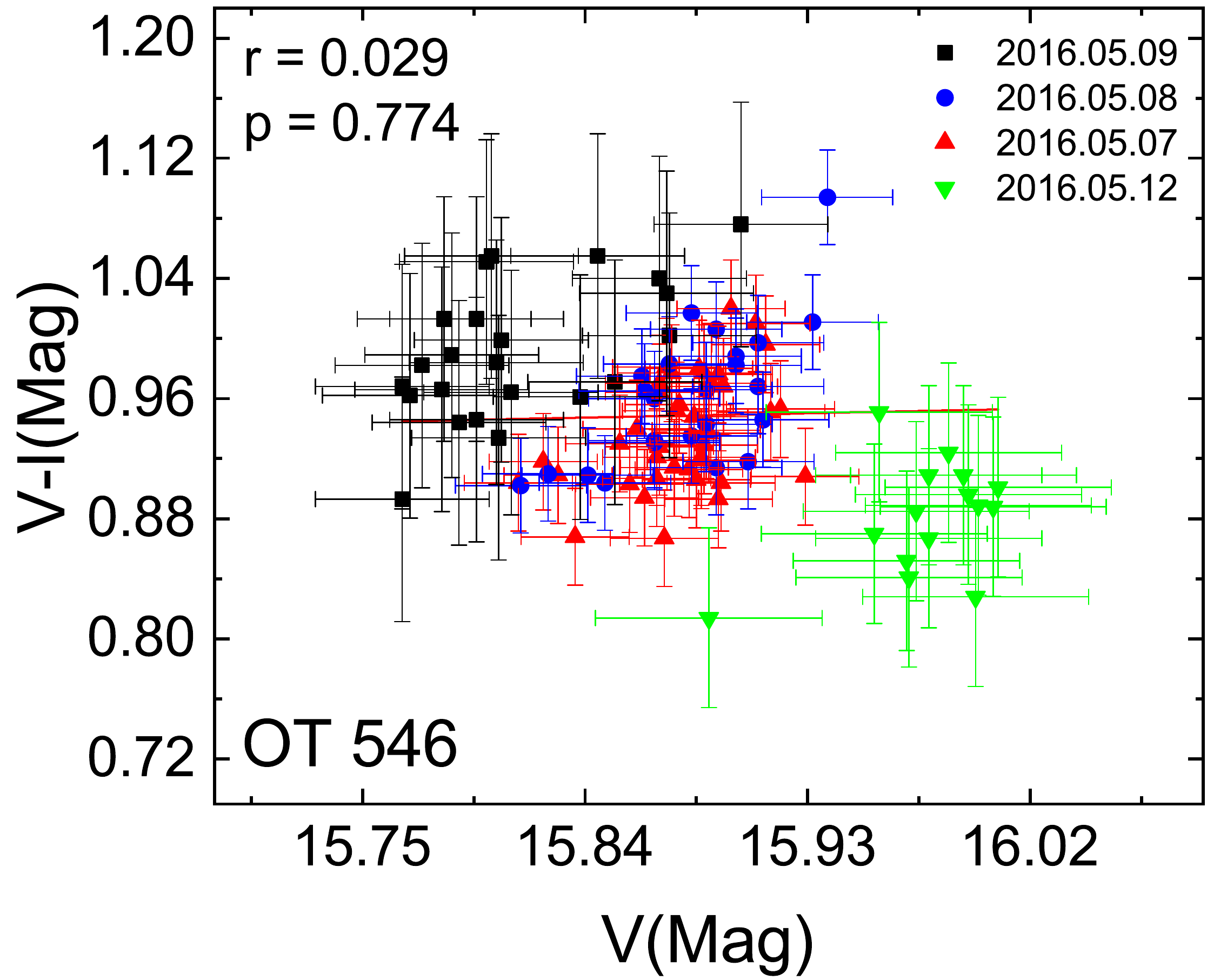}
\includegraphics[scale=.15]{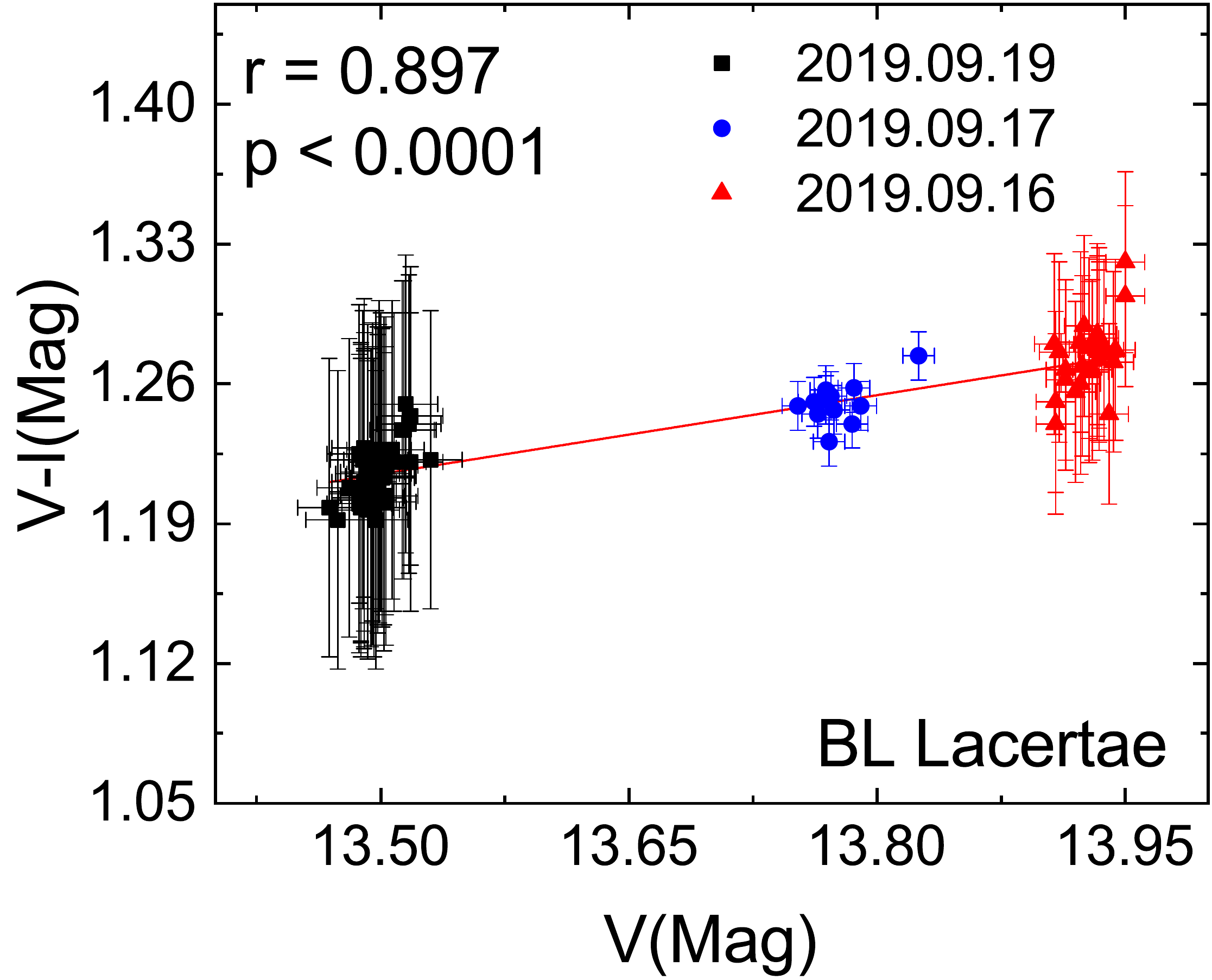}
\caption{Correlations between the $V - I$ index and $V$ magnitude for short-timescales and long-timescales. \label{}}
\end{figure*}

\begin{figure*}
\centering
\includegraphics[scale=.24]{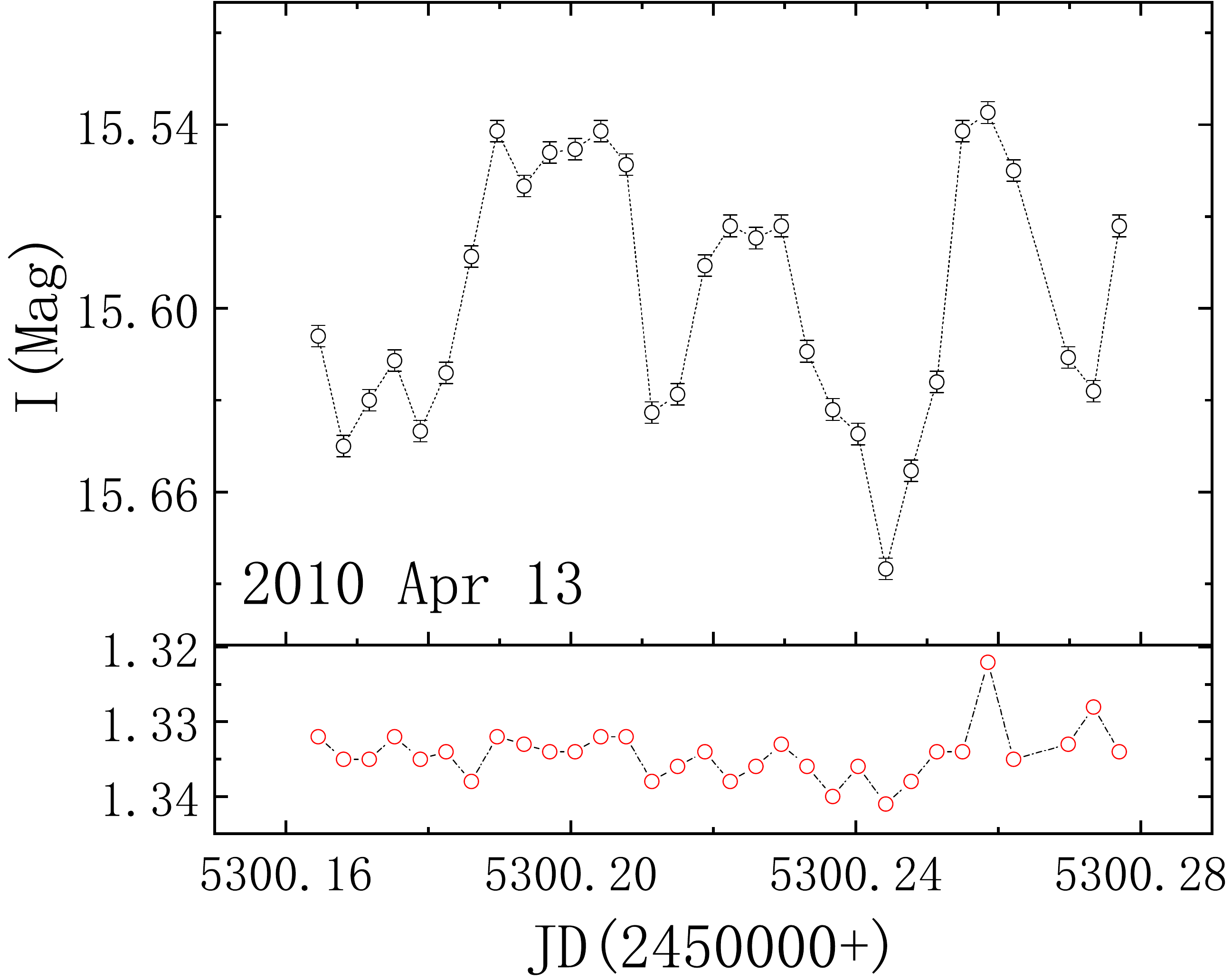}
\includegraphics[scale=.28]{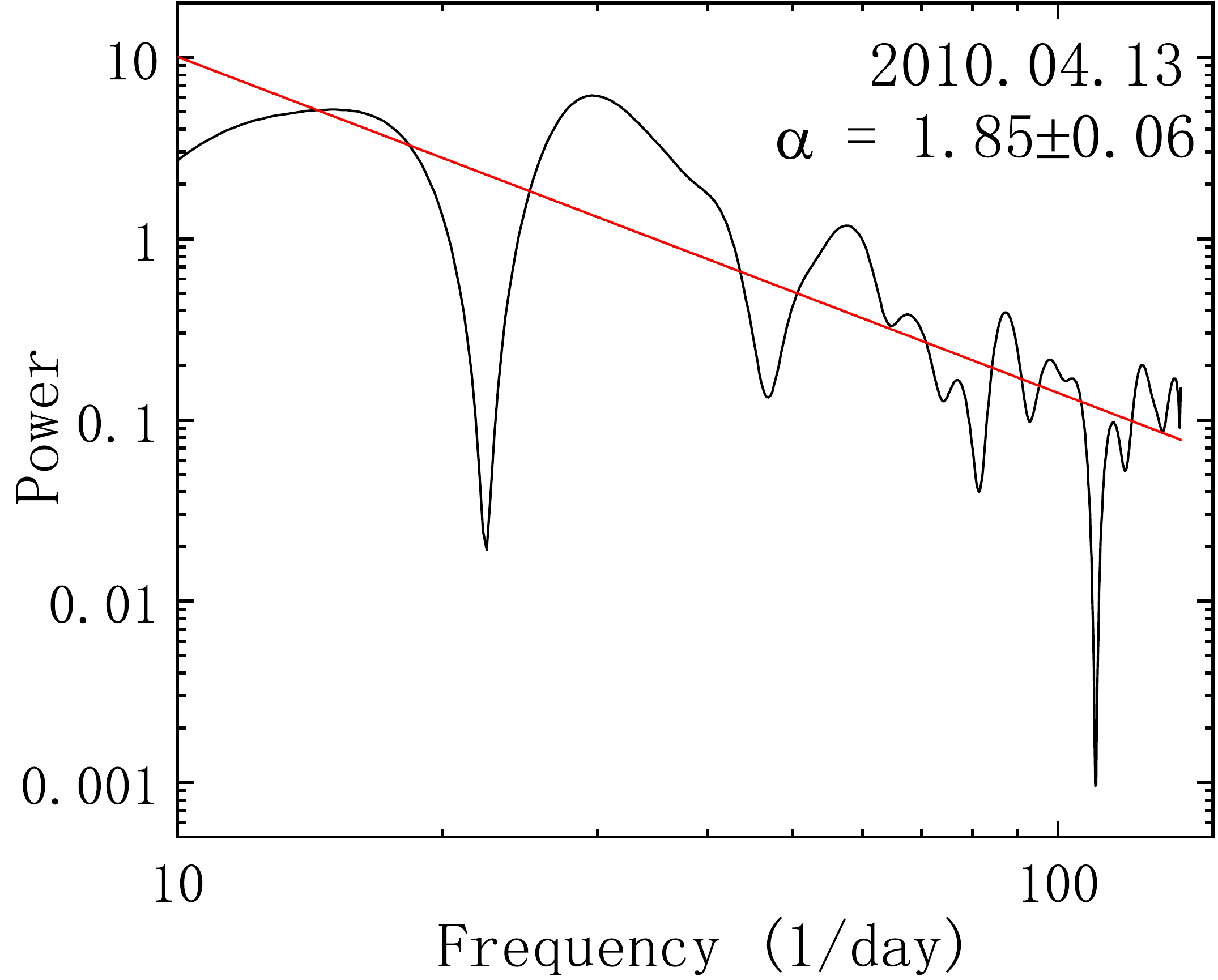}
\includegraphics[scale=.27]{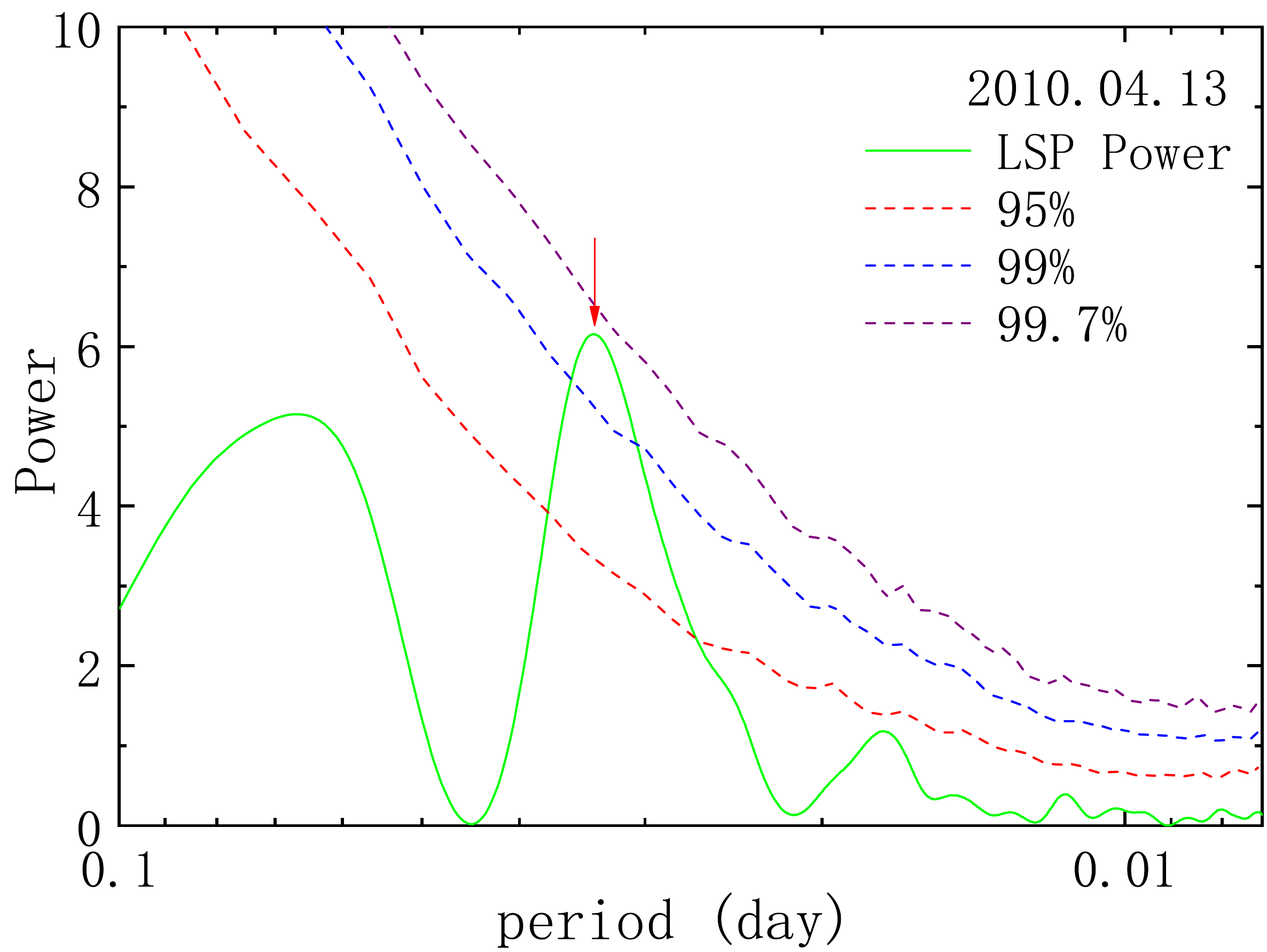}
\includegraphics[scale=.40]{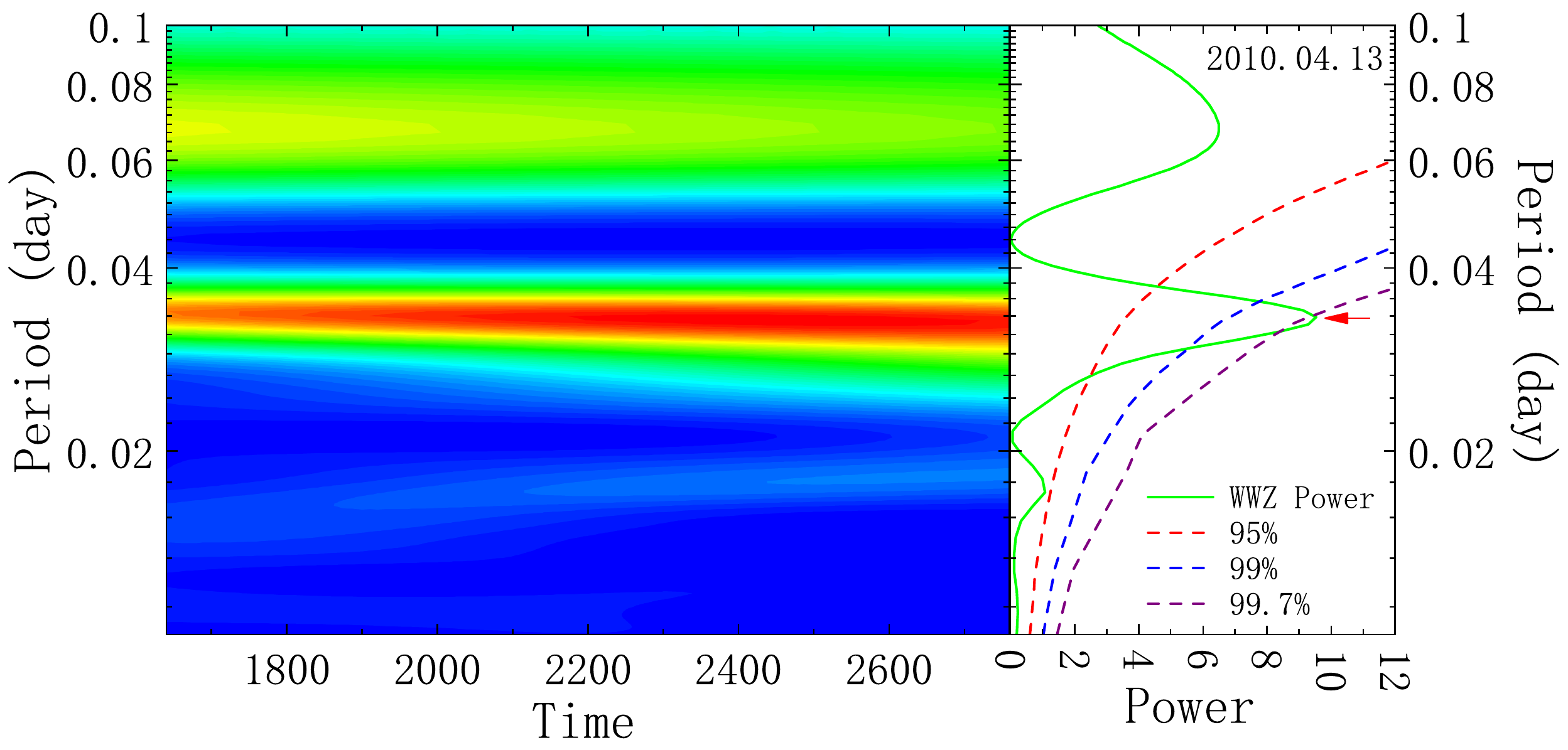}
\includegraphics[scale=.26]{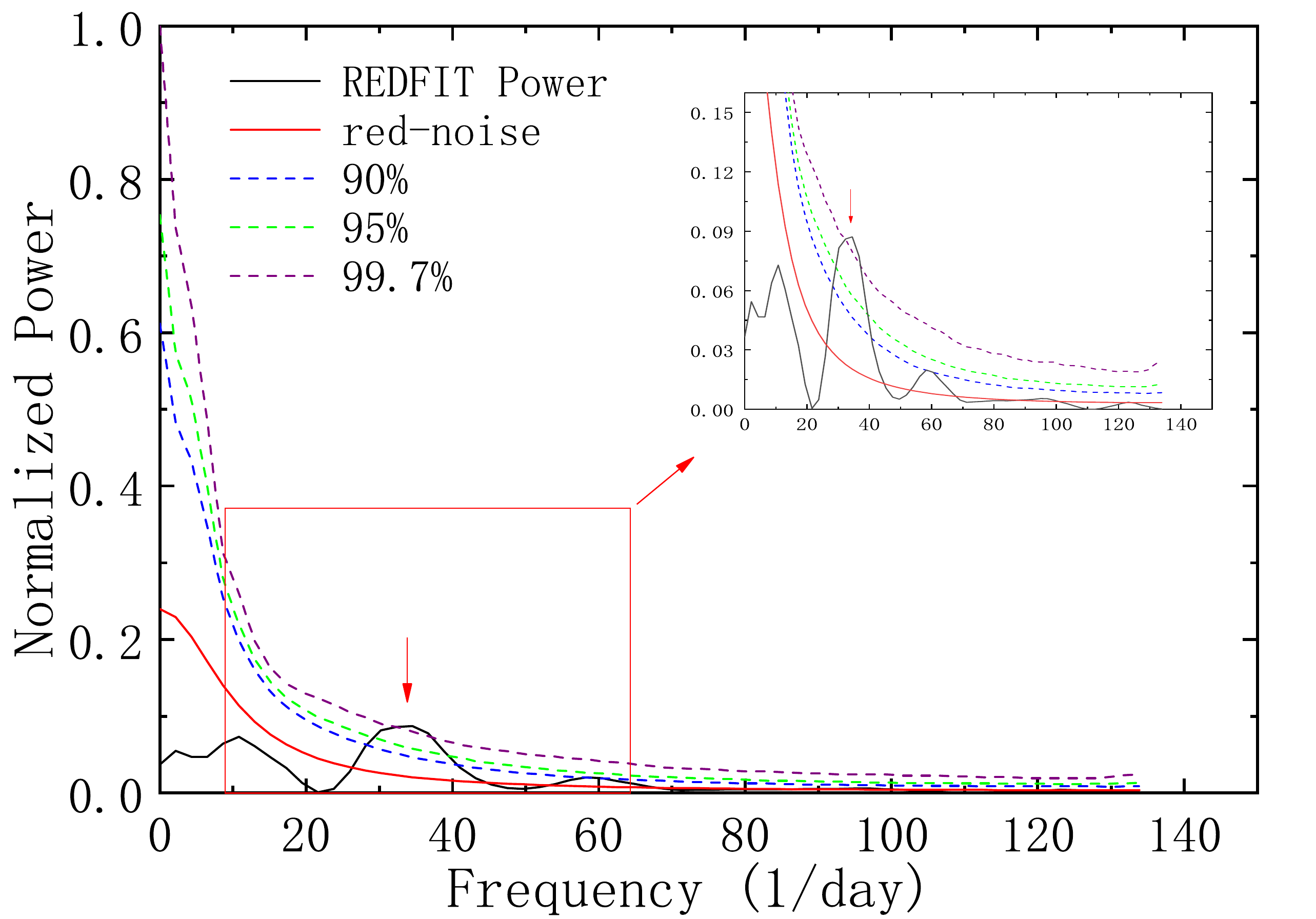}
\caption{The results of periodicity analysis of 1ES 1426+42.8 on April 13, 2010. Upper left panel: lightcurve of 1ES 1426+42.8 on April 13, 2010. Upper right panel: PSD of 1ES 1426+42.8; the solid lines denote the best-fitting power-law model of the underlying coloured noise. Middle left panel: corresponding power spectrum of LSP (green solid line); the red, blue, and purple dashed lines represent the confidence level of 95\%, 99\%, and 99.7\%, respectively. Middle right panel: 2D plane contour plot of the WWZ power of the light curve. The green solid line represents the time-averaged WWZ power; the red, blue, and purple dashed lines represent the confidence level of 95\%, 99\%, and 99.7\%, respectively. Bottom panel: corresponding result of REDFIT; the black line is the bias-corrected power spectra, the red line is the theoretical red-noise spectrum, and the blue, green, and purple dashed lines represent the confidence levels of 90\%, 95\%, and 99.7\%, respectively. \label{1ES 1426+42.8}}
\end{figure*}

\begin{figure*}
\centering
\includegraphics[scale=.22]{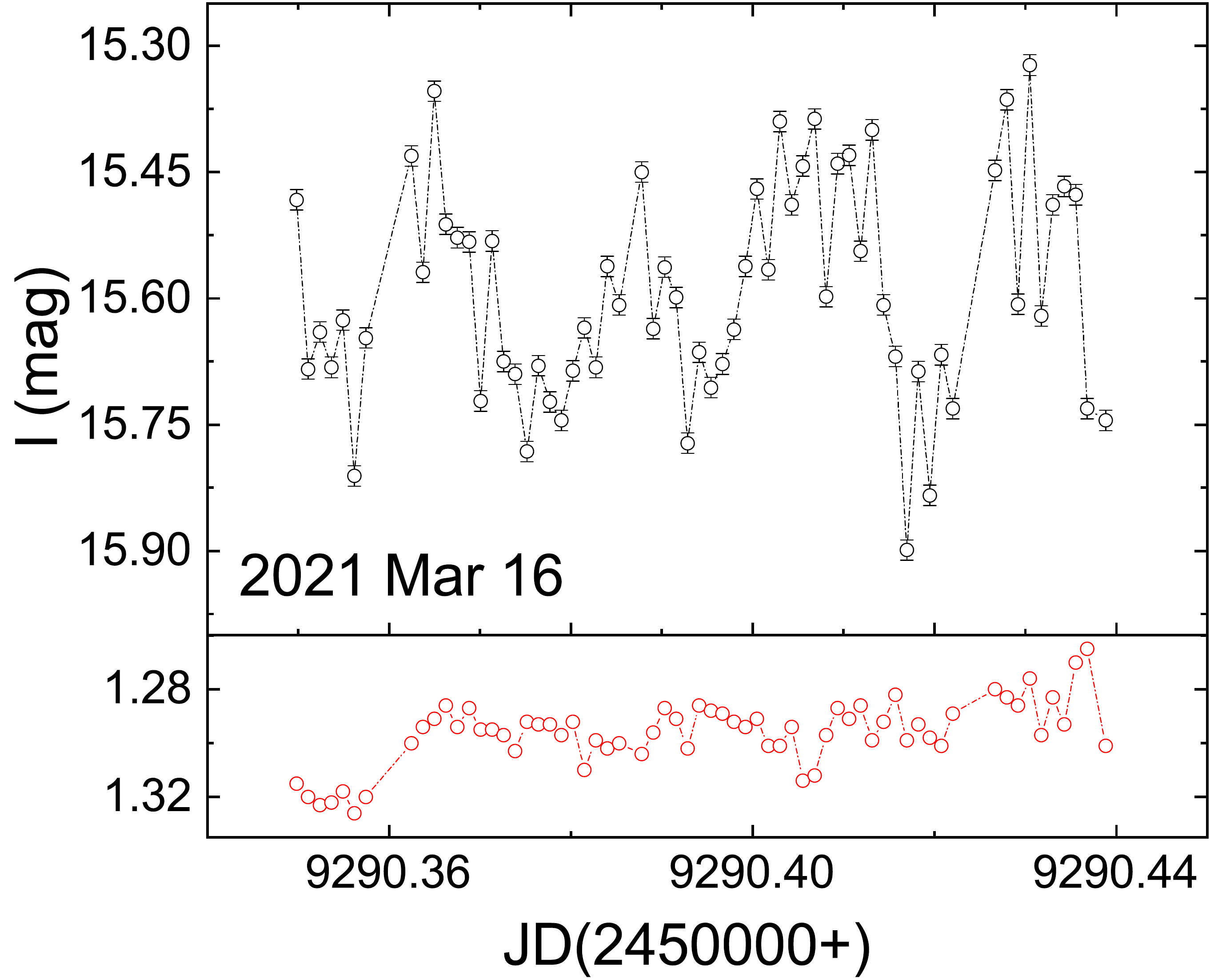}
\includegraphics[scale=.28]{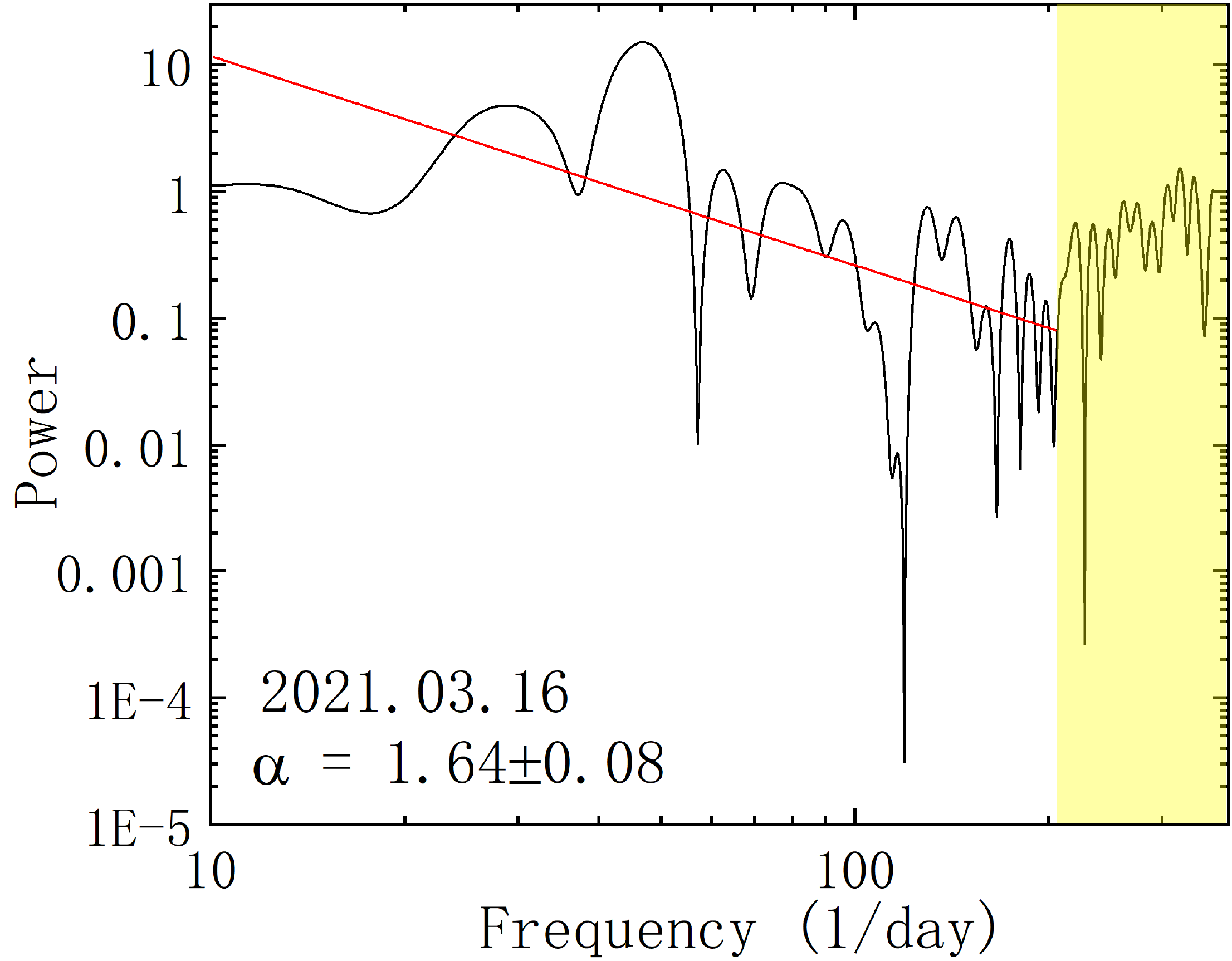}
\includegraphics[scale=.27]{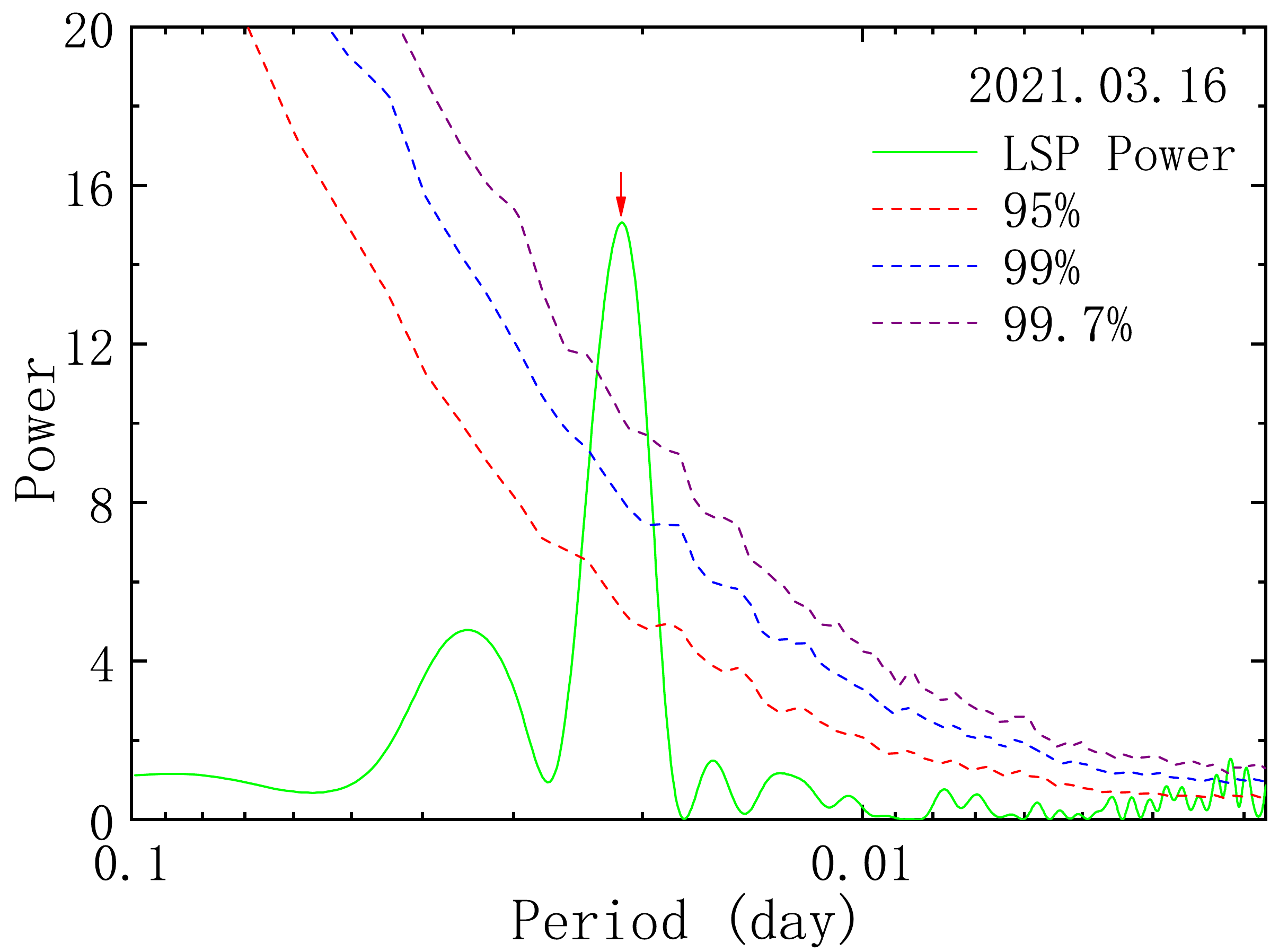}
\includegraphics[scale=.40]{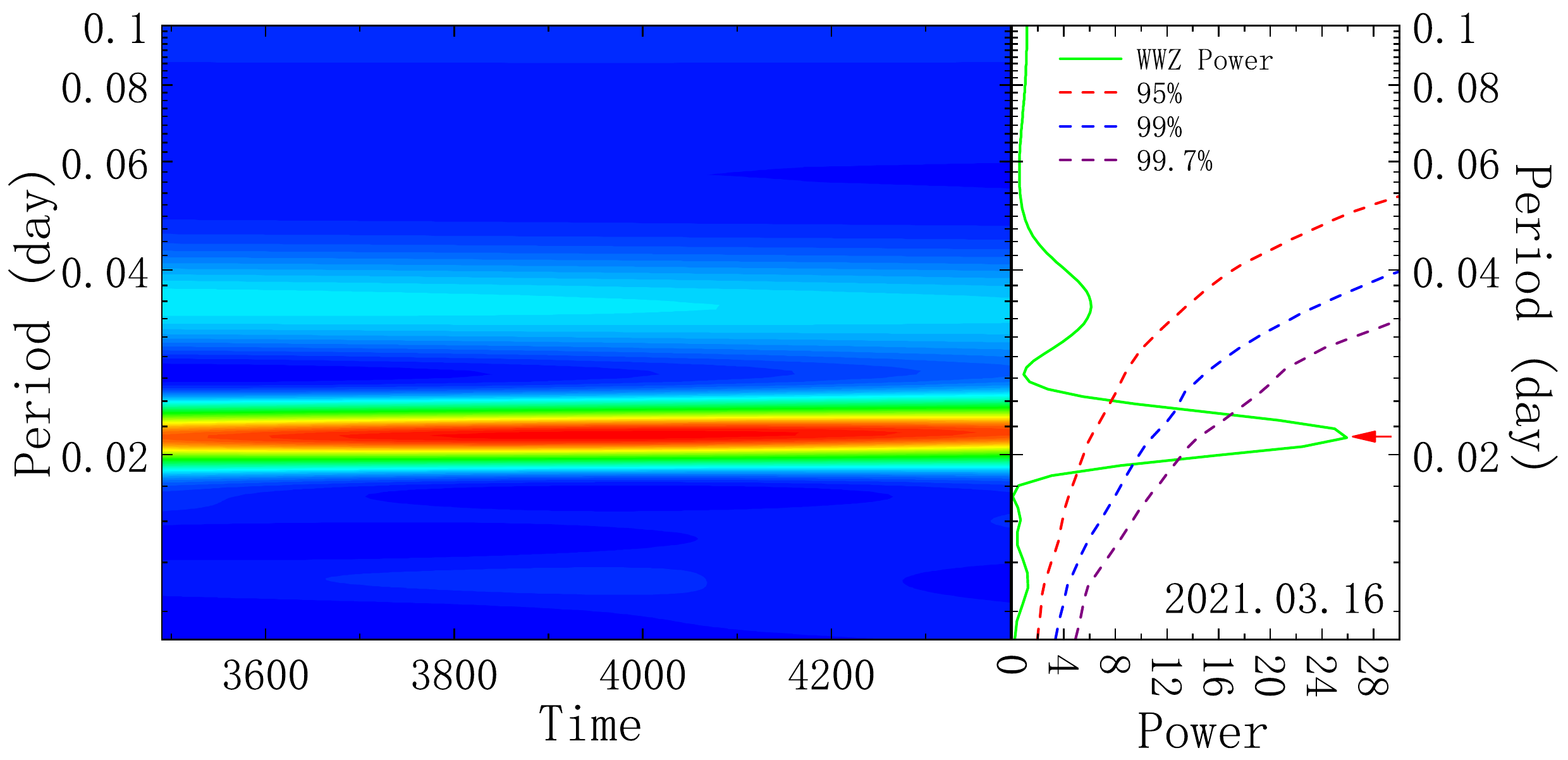}
\includegraphics[scale=.26]{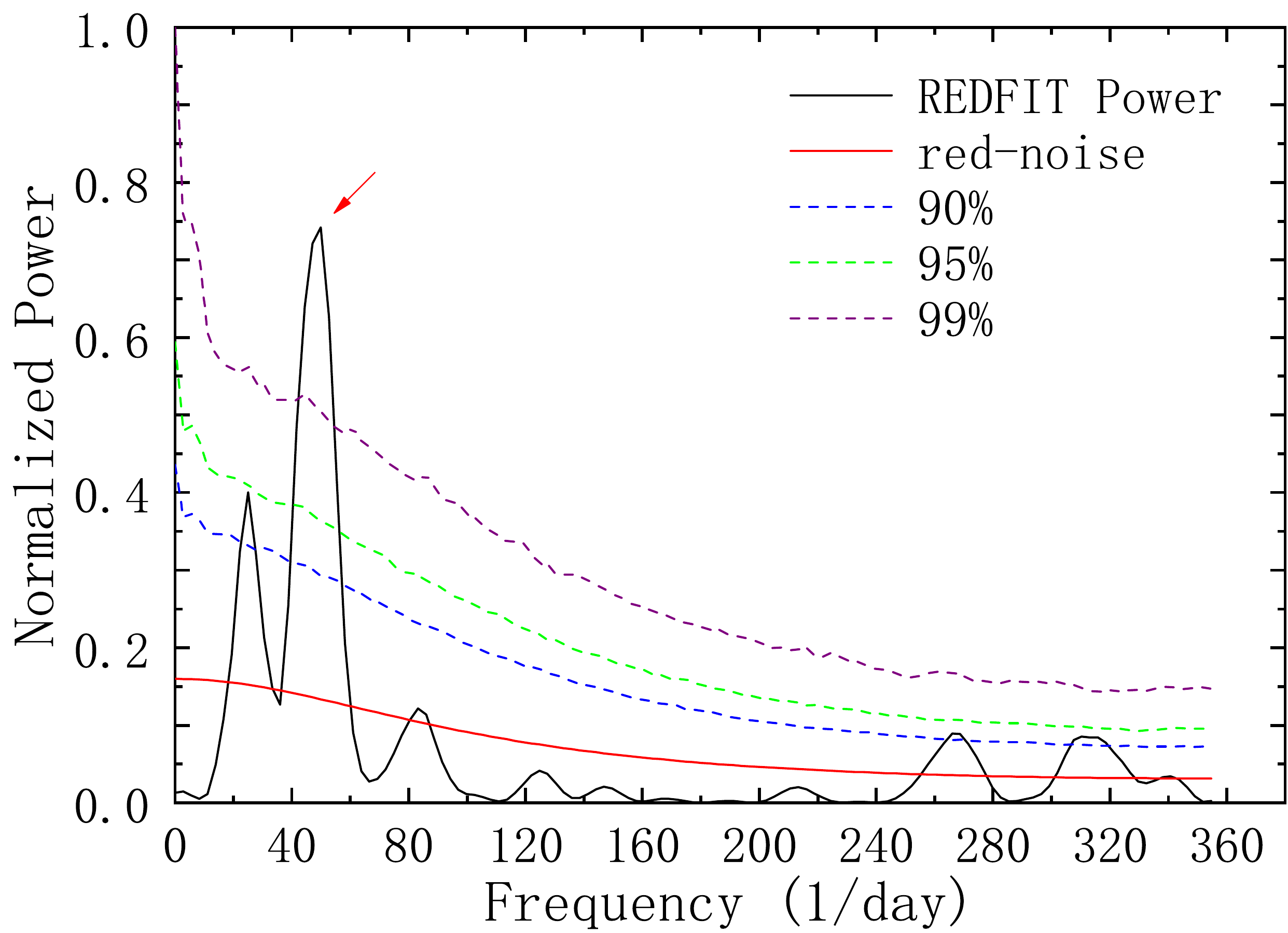}
\caption{The results of periodicity analysis of 1ES 1426+42.8 on March 16, 2021. The upper left panel is the lightcurve of 1ES 1426+42.8 on March 16, 2021. The upper right panel is the result of PSD. The middle left panel is the result of LSP method; The middle right panel is the result of WWZ method; The bottom panel is the result of REDFIT method. \label{}}
\end{figure*}

\section{Discussion}
\subsection{Optical Intraday Variability}
Many physical models have been proposed to explain variability, which can be divided into external mechanisms and internal mechanisms. The external mechanisms include interstellar scintillation and microgravity microlens models \citep{Hees87,Agar15}, and the internal mechanisms are related to relativistic jet activities and accretion disk instabilities \citep{Mang93,Chak93,Mars08}. The external mechanisms are not considered to explain optical IDVs \citep{Xion17}. Generally, the observed radiation in blazars is dominated by emission from the jet, which overwhelms the thermal emission from the accretion disk. The shock-in-jet model is often used to explain optical IDVs in blazars \citep{Mars85,Mars08,Xion16,Xion17}. The shocks propagate along the relativistic jets of plasma, sweeping emitting regions. If the emitting regions have large intrinsic changes, then a large variability on intraday timescale can be observed \citep{Mars85}. Therefore, the optical IDVs we detected are likely to be interpreted by the shock-in-jet model. In addition, the small optical variability could be attributed to turbulence behind a shock along the jet \citep{Agar15}, or the hot spots or disturbances in or above accretion disks \citep{Chak93,Mang93,Gaur12}. When the blazar is in the low state, the model based on instabilities of the accretion disks could give rise to IDVs because in the low state, any contribution from the jets is weak \citep{Rani11,Xion17}.

\subsection{Relations between Colours and Magnitudes}
Generally, BWB chromatic trend is dominant for most of the BL Lacs, while the RWB chromatic trend is usually observed in FSRQs \citep{Gu06,Dai09,Bind10}. The BWB chromatic trend is dominant for all of our objects on the intraday timescales. For short-term timescale, the BWB trends from ON 231, 3C 279, 1E 1458.8+2249, and BL Lacertae are strong, while OJ 287 and OT 546 have no chromatic trend. A moderate RWB trend of PKS 1510-089 on the short-term scale was found. The FSRQ 3C 279 showed the BWB chromatic trend in the long-term timescales.

The shock-in-jet model is most widely used to explain the BWB behavior \citep{Gupt08a,Xion16}. According to the shock-in-jet model, as the shock propagates down the jet, it strikes a region with a high electron population, and radiation at different wavelengths is produced from different distances behind the shock. High-energy photons from the synchrotron mechanism generally emerge faster and closer to the shock front than lower frequency radiation, thus causing the BWB behavior \citep{Agar15,Xion17}. Therefore, the detected BWB trends can be explained by the shock-in-jet model. In addition, the different relative contributions of the thermal versus non-thermal radiation to the optical emission may be responsible for the different trends of the colour index with brightness in FSRQs and BL Lac objects \citep{Gu06}. For FSRQ PKS 1510-089, a moderate RWB trend could be explained by a blend of jet and accretion components. As the target brightens, the red component of the jet dominates the radiation of the total flux. So a moderate RWB trend could be observed. The superposition of the different BWB trends on intraday timescales may explain the no chromatic trend in the two BL Lac objects OJ 287 and OT 546 \citep{Xion20}. In addition, the BWB trends may occur when electrons being accelerated to preferentially higher energies before radiative cooling, while the RWB trends may occur when the highest-energy electrons suffer the stronger radiative cooling or escape cooling \citep{Isle17,Xion20}.

\subsection{Intraday Periodicity}
We have searched for the QPO of 1ES 1426+42.8 in the I band light curve on April 13, 2010, using the LSP, WWZ, and REDFIT methods. The possible QPOs are $48.67 \pm 13.90$ minutes ($> 99\%$ confidence level) with LSP analysis, $47.23 \pm 11.21$ minutes ($> 3\sigma$) with WWZ analysis, and $41.62 \pm 14.25$ minutes ($> 3\sigma$) with REDFIT analysis, respectively. The QPOs ($> 3\sigma$) on March 16, 2021 are calculated with consistent periods by the three methods as $30.70 \pm 6.55$ minutes, $30.74 \pm 5.22$ minutes, and $28.88 \pm 8.66$ minutes, respectively. Therefore, based on our new observations, a possible intraday QPO (about 30 to 50 minutes) is found, which is rarely reported in previous literature.

For the periodic variability of blazar, some possible mechanisms are as follows: the close binary black hole system model \citep[e.g.,][]{Vill98}, the orbital motion of hot spots around the SMBH \citep[e.g.,][]{Brod06}, the global oscillations of the accretion disc \citep[e.g.,][]{Rubi05}, the precession of jet, or the rotating helical jet structure \citep[e.g.,][]{Vill99,Fan14}. In the binary black hole system model, the periodicity would be induced by the Keplerian orbital motion of a binary SMBH, thus producing the long-term or short-term QPOs in blazars \citep{Rome00}. According to this, the binary black hole system model is unlikely to explain the intraday QPO.

The detection of the periodic variability on the intraday timescale could be explained by the presence of a single dominating hot spot on the accretion disk \citep[e.g.,][]{Mang93,Chak93} or perhaps by pulsational modes in the disk \citep[e.g.,][]{Espa08}. The instability of a disk will lead to the periodicity of the blazar \citep{Ren21}. Based on the assumption that the QPO is related to the orbital timescale of a hot spot, spiral shocks, or other non-axisymmetric phenomena in the innermost portion of the rotating accretion disk, the SMBH mass can be estimated with \citep{Gupt09,Gupt19}
\begin{equation}
\frac {M_{\bullet}}{M_{\odot}} = \frac {\ 3.23 \times 10^4 \delta P}{{(r^{\ 3/2}+a)}{(1+z)}},
\label{eq:LebsequeIp3}
\end{equation}
where $\delta$ is the Doppler factor of the blazar, $P$ is the period of QPO in units of second, $r$ is the radius at the inner edge of accretion disk in units of $GM_{\bullet}/c^2$, $a$ is the SMBH spin parameter, and $z$ is the redshift. From Equation (12), we estimated the mass of SMBH as $10^{8.99^{+0.11}_{-0.15}} M_{\odot}$ for the Kerr black hole (with $r$ = 1.2, $a$ = 0.9982 and $P=48.67\pm13.90$ minutes) and $10^{8.19^{+0.11}_{-0.15}} M_{\odot}$ for the Schwarzschild black hole (with $r$ = 6.0, $a$ = 0 and $P=48.67\pm13.90$ minutes) in the case of 1ES 1426+42.8. Our black hole masses estimated by the period of QPO are within the ranges of black hole masses calculated by minimum timescale, which indicates that the intraday QPO caused by the perturbations on the accretion disk is possible.

\citet{Mara03} proposed that, for the low-luminosity blazars (BL Lacs), the jet luminosity is higher than the disk luminosity because of a very low radiative efficiency for the accretion disk. In the scenario of BL Lac 1ES 1426+42.8, a disk flux variation is unlikely to directly produce any detectable QPO. The fluxes from jets usually would swamp any disk fluxes. Thus, the QPOs of 1ES 1426+42.8 detected by us probably produce from the relativistic jets. It is worth noting that there is a connection between the jet and the accretion disk \citep{Dann20}. The perturbations caused by disk instabilities can drive changes in the mass flux entering the jets or the velocity (or density, magnetic field, etc.) of the jets, thus producing the relativistic shocks \citep{Wiit06}. When a relativistic shock advances along the helical structure of the jet, or the jet precesses/twists, variation in the Doppler boosting factor would amplify weak fluctuations, reduce timescales, and produce the quasi-periodic flux variations \citep{Came92,Vill99,Gupt19,Rani10}. Therefore, the perturbations from disks would probably produce actual QPOs, which are in turn amplified by the relativistic motions of jets. In addition, it is very likely that the intraday QPO is originated from jets, instead of accretion disks. As mentioned above, when a relativistic shock advances along the helical structure of the jet (or the jet precesses/twists), configured with high Lorentz factor and very small viewing angle, an intraday QPO could be observed.

\section{Conclusions}
We present quasi-simultaneous multi-colour photometric data of eight blazars (3244 data points) observed over a time span from 2010-2020. After analyzing flux variations, correlations between magnitudes and colours on different timescales, our main results are summarized as follows.

(1) Intraday variability (IDV) is detected in all eight sources of our sample. The IDV of OJ 287 is detected on one night. The IDV is found on 3 nights for ON 231, 2 nights for 3C 279, 1 night for 1ES 1426+42.8, 2 nights for 1E 1458.8+2249, 5 nights for PKS 1510-089, 1 night for OT 546 and 1 night for BL Lacertae.

(2) A BWB chromatic trend is dominant for all eight objects on intraday timescales. On the short timescales, the BWB trend is displayed in four targets (ON 231, 3C 279, 1E 1458.8+2249, and BL Lacertae). There is a target (3C 279) detected with the BWB trend on the long
timescales. A FSRQ (PKS 1510-089) is detected with the RWB trend.

(3) Based on the ACF analysis, the upper limits of black hole mass for three blazars are estimated using variability timescales. In the case of Kerr black holes, the black hole masses are $M_{\bullet} \lesssim 10^{8.30} M_{\odot}$ for ON 231, $M_{\bullet} \lesssim 10^{9.03} M_{\odot}$ for 1ES 1426+42.8, and $M_{\bullet} \lesssim 10^{9.01} M_{\odot}$ for PKS 1510-089, respectively. In the case of Schwarzschild black holes, the black hole masses are $M_{\bullet} \lesssim 10^{7.82} M_{\odot}$ for ON 231, $M_{\bullet} \lesssim 10^{8.55} M_{\odot}$ for 1ES 1426+42.8, and $M_{\bullet} \lesssim 10^{8.62} M_{\odot}$ for PKS 1510-089, respectively.

(4) On April 13, 2010, a potential QPO of $P=48.67\pm13.90$ minutes is found in 1ES 1426+42.8. The light curve on March 16, 2021 further confirms the existence of the QPO ($> 3\sigma$). The black hole mass is estimated as $10^{8.99^{+0.11}_{-0.15}} M_{\odot}$ for the Kerr black hole and $10^{8.19^{+0.11}_{-0.15}} M_{\odot}$ for the Schwarzschild black hole using the period of the QPO.

\section{acknowledgements} 
We thank the anonymous referee for the valuable comments and suggestions. This work is supported by the National Natural Science Foundation of China (grants 11863007, 12063005, 12063007, 11703078), the Yunnan Province Foundation (2019FB004), the Program for Innovative Research Team (in Science and Technology) in University of Yunnan Province (IRTSTYN), the Yunnan Local Colleges Applied Basic Research Projects (2019FH001-12), and National Astronomical Observatories Yunnan Normal University Astronomical Education Base. We acknowledge the science research grants from the China Manned Space Project with NO. CMS-CSST-2021-A06.

\section{Data availability}
The data underlying this article will be shared on reasonable request to the corresponding author.
Sample of the all observed data will be available electronically at Vizier.

~\\
\appendix{APPENDIX}
\section{The entirety of Table 2, 3}

In electronic form only.


\begin{thebibliography}{90}
\expandafter\ifx\csname natexlab\endcsname\relax\def\natexlab#1{#1}\fi
\bibitem[Abdo et al.(2010a)]{Abdo10a} Abdo A. A. et al., 2010a, ApJ, 716, 30
\bibitem[Abdo et al.(2010b)]{Abdo10b} Abdo A. A. et al., 2010b, ApJ, 715, 429
\bibitem[Abramowicz \& Nobili(1982)]{Abra82} Abramowicz M. A., Nobili L., 1982, Nature, 300, 506
\bibitem[Agarwal et al.(2016)]{Agar16} Agarwal A. et al., 2016, MNRAS, 455, 680
\bibitem[Agarwal \& Gupta(2015)]{Agar15} Agarwal A., Gupta A. C., 2015, MNRAS, 450, 541
\bibitem[Ajello et al.(2017)]{Ajel17} Ajello M. et al., 2017, ApJS, 232, 18
\bibitem[Ajello et al.(2020)]{Ajel20} Ajello M. et al., 2020, ApJ, 892, 105
\bibitem[Alexander(1997)]{Alex97} Alexander T., 1997, in Astronomical Time Series, ed. D. Maoz A. Sternberg, \& E. M. Leibowitz (Dordrecht: Kluwer), 163
\bibitem[Angel \& Stockman(1980)]{Ange80} Angel J. R. P., Stockman H. S., 1980, ARA\&A, 18, 321
\bibitem[Bai et al.(1998)]{Bai98} Bai J. M., Xie G. Z., Li K. H., Zhang X., Liu W. W., 1998, A\&AS, 132, 83
\bibitem[Bonning et al.(2012)]{Bonn12} Bonning E. et al., 2012, ApJ, 756, 13
\bibitem[Bottcher et al.(2013)]{Bott13} Bottcher M., Reimer A., Sweeney K., Prakash A., 2013, ApJ, 768, 54
\bibitem[Bhatta(2017)]{Bhat17} Bhatta G., 2017, ApJ, 847, 7
\bibitem[Bhatta(2021)]{Bhat21} Bhatta G., 2021, ApJ, 923, 7
\bibitem[Bindu et al.(2010)]{Bind10} Bindu R. et al., 2010, MNRAS, 404, 1992
\bibitem[Broderick \& Loeb(2006)]{Brod06} Broderick A. E., Loeb A., 2006, MNRAS, 367, 905
\bibitem[Camenzind \& Krockenberger(1992)]{Came92} Camenzind M., Krockenberger M., 1992, A\&A, 255, 59
\bibitem[Carini et al.(1992)]{Cari92} Carini M. T., Miller H. R., Noble J. C., Goodrich B. D., 1992, AJ, 104, 15
\bibitem[Chatterjee et al.(2012)]{Chat12} Chatterjee R. et al., 2012, ApJ, 749, 191
\bibitem[Chakrabarti \& Wiita(1993)]{Chak93} Chakrabarti S. K., Wiita P. J., 1993, ApJ, 411, 602
\bibitem[Cohen(1988)]{Cohe88} Cohen J., 1988, Statsitical Power Analysis for the Behavioral Sciences (2nd ed.; New York: Academic Press)
\bibitem[Costamante et al.(2003)]{Cost03} Costamante L., Aharonian F., Ghisellini G., Horns D., 2003, New Astron. Rev., 47, 677
\bibitem[Dai et al.(2001)]{Dai01} Dai B. Z. et al., 2001, AJ, 122, 2901
\bibitem[Dai et al.(2009)]{Dai09} Dai B. Z. et al., 2009, MNRAS, 392, 1181
\bibitem[Dai et al.(2015)]{Dai15} Dai B. Z. et al., 2015, ApJS, 218, 18
\bibitem[Dannen et al.(2020)]{Dann20} Dannen R. C., Proga D., Waters T., Dyda S., 2020, ApJ, 893, L34
\bibitem[Dermer et al.(2012)]{Derm12} Dermer C. D., Murase K., Takami H., 2012, ApJ, 755, 147
\bibitem[de Diego(2010)]{de10} de Diego J. A., 2010, AJ, 139, 1269
\bibitem[Espaillat et al.(2008)]{Espa08} Espaillat C., Bregman J., Hughes P., Lloyd-Davies E., 2008, ApJ, 679, 182
\bibitem[Esposito et al.(2015)]{Espo15} Esposito V. et al., 2015, A\&A, 576, 122
\bibitem[Fan(1995)]{Fan95} Fan J. H., 1995, Ap\&SS, 229, 157
\bibitem[Fan(1999)]{Fan99} Fan J. H., 1999, MNRAS, 308, 1032
\bibitem[Fan et al.(2009)]{Fan09} Fan J. H. et al., 2009, ApJS, 181, 466
\bibitem[Fan et al.(2014)]{Fan14} Fan J. H. et al., 2014, ApJS, 213, 26
\bibitem[Fiorucci et al.(1998)]{Fior98} Fiorucci M., Tosti G., Rizzi N., 1998, PASP, 110, 105
\bibitem[Foster(1996)]{Fost96} Foster G., 1996, AJ, 112, 1709
\bibitem[Fossati et al.(1998)]{Foss98} Fossati G., Maraschi L., Celotti A., Comastri A., Ghisellini G., 1998, MNRAS, 299, 433
\bibitem[Gaur et al.(2010)]{Gaur10} Gaur H., Gupta A. C., Lachowicz P., Wiita P. J., 2010, ApJ, 718, 279
\bibitem[Gaur et al.(2012)]{Gaur12} Gaur H. et al., 2012, MNRAS, 420, 3147
\bibitem[Ghisellini et al.(2010)]{Ghis10}Ghisellini G. et al., 2010, MNRAS, 402, 497
\bibitem[Giveon et al.(1999)]{Give99} Giveon U., Maoz D., Kaspi S., Netzer H., Smith P. S., 1999, MNRAS, 306, 637
\bibitem[Gu et al.(2006)]{Gu06} Gu M. F., Lee C. U., Pak S., Yim H. S., Fletcher A. B., 2006, A\&A, 450, 39
\bibitem[Gu \& Ai(2011)]{Gu11} Gu M. F., Ai Y. L., 2011, A\&A, 528, 95
\bibitem[Guo et al.(2014)]{Guo14} Guo D., Hu S., Chen X., 2014, J. Astrophys. Astron., 35, 283
\bibitem[Gupta et al.(2008a)]{Gupt08a} Gupta A. C., Deng W. G., Joshi U. C., Bai J. M., Lee M. G., 2008a, New Astron., 13, 375
\bibitem[Gupta et al.(2008b)]{Gupt08b} Gupta A. C., Fan J. H., Bai J. M., Wagner S. J., 2008b, AJ, 135, 1384
\bibitem[Gupta et al.(2009)]{Gupt09} Gupta A. C., Srivastava A. K., Wiita P. J., 2009, ApJ, 690, 216
\bibitem[Gupta et al.(2019)]{Gupt19} Gupta A. C. et al., 2019, MNRAS, 484, 5785
\bibitem[Heeschen et al.(1987)]{Hees87} Heeschen D. S., Krichbaum T., Schalinski C. J., Witzel A., 1987, AJ, 94, 1493
\bibitem[Heidt \& Wagner(1996)]{Heid96} Heidt J., Wagner S. J., 1996, A\&A, 305, 42
\bibitem[Heidt \& Wagner(1998)]{Heid98} Heidt J., Wagner S. J., 1998, A\&A, 329, 853
\bibitem[Hong et al.(2018)]{Hong18} Hong S., Xiong D., Bai J., 2018, AJ, 155, 31
\bibitem[Hu et al.(2014)]{Hu14} Hu S. M., Chen X., Guo D. F., Jiang Y. G., Li K., 2014, MNRAS, 443, 2940
\bibitem[Isler et al.(2017)]{Isle17} Isler J. C. et al., 2017, ApJ, 844, 107
\bibitem[Joshi et al.(2011)]{Josh11} Joshi R., Chand H., Gupta A. C., Wiita P. J., 2011, MNRAS, 412, 2717
\bibitem[Katajainen et al.(2000)]{Kata00} Katajainen S. et al., 2000, A\&AS, 143, 357
\bibitem[Kataoka et al.(2008)]{Kata08} Kataoka J. et al., 2008, ApJ, 672, 787
\bibitem[Kinman(1976)]{Kinm76} Kinman T. D., 1976, ApJ, 205, 1
\bibitem[Kurtanidze et al.(2009)]{Kurt09} Kurtanidze O. M. et al., 2009, in Wang W. M., Yang Z. Q., Luo Z. J., Chen, Z., eds, ASP Conf. Ser. Vol. 408, The Starburst-AGN Connection. Astron. Soc. Pac., San Francisco, p.266
\bibitem[Lahteenmaki \& Valtaoja(1999)]{Laht99} Lahteenmaki A., Valtaoja E., 1999, ApJ, 521, 493
\bibitem[Lachowicz et al.(2009)]{Lach09} Lachowicz P., Gupta A. C., Gaur H., Wiita P. J., 2009, A\&A, 506, L17
\bibitem[Leonardo et al.(2009)]{Leon09} Leonardo E., et al., 2009, arXiv:0907.0959
\bibitem[Li et al.(2016)] {Li16} Li H. Z., Jiang Y. G., Guo D. F., Chen X., Yi T. F., 2016, PASP, 128, 074101
\bibitem[Li et al.(2021a)]{Li21a} Li T., Wu J. H., Meng N. K., Dai Y., Zhang X. Y., 2021a, RAA, 21, 259
\bibitem[Li et al.(2021b)]{Li21b} Li X. P. et al., 2021b, JApA, 42, 92
\bibitem[Liu et al.(2008)]{Liu08} Liu H. T., Bai J. M., Zhao X. H., Ma L., 2008, ApJ, 677, 884
\bibitem[Liu \& Bai(2015)]{Liu15} Liu H. T., Bai J. M., 2015, AJ, 149, 191
\bibitem[Liao et al.(2014)]{Liao14} Liao N. H. et al., 2014, ApJ, 783, 83
\bibitem[Lomb(1976)]{Lomb76} Lomb N. R., 1976, Ap\&SS, 39, 447
\bibitem[Mangalam \& Wiita(1993)]{Mang93} Mangalam A. V., Wiita P. J., 1993, ApJ, 406, 420
\bibitem[Massaro et al.(1999)]{Mass99} Massaro E. et al., 1999, A\&A, 342, 49
\bibitem[Massaro et al.(2003)]{Mass03} Massaro E. et al., 2003, A\&A, 399, 33
\bibitem[Marcha et al.(1996)]{Marc96} Marcha M. J. M., Browne I. W. A., Impey C. D., Smith P. S., 1996, MNRAS, 281, 425
\bibitem[Marscher et al.(2008)]{Mars08} Marscher A. P. et al., 2008, Nature, 452, 966
\bibitem[Marscher \& Gear(1985)]{Mars85} Marscher A. P., Gear W. K., 1985, ApJ, 298, 114
\bibitem[Maraschi \& Tavecchio(2003)]{Mara03} Maraschi L., Tavecchio F., 2003, ApJ, 593, 667
\bibitem[Miller \& Hawley(1977)]{Mill77} Miller J. S., Hawley S. A., 1977, ApJ, 212, 47
\bibitem[Miller \& Noble(1996)]{Mill96} Miller H. R., Noble J. C., 1996, in Miller, H. R., Webb, J. R., \& Noble, J. C., eds, ASP Conf. Ser. Vol. 110, Blazar Continuum Variability. Astron. Soc. Pac., San Francisco, p.17
\bibitem[Miller et al.(1989)]{Mill89} Miller H. R., Carini M. T., Goodrich B. D., 1989, Nature, 337, 627
\bibitem[Oke(1978)]{Oke78} Oke J. B., 1978, ApJ, 219, L97
\bibitem[Pica et al.(1988)]{Pica88} Pica A. J. et al., 1988, AJ, 96, 1215
\bibitem[Pollock et al.(1979)]{Poll79} Pollock J. T. et al., 1979, AJ, 84, 1658
\bibitem[Press et al.(1992)]{Pres92} Press W. H., Teukolsky S. A., Vetterling W. T., Flannery B. P., 1992, Numeric. Recipies in FORTRAN (2nd ed.; Cambridge: Cambridge Univ. Press)
\bibitem[Qian \& Tao(2003)]{Qian03} Qian B. C., Tao J., 2003, PASP, 115, 490
\bibitem[Rani et al.(2010)]{Rani10} Rani B., Gupta A. C., Joshi U. C., Ganesh S., Wiita P. J., 2010, ApJ, 719, L153
\bibitem[Rani et al.(2011)]{Rani11} Rani B., Gupta A. C., Joshi U. C., Ganesh S., Wiita P. J., 2011, MNRAS, 413, 2157
\bibitem[Ren et al.(2021)]{Ren21} Ren G. W. et al., 2021, MNRAS, 506, 3791
\bibitem[Romero et al.(2000)]{Rome00} Romero G. E., Chajet L., Abraham Z., Fan. J. H., 2000, A\&A, 360, 57
\bibitem[Rubio-Herrera \& Lee(2005)]{Rubi05} Rubio-Herrera E., Lee W. H., 2005, MNRAS, 357, 31
\bibitem[Sandrinelli et al.(2016)]{Sand16} Sandrinelli A., Covino S., Dotti M., Treves A., 2016, AJ, 151, 54
\bibitem[Sagar et al.(2004)]{Saga04} Sagar R. et al., 2004, MNRAS, 348, 176
\bibitem[Scargle(1982)]{Scar82} Scargle J. D., 1982, ApJ, 263, 835
\bibitem[Schulz \& Mudelsee(2002)]{Schu02} Schulz M., Mudelsee M., 2002, CG, 28, 421
\bibitem[Sillanpaa et al.(1988)]{Sill88} Sillanpaa A., Haarala S., Valtonen M. J., Sundelius B., Byrd G. G., 1988, ApJ, 325, 628
\bibitem[Stocke et al.(1991)]{Stoc91} Stocke J. T., Case J., Donahue M., Shull J. M., Snow T. P., 1991, ApJ, 374, 72
\bibitem[Takalo \& Sillanpaa(1989)]{Taka89} Takalo L. O., Sillanpaa A., 1989, A\&A, 218, 45
\bibitem[Timmer \& Koenig(1995)]{Timm95}Timmer J., Koenig M., 1995, A\&A, 300, 707
\bibitem[Tosti et al.(1998)]{Tost98} Tosti G. et al., 1998, A\&AS, 130, 109
\bibitem[Urry \& Padovani(1995)]{Urry95} Urry C. M., Padovani P., 1995, PASP, 107, 803
\bibitem[Urry et al.(1993)]{Urry93} Urry C. M. et al., 1993, ApJ, 411, 614
\bibitem[Uttley et al.(2002)]{Uttl02} Uttley P., McHardy I. M., Papadakis I. E., 2002, MNRAS, 332, 231
\bibitem[Vagnetti et al.(2003)]{Vagn03} Vagnetti F., Trevese D., Nesci R., 2003, ApJ, 590, 123
\bibitem[Valtonen et al.(2016)]{Valt16} Valtonen M. et al., 2016, DDA meeting, American Astronomical Society, id.302.06
\bibitem[Vaughan(2005)]{Vaug05} Vaughan S., 2005, A\&A, 431, 391
\bibitem[Villforth et al.(2010)]{Vill10} Villforth C. et al., 2010, MNRAS, 402, 2087
\bibitem[Villata et al.(1998)]{Vill98} Villata M., Raiteri C. M., Sillanpaa A., Takalo L. O., 1998, MNRAS, 293, L13
\bibitem[Villata \& Raiteri(1999)]{Vill99} Villata M., Raiteri C. M., 1999, A\&A, 347, 30
\bibitem[Wagner \& Witzel(1995)]{Wagn95} Wagner S. J., Witzel A., 1995, ARA\&A, 33, 163
\bibitem[Wang et al.(2014)]{Wang14} Wang J. Y., An T., Baan W. A., Lu X. L., 2014, MNRAS, 443, 58
\bibitem[Webb et al.(1990)]{Webb90} Webb J. R. et al., 1990, AJ, 100, 1452
\bibitem[Weistrop(1973)]{Weis73} Weistrop D., 1973, Nature Physical Science, 241, 157
\bibitem[Weistrop et al.(1985)]{Weis85} Weistrop D., Shaffer D. B., Hintzen P., Romanishin W., 1985, ApJ, 292, 614
\bibitem[Wiita(2006)]{Wiit06} Wiita P. J., 2006, arXiv:astro-ph/0603728
\bibitem[Woo \& Urry(2002)]{Woo02} Woo J. H., Urry C. M., 2002, ApJ, 581, L5
\bibitem[Wu et al. (2005)]{Wu05} Wu J. et al., 2005, MNRAS, 361, 155
\bibitem[Wu et al. (2006)]{Wu06} Wu J. et al., 2006, AJ, 132, 1256
\bibitem[Wu et al. (2009)]{Wu09} Wu Z. Z., Gu M. F., Jiang D. R., 2009, RAA, 9, 168
\bibitem[Xie et al.(1999)]{Xie99} Xie G. Z., Li K. H., Zhang X., Bai J. M., Liu W. W., 1999, ApJ, 522, 846
\bibitem[Xie et al.(2001)]{Xie01} Xie G. Z. et al., 2001, ApJ, 548, 200
\bibitem[Xie et al.(2002)]{Xie02} Xie G. Z. et al., 2002, MNRAS, 334, 459
\bibitem[Xiong et al.(2016)]{Xion16} Xiong D. et al., 2016, ApJS, 222, 24
\bibitem[Xiong et al.(2017)]{Xion17} Xiong D. et al., 2017, ApJS, 229, 21
\bibitem[Xiong et al.(2020)]{Xion20} Xiong D. et al., 2020, ApJS, 247, 49
\bibitem[Yang et al.(2020)]{Yang20} Yang X. et al., 2020, PASP, 132, 044101
\bibitem[Zhang et al.(2004)]{Zhan04} Zhang X. et al., 2004, AJ, 128, 1929
\bibitem[Zwicky(1966)]{Zwic66} Zwicky F., 1966, ApJ, 143, 192

\end{thebibliography}
\end{document}